\documentclass[12pt,english,amsmath,amssymb]{article}
\usepackage[latin9]{inputenc}
\usepackage[letterpaper]{geometry}
\usepackage{amsmath}
\usepackage{amssymb}
\usepackage{graphicx}
\usepackage{esint}
\usepackage[authoryear]{natbib}

\makeatletter
\newcommand{\lyxaddress}[1]{
\par {\raggedright #1
\vspace{1.4em}
\noindent\par}
}

\newcommand\Real{\mbox{Re}} 
\newcommand\Imag{\mbox{Im}} 

\newcommand{\Ma}{\text{Ma}}
\newcommand{\Bo}{\text{Bo}}
\usepackage{babel}
\usepackage{esint}
\usepackage{siunitx}
\@addtoreset{equation}{section}
\numberwithin{equation}{section}

\makeatother

\usepackage{babel}
\begin{document}

\title{Surfactant and gravity dependent inertialess instability of two-layer
Couette flows and its nonlinear saturation}

\author{Alexander L. Frenkel\textsuperscript{1} and David Halpern\textsuperscript{1}}
\maketitle

\lyxaddress{\textsuperscript{1}Department of Mathematics, University of Alabama,
Tuscaloosa AL 35487, USA}

\section*{Abstract}

A horizontal channel flow of two immiscible fluid layers with different
densities, viscosities and thicknesses, subject to vertical gravitational
forces and with an insoluble surfactant monolayer present at the interface,
is investigated. The base Couette flow is driven by the uniform horizontal
motion of the channel walls. Linear and nonlinear stages of the (inertialess)
surfactant and gravity dependent long-wave instability are studied
using the lubrication approximation, which leads to a system of coupled
nonlinear evolution equations for the interface and surfactant disturbances.
The (inertialess) instability is a combined result of the surfactant
action characterized by the Marangoni number $\text{Ma}$ and the
gravitational effect corresponding to the Bond number $\text{Bo}$
that ranges from $-\infty$ to $\infty$. The other parameters are
the top-to-bottom thickness ratio $n$, which is restricted to $n\ge1$
by a reference frame choice, the top-to-bottom viscosity ratio $m$,
and the base shear rate $s$.

The linear stability is determined by an eigenvalue problem for the
normal modes, where the complex eigenvalues (determining growth rates
and phase velocities) and eigenfunctions (the amplitudes of disturbances
of the interface, surfactant, velocities, and pressures) are found
analytically by using the smallness of the wavenumber. For each wavenumber,
there are two active normal modes, called the surfactant and the robust
modes. The robust mode is unstable when $\text{Bo}/\text{Ma}$ falls
below a certain value dependent on $m$ and $n$. The surfactant branch
has instability for $m<1$, and any $\text{Bo}$, although the range
of unstable wavenumbers decreases as the stabilizing effect of gravity
represented by $\text{Bo}$ increases. Thus, for certain parametric
ranges, even arbitrarily strong gravity cannot completely stabilize
the flow.

The correlations of vorticity-thickness phase differences with instability,
present when gravitational effects are neglected, are found to break
down when gravity is important. The physical mechanisms of instability
for the two modes are explained with vorticity playing no role in
them. This is in marked contrast to the dynamical role of vorticity
in the mechanism of the well-known Yih instability due to effects
of inertia, and is contrary to some earlier literature.

Unlike the semi-infinite case that we previously studied, a small-amplitude
saturation of the surfactant instability is possible in the absence
of gravity. For certain $(m,n)$-ranges, the interface deflection
is governed by a decoupled Kuramoto-Sivashinsky equation, which provides
a source term for a linear convection-diffusion equation governing
the surfactant concentration. When the diffusion term is negligible,
this surfactant equation has an analytic solution which is consistent
with the full numerics. Just like the interface, the surfactant wave
is chaotic, but the ratio of the two waves turns out to be constant.

\section{Introduction}

\label{sec:Intro}

Flows of fluid films occur frequently in nature and industry. (For
recent reviews, see e.g. \citet{Oron1997,CrasterMatar09}.) Instabilities
of multifluid film flows are of considerable interest (\citet{Joseph1993}).
Such instabilities can be significantly influenced by interfacial
surfactants.

Surfactants are surface active compounds that reduce the surface tension
between two fluids, or between a fluid and a solid. \citet{Frenkel2002}
(hereafter referred to as FH) and \citet{Halpern2003} (hereafter
referred to as HF) uncovered a new instability due to interfacial
surfactants: certain stable surfactant-free Stokes flows become unstable
if an interfacial surfactant is introduced. For this, the interfacial
shear of velocity must be nonzero; in particular, this instability
disappears if the basic flow is stopped. In contrast to the well-known
instability of two viscous fluids (\citet{Yih1967}) which needs inertia
effects for its existence, the new instability may exist in the absence
of fluid inertia. With regard to multi-fluid channel flows, this instability
has been further studied in such papers as \citet{Blyth2004b}, \citet{Pozrikidis2004a},
\citet{Blyth2004a}, \citet{Frenkel2005}, \citet{Wei2005c}, \citet{Frenkel2006},
\citet{Halpern2008}, \citet{Bassom2010}, \citet{JIE2010}, \citet{Kalogirou2012},
\citet{Samanta2013}, \citet{kalogirou2016} and \citet{picardo2016}.

For simplicity, consideration in FH and HF was restricted to flows
whose stability properties did not depend on gravity. The same is
true for the further studies mentioned above. The stability effects
of gravity in multi-fluid horizontal systems without surfactants were
investigated since as long ago as the fundamental work of Lord \citet{Rayleigh1900}.
 Gravity is stabilizing when the lighter fluid layer is on top of
the heavier fluid layer, or destabilizing when heavier fluid is above
the lighter fluid. The latter is the well-known Rayleigh-Taylor
instability (RTI) that has been studied extensively (see, e.g. the
classical book by \citet{Chandrasekhar1961}). Recent reviews of
the RTI and its numerous important applications are given in \citet{Kull1991}.
The combination of RTI with various viscous, inertial and nonlinear
effects in two-fluid channel flows was studied in such papers as \citet{Babchin1983a},
\citet{Hooper1985a} and \citet{Yiantsios1988}. Some industrial situations
where surfactant and gravity effects are both relevant in oil recovery
were studied e.g. in \citet{Hirasaki2004}.

In this paper, we study the interplay between the inertialess effects
due to surfactants and gravity in Couette flows of two incompressible
Newtonian liquids in a horizontal channel. Both linear and nonlinear
stability is investigated. One can expect a rich landscape of stability
properties, especially since, even in the absence of gravity, there
are two active normal modes for each wavenumber of infinitesimal disturbances,
corresponding to the two interfacial functions: the interface displacement
and the interfacial surfactant concentration (FH, HF). Their growth
rates are given by a (complex) quadratic equation, and hence in many
instances numerical results may enjoy analytic (asymptotic) corroboration.
The linear stability properties of two-layer Couette flows for arbitrary
wavelength with both interfacial surfactant and gravitational effects
were the subject of the dissertation by \citet{schweiger2013gravity},
and will be further investigated elsewhere. On the other hand, the
nonlinear lubrication approximation equations were obtained in \citet{Blyth2004b}
for the long-wave disturbances of these flows, although only zero-gravity
results were given in that paper. The linearized lubrication-approximation
approach was used also in \citet{Wei2005c} for the no-gravity case
to offer a mechanism of long-wave instability. In this paper, we re-derive,
with certain modifications, the fore-mentioned system of two nonlinear
lubrication-approximation equations coupling the interface location
and the interfacial surfactant concentration for the Couette flows
with the insoluble surfactant and gravity, provided that the characteristic
length-scale of the flow disturbances is much larger than the thicknesses
of both fluid layers. The linear system of equations coupling the
surfactant and gravity follows as the limit of (long) infinitesimal
waves. It is also of interest to determine, to the two leading orders
in the long-wave parameter, which are allowed by the lubrication approximation,
the complete set of eigenfunctions of the eigenvalue problem for the
normal modes including the velocities and pressures, and, based on
these, to clarify the mechanisms of instability for the two normal
modes. The inclusion of gravity may be expected to clarify the limitations
of the conclusions obtained by studying the flow in the absence of
gravity and to observe new linear and nonlinear effects.

Concerning the linear stability, in the present paper, we concentrate
on the parametric thresholds of instability. The latter turn out to
be determined by the leading-order of the small wavenumber expansion,
which allows neglecting the higher order capillary effects. However,
we include these effects in investigating the nonlinear stages of
the instability. A natural question concerning the interaction of
gravity and the surfactants is whether sufficiently strong gravitational
forces can always suppress the linear instability caused by surfactants.
On the other hand, one can ask if surfactants can suppress the Rayleigh-Taylor
instability. These questions are answered below.

The nonlinear saturation of the surfactant instability was studied
before for the case of one layer being infinitely thick, and it was
shown that it is impossible to have the saturated amplitudes small
for both surfactant and interface displacement (\citet{Frenkel2006}).
For the finite thickness ratio, the limited nonlinear simulations
in \citet{Blyth2004b} featured the same property, but the question
remained if it holds in all cases. We investigate this below in a
more systematic way (which shows that both surfactant disturbance
and interface displacement can be small in some saturated regimes).

The paper is organized as follows. In section \ref{sec:GovEqnsStabForm},
the general stability problem is formulated. In section \ref{sec:Lubrication-approximation},
the nonlinear and linear systems of governing equations are obtained
using the lubrication approximation approach. The long-wave growth
rates and instability thresholds are considered in section \ref{sec:Increments, growth rates}.
In section \ref{sec:phase differences} we study the surfactant-thickness
and vorticity-thickness phase differences in connection with their
purported significance for (in)stability. In section \ref{sec:mechanical},
we uncover the physical mechanisms of instability for the different
branches of normal modes. Also, the eigenfunctions of the normal modes
are discussed in connection with deriving growth rates and wave velocities
using the integral form of conservation laws of liquid and surfactant.
In section \ref{sec:nonlinear}, the nonlinear evolution of disturbances
is studied, including weakly, almost-weakly, and strongly nonlinear
regimes. Finally, section \ref{sec:Conclusions} contains summary,
discussion, and concluding remarks. Some more technical information
is delegated to Appendix \ref{sec:coefficient}, while Appendix \ref{sec:Augmented-lubrication-theory}
is concerned with the next approximation refining the lubrication
theory. Appendix \ref{sec:Eigenfunctions} gives the complete collection
of the normal-mode eigenfunctions.

\section{General problem framework}

\label{sec:GovEqnsStabForm}

\begin{figure}
\centering{}\includegraphics[width=0.95\textwidth]{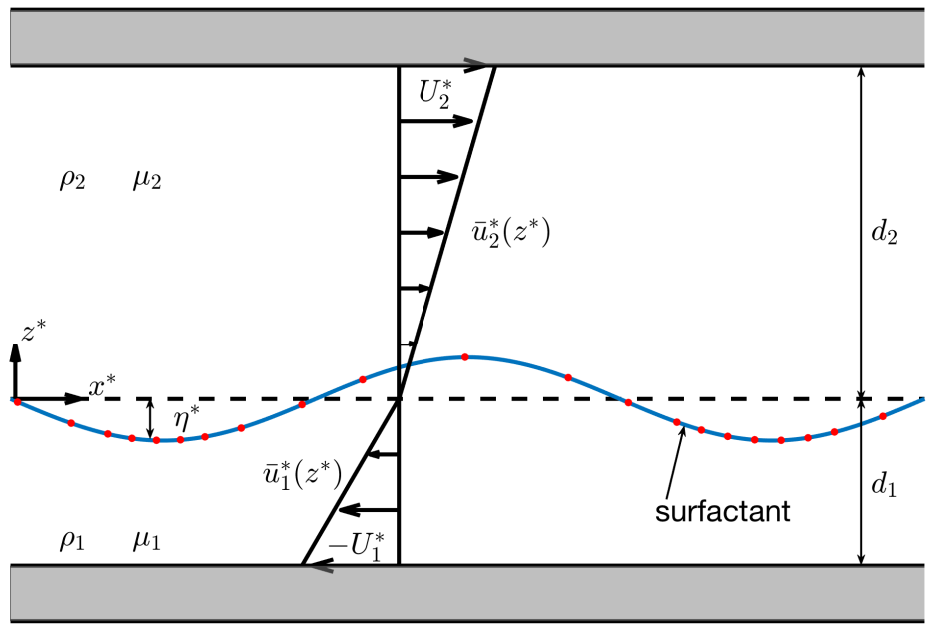}\protect\protect\caption{Sketch of a disturbed two-layer Couette flow of two horizontal liquid
layers with different thicknesses, viscosities, and mass densities.
The insoluble surfactant monolayer is located at the interface and
is indicated by the dots. The (spanwise) uniform gravity field with
a constant acceleration $g$ is not shown. \label{fig:FigDefinitionSketch}
\ }
\end{figure}

The formulation used in this paper is similar to that of HF; however,
gravitational effects, which were absent in HF, play an active role
here. Two immiscible fluid layers with different densities, viscosities
and thicknesses are bounded by two infinite horizontal plates, a distance
$d=d_{1}+d_{2}$ apart, with the top plate moving at a constant relative
velocity $U^{*}$, as shown in figure \ref{fig:FigDefinitionSketch}.
The vertical coordinate is denoted $z^{\ast}$, and we choose $z^{\ast}=0$
at the base liquid-liquid interface. (We use the symbol $^{\ast}$
to indicate a dimensional quantity.) The top plate is located at $z^{\ast}=d_{2}$
and the bottom plate is located at $z^{\ast}=-d_{1}$. The horizontal\ $x^{\ast}$-axis
is streamwise. At the interface, the surface tension, $\sigma^{\ast}$,
depends on the concentration of the insoluble surfactant monolayer,
$\Gamma^{\ast}$. The basic flow is driven by the steady motion of
the top plate. If the frame of reference is fixed at the liquid-liquid
interface, the velocity of the bottom plate is denoted $-U_{1}^{*}$,
and that of the top plate is $U_{2}^{*}$, then clearly $U_{1}^{*}+U_{2}^{*}=U^{*}$. In
the base state, the horizontal velocity profiles are linear in $z^{\ast}$,
the interface is flat, and the surfactant concentration is uniform.
Once disturbed, the surfactant concentration, $\Gamma^{*}(x^{*},\;t^{*})$,
is no longer uniform, and there is a varying deflection of the interface,
$\eta^{\ast}(x^{\ast},t^{\ast})$, where $t^{\ast}$ is time.

The continuity equation and the Navier-Stokes momentum equations govern
the fluid motion in the two layers (with $j=1$ for the bottom liquid
layer and $j=2$ for the top liquid layer). \ They are 
\begin{equation}
\quad\nabla^{\ast}\cdot\boldsymbol{v}_{j}^{\ast}=0\text{,}\label{oneCon}
\end{equation}
\begin{equation}
\rho_{j}\left(\frac{\partial\boldsymbol{v}_{j}^{\ast}}{\partial t^{\ast}}+\boldsymbol{v}_{j}^{\ast}\cdot\nabla^{\ast}\boldsymbol{v}_{j}^{\ast}\right)=-\nabla^{\ast}p_{j}^{\ast}+\mu_{j}\nabla^{\ast2}\boldsymbol{v}_{j}^{\ast}-\rho_{j}g\mathbf{\hat{z}}\text{,}\label{twoNS}
\end{equation}
where $\rho_{j}$ is the density, $\boldsymbol{v}_{j}^{\ast}=(u_{j}^{\ast},w_{j}^{\ast})$
is the fluid velocity vector with horizontal component $u_{j}^{\ast}$
and vertical component $w_{j}^{\ast}$, the operator vector $\nabla^{\ast}=(\partial/\partial x^{\ast},\partial/\partial z^{\ast})$,
$p_{j}^{\ast}$ is the pressure, $\mu_{j}$ is the viscosity, $g$
is the gravity acceleration, and $\mathbf{\hat{z}}$ is the unit vector
in the upward $z^{\ast}$ direction. \ 

At the plates, $z^{\ast}=-d_{1}$ and $z^{\ast}=d_{2}$, the no-slip
boundary conditions require 
\begin{equation}
u_{1}^{\ast}\left(-d_{1}\right)=-U_{1}^{\ast}\text{, }w_{1}^{\ast}(-d_{1})=0\text{, }u_{2}^{\ast}(d_{2})=U_{2}^{\ast}\text{, }w_{2}^{\ast}(d_{2})=0\text{.}\label{threeBCplate}
\end{equation}
There are several boundary conditions at the interface. \ The velocities
of the layers must be equal: 
\begin{equation}
\lbrack\boldsymbol{v}^{\ast}]_{1}^{2}=0.\label{fourVelCon}
\end{equation}
Taking into account the spatial variation of surface tension and the
capillary jump between the viscous normal stresses, the balances of
the tangential and normal stresses are, respectively, 
\begin{equation}
\frac{1}{1+\eta_{x^{\ast}}^{\ast2}}\left[(1-\eta_{x^{\ast}}^{\ast2})\mu(u_{z^{\ast}}^{\ast}+w_{x^{\ast}}^{\ast})+2\eta_{x^{\ast}}^{\ast}\mu(w_{z^{\ast}}^{\ast}-u_{x^{\ast}}^{\ast})\right]_{1}^{2}=-\frac{\sigma_{x^{\ast}}^{\ast}}{(1+\eta_{x^{\ast}}^{\ast2})^{1/2}}\text{,}\label{fiveTan}
\end{equation}
\begin{equation}
\lbrack(1+\eta_{x^{\ast}}^{\ast2})p^{\ast}-2\mu(\eta_{x^{\ast}}^{\ast2}u_{x^{\ast}}^{\ast}-\eta_{x^{\ast}}^{\ast}(u_{z^{\ast}}^{\ast}+w_{x^{\ast}}^{\ast})+w_{z^{\ast}}^{\ast})]_{1}^{2}=\frac{\eta_{x^{\ast}x^{\ast}}^{\ast}}{(1+\eta_{x^{\ast}}^{\ast2})^{1/2}}\sigma^{\ast}\text{,}\label{sixNorm}
\end{equation}
where the subscripts indicate partial differentiation. \ The kinematic
interfacial condition is 
\begin{equation}
\eta_{t^{\ast}}^{\ast}=w^{\ast}-u^{\ast}\eta_{x^{\ast}}^{\ast}\text{.}\label{sevenKin}
\end{equation}
The surface concentration of the insoluble surfactant on the interface,
$\Gamma^{\ast}(x^{\ast},t^{\ast})$, obeys the transport equation
\begin{equation}
\frac{\partial}{\partial t^{\ast}}(H^{\ast}\Gamma^{\ast})+\frac{\partial}{\partial x^{\ast}}\left(H^{\ast}\Gamma^{\ast}u^{\ast}\right)=D_{\Gamma}\frac{\partial}{\partial x^{\ast}}\left(\frac{1}{H^{\ast}}\frac{\partial\Gamma^{\ast}}{\partial x^{\ast}}\right)\text{,}\label{eightSurf}
\end{equation}
where $H^{\ast}=(1+\eta_{x^{\ast}}^{\ast2})^{1/2}$, $u^{\ast}=u^{\ast}(x^{\ast},\eta^{\ast}(x^{\ast},t^{\ast}),t^{\ast})$,
and $D_{\Gamma}$ is the surface molecular diffusivity of the insoluble
surfactant (\citet{Halpern2003}). We assume that the dependence of
surface tension on the surfactant concentration given by the well-known
Langmuir isotherm relation (Edwards et al. (1991)) which becomes the
linear Gibbs isotherm when the surfactant concentration $\Gamma^{*}$
is much smaller than the maximum packing value $\Gamma_{\infty}$.
Then we can write 
\begin{equation}
\sigma^{\ast}=\sigma_{0}-RT(\Gamma^{\ast}-\Gamma_{0})\text{,}\label{nineSigmaStar}
\end{equation}
where $\sigma_{0}$ is the base surface tension, $R$ is the universal
gas constant and $T$ is the absolute temperature.

These governing equations are invariant under a transformation corresponding
to looking at the flow in the ``upside down'' way, reversing the
direction of the $z$-axis (and of the $x$-axis as well). This is
discussed in detail below in \ref{sec:Lubrication-approximation}.

We introduce the following dimensionless variables: 
\[
(x,z,\eta)=\frac{(x^{\ast},z^{\ast},\eta^{\ast})}{d_{1}}\text{, }t=\frac{t^{\ast}}{d_{1}\mu_{1}/\sigma_{0}}\text{, }\boldsymbol{v}_{j}=(u_{j},w_{j})=\frac{(u_{j}^{\ast},w_{j}^{\ast})}{\sigma_{0}/\mu_{1}}\text{,}
\]
\begin{equation}
\text{ }p_{j}=\frac{p_{j}^{\ast}}{\sigma_{0}/d_{1}}\,\text{, }\Gamma=\frac{\Gamma^{\ast}}{\Gamma_{0}}\text{, }\sigma=\frac{\sigma^{\ast}}{\sigma_{0}}\text{.}\label{scales}
\end{equation}
(Similar to FH and HF, using the velocity scale $\sigma_{0}/\mu_{1}$,
rather than the plate speed, allows one to include into consideration
the case of zero plate velocity corresponding to the absence of base
flow.) The continuity equation and the Navier-Stokes momentum equations
are, respectively, 
\begin{equation}
\quad\nabla\cdot\boldsymbol{v}_{j}=0\text{,}\label{dCon}
\end{equation}
\begin{equation}
\frac{Re_{j}}{Ca_{j}}\left(\frac{\partial\boldsymbol{v}_{j}}{\partial t}+\boldsymbol{v}_{j}\cdot\nabla\boldsymbol{v}_{j}\right)=-\nabla p_{j}+m_{j}\nabla^{2}\boldsymbol{v}_{j}-\text{Bo}_{j}\mathbf{\hat{z}}\text{,}\label{dNavSto}
\end{equation}
where the vector operator $\nabla:=(\partial/\partial x,\partial/\partial z),$
$\text{Re}_{j}:=U_{j}^{*}d_{1}/\mu_{1}$ is the Reynolds number, $Ca_{j}:=U_{j}^{\ast}\mu_{1}/\sigma_{0}$
is the capillary number, $m_{j}:=\mu_{j}/\mu_{1}$ - where-from $m_{1}=1$
and $m_{2}=\mu_{2}/\mu_{1}=:m$ is the viscosity ratio, $\text{Bo}_{j}:=\rho_{j}gd_{1}^{2}/\sigma_{0}$
is the layer Bond number, and $\mathbf{\hat{z}}$ is the unit vector
of the $z$-axis. The plate boundary conditions are 
\begin{equation}
u_{1}\left(-1\right)=-Ca_{1}\text{, }w_{1}(-1)=0\text{, }u_{2}(n)=Ca_{2}\text{, }w_{2}(n)=0,\label{PlatesNoSlipNoPen}
\end{equation}
where $n=$\ $d_{2}/d_{1}$ is the thickness ratio of the liquid
layers. \ Without loss of generality, by appropriately directing
the $z$-axis, we obtain $n\geq1$. Note that this allows for negative
as well as positive values of $g$. The interfacial conditions for
the velocities, the tangential stresses, and the normal stresses are,
respectively, 
\begin{equation}
\lbrack\boldsymbol{v}]_{1}^{2}=0\text{,}\label{ContinuousVatInterfaceBC}
\end{equation}
\begin{equation}
\frac{1}{1+\eta_{x}^{2}}\left[(1-\eta_{x}^{2})\frac{\mu}{\mu_{1}}(u_{z}+w_{x})+2\eta_{x}\frac{\mu}{\mu_{1}}(w_{z}-u_{x})\right]_{1}^{2}=-\frac{\sigma_{x}}{(1+\eta_{x}^{2})^{1/2}}\text{,}\label{TanStressBC}
\end{equation}
and 
\begin{equation}
\lbrack(1+\eta_{x}^{2})p-2\frac{\mu}{\mu_{1}}(\eta_{x}^{2}u_{x}-\eta_{x}(u_{z}+w_{x})+w_{z})]_{1}^{2}=\frac{\eta_{xx}}{(1+\eta_{x}^{2})^{1/2}}\sigma,\label{NrmStressBC}
\end{equation}
where $[A]_{1}^{2}=A_{2}-A_{1}$ denotes the jump in $A$ across the
interface $z=\eta(t,x)$. The surfactant transport equation is (see
HF) 
\begin{equation}
\frac{\partial}{\partial t}(H\Gamma)+\frac{\partial}{\partial x}\left(H\Gamma u\right)=\frac{1}{\text{Pe}}\frac{\partial}{\partial x}\left(\frac{1}{H}\frac{\partial\Gamma}{\partial x}\right)\text{,}\label{dSurfConc}
\end{equation}
where $H=(1+\eta_{x}^{2})^{1/2}$, $u=u(t,x,\eta(t,x))$ and $\textrm{Pe}^{-1}$
$=D_{\Gamma}\mu_{1}/\sigma_{0}d_{1}$ is the inverse surface Péclet
number, the dimensionless representation of the surface molecular
diffusivity $D_{\Gamma}$ of the insoluble surfactant. Usually, the
latter is small and the surfactant diffusion term is negligible.\ The
kinematic boundary condition is 
\begin{equation}
\eta_{t}=w-u\eta_{x}\text{,}\label{dKineBC}
\end{equation}
and the dimensionless form of the equation of state for the surface
tension, (\ref{nineSigmaStar}), is 
\begin{equation}
\sigma=1-\text{Ma}(\Gamma-1)\text{.}\label{sigmastar}
\end{equation}
where $\text{Ma}:=RT\Gamma_{0}/\sigma_{0}$ is the Marangoni number.
It is easy to see that the Marangoni number can be written as $\text{Ma}=(\sigma_{c}-\sigma_{0})/\sigma_{0}$,
where $\sigma_{c}$ is the surface tension in the absence of surfactant.
Usually $\sigma_{c}-\sigma_{0}\ll\sigma_{0}$ since we are restricted
to the linear part of the isotherm (see, for e.g, figure 2 in \citet{Mensire2016}).
This implies the range of Marangoni numbers to be 
\begin{equation}
0<\text{Ma}\ll1.\label{eq:Marangoni range}
\end{equation}
The dimensionless velocity field of the basic Couette flow, with a
flat interface, $\eta=0$, uniform surface tension, $\bar{\sigma}=1$,
and corresponding surfactant concentration, $\bar{\Gamma}=1$ (where
the over-bar indicates a base quantity), is 
\begin{equation}
\bar{u}_{1}(z)=sz,\;\bar{w}_{1}=0\text{, }\text{and }\bar{p}_{1}=-\text{Bo}_{1}z\text{ \ \ for \ }-1\leq z\leq0\text{,}\label{u1w1p1BSprofiles}
\end{equation}
\begin{equation}
\bar{u_{2}}(z)=\frac{s}{m}z,\;\bar{w}_{2}=0\text{, and }\bar{p}_{2}=-\text{Bo}_{2}z\text{ \ \ for \ }0\leq z\leq n\text{.}\label{u2w2p2BSprofiles}
\end{equation}
The constant $s$ is used to characterize the flow in place of the
relative velocity of the plates, and represents the base interfacial
shear rate of the bottom layer, $s=D\bar{u}_{1}(0)$, where $D=d/dz$.\ Clearly,
$s=Ca_{1}$, while $Ca_{2}=sn/m$, and thus $\mu_{1}U^{*}/\sigma_{0}=s(1+n/m)$.
To estimate the range of $s$, note that for $\sigma_{0}\sim10$ (in
cgs units), fairly large viscosity $\mu_{1}\sim10$ , and $U_{1}^{*}\sim1$,
we obtain $s\sim1$. This implies that in practice 
\begin{equation}
0\le s\le1.\label{eq:s range}
\end{equation}

The disturbed state with small deviations (indicated by the top tilde,
$^{\sim}$) from the base flow is given by 
\begin{equation}
\eta=\tilde{\eta}\text{,}\ u_{j}=\bar{u}_{j}+{\tilde{u}}_{j}\text{, }w_{j}=\tilde{w}_{j}\text{, }p_{j}=\bar{p}_{j}+\tilde{p}_{j}\text{, }\Gamma=\bar{\Gamma}+\tilde{\Gamma}\text{.}\label{transformations}
\end{equation}

\section{Lubrication approximation\label{sec:Lubrication-approximation}}

We will use the lubrication approximation, assuming that the characteristic
horizontal length-scale $L$ of the disturbances is much larger than
the thicknesses of both layers. The equations were derived before
in \citet{Blyth2004b} for an inclined channel. We find it convenient
to briefly re-derive them for our horizontal-channel case and somewhat
different coordinate and non-dimensionalization choices.

It is well known that in this approximation the pressure disturbances
are independent of the vertical coordinate, and the horizontal velocities
satisfy the second order differential equation 
\begin{equation}
D^{2}u{}_{j}=\frac{1}{m_{j}}p_{jx},\label{eq:uj momentum}
\end{equation}
where we have dropped the tildes in the notations for the disturbances
(for the sake of brevity). (As will become clear later, this equation
combines the orders $1/L$ and $1/L^{2}$ (corresponding to its real
and imaginary parts) relative to the interface displacement $\eta$
(See Appendix \ref{sec:Eigenfunctions}). The general solution satisfying
the no-slip conditions at the plates is 
\begin{equation}
u_{j}=\frac{1}{2m_{j}}p_{jx}(z^{2}-n_{j}^{2})+A_{j}(z-n_{j})\label{eq:uj vel}
\end{equation}
where the functions $A_{j}$ are independent of $z$ and may be interpreted
as vorticity components; they will be determined later on. In this
formula and below, by definition, $n_{j}$ has the values $n_{1}=-1$
and $n_{2}=n$. The vertical velocity disturbance is determined by
the continuity equation (\ref{dCon}) 
\begin{equation}
Dw_{j}=-u_{jx}.\label{eq:incompress}
\end{equation}
The general solutions satisfying the zero velocity conditions at the
plates are then 
\begin{equation}
w_{j}=\frac{1}{6m_{j}}(-z^{3}+3n_{j}^{2}z-2n_{j}^{3})p_{jxx}-\frac{1}{2}\left(z-n_{j}\right)^{2}A_{jx}.\label{eq:wj vel}
\end{equation}
The normal stress condition (\ref{NrmStressBC}) yields 
\begin{equation}
\Pi[\eta,\Gamma]:=p_{1}-p_{2}=\text{Bo}\eta-\sigma\eta_{xx}\label{eq:Pi}
\end{equation}
where 
\begin{equation}
\text{Bo}:=\frac{(\varrho_{1}-\varrho_{2})gd_{1}^{2}}{\sigma_{0}}\label{eq:Bond number definition}
\end{equation}
is the Bond number (equal to the difference of the Bond numbers of
the layers, $\text{Bo}_{1}-\text{Bo}_{2}$), and we write $\sigma$
in the form $\sigma=1-\text{Ma}\Gamma$, where $\Gamma:=\tilde{\Gamma},$
the disturbance of the surfactant concentration. Note that clearly
a positive $\text{Bo }$ corresponds to a gravity force acting in
the direction from the lighter to the heavier fluid, and the negative
$\text{Bo }$ corresponds to the opposite direction of the gravity
forces. In the latter configuration, gravity has a destabilizing effect
corresponding to the Rayleigh-Taylor instability. To estimate the
range of the Bond number, for the Earth's gravity, $g\approx10^{3}$
(in cgs units), $\rho_{1}\sim\rho_{2}\sim1$, $\sigma_{0}\sim10$,
we obtain $|\text{Bo}|\sim10^{2}$ for $d_{1}\sim1$ and $\text{Bo}\sim1$
for $d_{1}\sim10^{-1}$. Also, $\text{Bo}\ll1$ for small density
contrasts, \textbar{}$\rho_{1}-\rho_{2}|\ll1$, or even for $|\rho_{1}-\rho_{2}|\sim1$
under microgravity conditions. (Note that although $\eta_{xx}\sim\eta/L^{2}$
where $L$ is assumed large, the two terms of equation (\ref{eq:Pi})
are comparable for $\text{Bo}/\sigma$ of the order $1/L^{2}$. In
fact, we find below in section 6 that in the nonlinear evolution,
the lengthscale $L$ might develop to be $L=O(\sqrt{\sigma/\text{Bo}})$
or, since it is assumed that the change of surface tension by the
surfactant is small, so that $\sigma=O(1)$, we have $L=O(\text{Bo}^{-1/2})$
(see, e.g., the discussion on page 42, between equations (7.4) and
(7.5)).)The tangential stress condition (\ref{TanStressBC}) yields
\begin{equation}
Du_{1}-mDu_{2}(=\sigma_{x})=-\text{Ma}\Gamma_{x}.\label{eq:tangential stress lubrication}
\end{equation}
Hence we can eliminate $p_{2x}$ and $A_{2}:$ 
\begin{align}
p_{2x} & =p_{1x}-\Pi_{x},\label{eq:p2 in terms of p1}\\
A_{2} & =\frac{1}{m}\left(A_{1}+\text{Ma}\Gamma_{x}+\Pi_{x}\eta\right).\label{eq:A2 in terms of A1}
\end{align}
We substitute these into the expressions for $u_{2}$ and $w_{2}$,
and apply the continuity of velocity conditions, (\ref{ContinuousVatInterfaceBC}),
at the interface $z=\eta$, that is, $\bar{u}_{1}+u_{1}=\bar{u}_{2}+u_{2}$,
or 
\begin{equation}
u_{2}-u_{1}=\frac{m-1}{m}s\eta,\label{eq:continuityofu}
\end{equation}
and 
\begin{equation}
w_{1}=w_{2},\label{eq:continuityofw}
\end{equation}
to obtain the following system of equations for $p_{1x}$ and $A_{1}$:
\begin{align*}
\left(-m+n^{2}+(m-1)\eta^{2}\right)p_{1x}+2\left(m+n+(m-1)\eta\right)A_{1}\\
=-2(m-1)s\eta+2(-n+\eta)\text{Ma}\Gamma_{x}+(-n+\eta)^{2}\Pi_{x}
\end{align*}
and 
\begin{align*}
\left(2(m+n^{3})+3(m-n^{2})\eta-(m-1)\eta^{3}\right)p_{1x}+3\left(-m+n^{2}-2(m+n)\eta-(m-1)\eta^{2}\right)A_{1}\\
=3(m-1)s\eta^{2}-3(-n+\eta)^{2}\text{Ma}\Gamma_{x}-2(-n+\eta)^{3}\Pi_{x}+C(t).
\end{align*}
Note that the second equation contains an arbitrary function $C(t)$
that does not depend on $x$, obtained by integrating (\ref{eq:continuityofw})
which contains the derivatives $p_{1xx}$ and $A_{1x}$. Solving this
linear system, we can express $p_{1x}$ and $A_{1}$, and therefore
all the velocities, in terms of $\eta,$ $\Gamma$, $\Pi[\eta,\Gamma]$
and $C(t)$: 
\begin{align}
p_{1x} & =-\frac{6(m-1)s}{{\cal D}}\left[m-n^{2}+(m+n)\eta\right]\eta+\frac{6m(n+1)}{{\cal D}}(1+\eta)(-n+\eta)\text{Ma}\Gamma_{x}\nonumber \\
 & -\frac{\left(n-\eta\right)^{2}}{{\cal D}}\left[-(3+4n)m-n^{2}-2(m-n+2mn)\eta+(m-1)\eta^{2}\right]\Pi_{x}\nonumber \\
 & +\frac{2}{{\cal D}}\left(m+n+(m-1)\eta\right)C(t)\label{eq:p1x}
\end{align}
and 
\begin{align}
A_{1} & =-\frac{(m-1)s}{{\cal D}}\eta\left[4(m+n^{3})+3(m-n^{2})\eta+(m-1)\eta^{3}\right]\nonumber \\
 & +\frac{\left(\eta-n\right)}{{\cal D}}\left[(4+3n)m+n^{3}+3(m-n^{2})\eta-3(m-1)n\eta^{2}+(m-1)\eta^{3}\right]\text{Ma}\Gamma_{x}\nonumber \\
 & +\frac{\left(\eta-n\right)^{2}}{{\cal D}}\left[2m(n+1)+(m-n^{2})\eta-2(m-1)n\eta^{2}+(m-1)\eta^{3}\right]\Pi_{x}\nonumber \\
 & -\frac{\left(-(m-n^{2})+(m-1)\eta^{2}\right)}{{\cal D}}C(t).\label{eq:A1}
\end{align}
Here the determinant ${\cal D}$ is 
\[
{\cal D}=(m-1)^{2}\eta^{4}+4(m-1)(m+n)\eta^{3}+6(m-1)(m-n^{2})\eta^{2}+4(m-1)(m+n^{3})\eta+\psi,
\]
where constants $\phi$ and $\psi$ are defined as follows: 
\begin{equation}
\varphi={n}^{3}+3\,{n}^{2}+3\,mn+m\text{}\label{phi}
\end{equation}
and 
\begin{equation}
\psi={n}^{4}+4\,m{n}^{3}+6\,m{n}^{2}+4\,mn+{m}^{2}\text{.}\label{psi}
\end{equation}
The function $C(t)$ is obtained by the boundary conditions in $x$.
We adopt the condition of periodicity of pressure over the $x-$interval
of length $\Lambda$ similar to \citet{Blyth2004b}. (The periodic
boundary conditions pertain to closed flows such as the Couette flow
in a circular toroidal channel with a rectangular cross-section of
\textbackslash{}citet\{Barthelet1995\}. For rectilinear channels,
we expect solutions which are largely independent of the boundary
conditions at the channel ends if the channel is sufficiently long.)
Then 
\[
\int_{0}^{\Lambda}p_{1x}\;dx=0.
\]
From this equation one obtains an explicit expression for $C(t)$
in terms of the integrals over that interval. Thus, $C(t)$ is a functional
of $\eta$ and $\Gamma$.

To obtain the evolution equations, we substitute the velocity field,
(\ref{eq:uj vel}) and (\ref{eq:wj vel}), into the kinematic boundary
condition (\ref{dKineBC}) and the surfactant transport equation (\ref{dSurfConc}):
\begin{align}
\eta_{t}+\left[\frac{s}{2}\eta^{2}+\frac{1}{2}\left(1+\eta\right)^{2}\left(-\frac{1}{3}\left(2-\eta\right)p_{1x}+A_{1}\right)\right]_{x} & =0,\label{eq:eta evolution equation}\\
\Gamma_{t}+\left[\left(s\eta-\frac{1}{2}\left(1-\eta^{2}\right)p_{1x}+\left(1+\eta\right)A_{1}\right)\left(1+\Gamma\right)\right]_{x} & =0,\label{eq:Gamma evolution equation}
\end{align}
where $p_{1x}$ and $A_{1}$ are given by (\ref{eq:p1x}) and (\ref{eq:A1})
respectively, and we have omitted the tilde from the disturbance of
surfactant. We will solve this system of evolution equations, (\ref{eq:eta evolution equation})
and (\ref{eq:Gamma evolution equation}), numerically, when we discuss
nonlinear results in section \ref{sec:nonlinear}.

The regimes in which $\eta$ and $\Gamma$ are much smaller than unity
may be described by weakly nonlinear equations which are obtained
from (\ref{eq:eta evolution equation}) and (\ref{eq:Gamma evolution equation})
by neglecting those nonlinear terms which are clearly smaller than
some other terms: 
\begin{eqnarray}
\nonumber \\
 & \eta_{t} & +sN_{1}\eta\eta_{x}-\frac{2(m-1)n^{2}(n+1)s}{\psi}\eta_{x}-\frac{(m+n)n^{3}\Bo}{3\psi}\eta_{xx}\nonumber \\
 &  & +\frac{(m+n)n^{3}}{3\psi}\eta_{xxxx}+\frac{n^{2}(m-n^{2})\Ma}{2\psi}\varGamma_{xx}=0\label{eq:weak-eta-evolution}
\end{eqnarray}
and 
\begin{eqnarray}
\nonumber \\
 & \varGamma_{t} & +sN_{2}\eta\eta_{x}+\frac{(n+1)\phi s}{\psi}\eta_{x}-\frac{n^{2}(n^{2}-m)\Bo}{2\psi}\eta_{xx}+\frac{n^{2}(n^{2}-m)}{2\psi}\eta_{xxxx}\nonumber \\
 &  & -\frac{n(m+n^{3})\Ma}{\psi}\varGamma_{xx}=0,\label{eq:weak-surfactant-evolution}
\end{eqnarray}
where 
\begin{align*}
N_{1} & =1+\frac{(m-1)}{\psi}\left[-m+4n-3n^{2}-8n^{3}\right]+\left[\frac{4(m-1)n}{\psi}\right]^{2}(m+n^{3})(n+1),\\
N_{2} & =\frac{2(m-1)}{\psi}\left[3n(n+1)-4(m+n^{3})\right]+8\left[\frac{(m-1)}{\psi}\right]^{2}(m+n^{3})(m+3n^{2}+4n^{3}).
\end{align*}
We need to keep the nonlinear term in equation (\ref{eq:weak-eta-evolution}).
Even though it appears that it can be neglected as compared to the
term with the linear term $\eta_{x}$, the latter will be eliminated
by a coordinate change, $x\rightarrow x+Vt$ where $V$ is the coefficient
of $\eta_{x}$ in (\ref{eq:weak-eta-evolution}). However, we will
discard the nonlinear term from equation (\ref{eq:weak-surfactant-evolution}),
since in the latter the larger (by a factor $1/\eta$) linear $\eta_{x}$
term is not eliminated by this coordinate change. Also, calculating
$C(t)$ (from the spatial periodicity of the pressure; see the paragraph
that follows equation (\ref{psi})), one finds that $C(t)$ is proportional
to the average of $\eta^{2}$. (Actually, it is not difficult to see
that in the limit of very long channels the same conclusion holds
for other boundary conditions with the end-point data being arbitrary
but bounded functions of time.) So its contribution in equations (\ref{eq:eta evolution equation})
and (\ref{eq:Gamma evolution equation}) for small $\eta$ and $\Gamma$
is at most of the orders $\eta^{2}\eta_{x}$ and $\eta^{2}\Gamma_{x}$,
and therefore is neglected in the system (\ref{eq:weak-eta-evolution})-(\ref{eq:weak-surfactant-evolution})
in comparison with the nonlinear terms. These weakly nonlinear evolution
equations do not imply any restrictions on the parameters beyond the
lubrication approximation, and are different from the equations of
\citet{Frenkel2006}, \citet{Bassom2010} and \citet{Kalogirou2012},
which assumed a small aspect ratio $1/n$ (and in some cases other
restrictions on the parameters). In particular, in those papers the
lubrication approximation was used for the thin layer only.

We remind the reader that in view of our derivation, it is clear that
the system of equations (\ref{eq:weak-eta-evolution})-(\ref{eq:weak-surfactant-evolution})
allows for relative errors of the order $O(1/L^{2})$, where $L$
is the characteristic lengthscale of solutions which is assumed to
be large. The same is true of the strongly nonlinear system (\ref{eq:eta evolution equation})-(\ref{eq:Gamma evolution equation}).

Below, we encounter weakly nonlinear regimes in which saturated $\eta$
is small but $\Gamma$ is not small. Then, from equation (\ref{eq:Gamma evolution equation})
we see that a nonlinear term of the form $N_{3}s(\eta\Gamma){}_{x}$,
should be added into the transport equation (\ref{eq:weak-surfactant-evolution})
where 
\begin{equation}
N_{3}=\frac{(n+1)\phi}{\psi},\label{eq:N3coeff}
\end{equation}
the coefficient of $\eta_{x}$ in (\ref{eq:weak-surfactant-evolution}).

From the system (\ref{eq:weak-eta-evolution}) and (\ref{eq:weak-surfactant-evolution}),
we can obtain the linear stability equations for the normal modes
of disturbances, 
\begin{equation}
(\eta\text{, }\Gamma)=[h\text{, }G]e^{i\alpha x+\gamma t},\label{eq:ModesNormal}
\end{equation}
where $\alpha$ is the wavenumber of the disturbance, $G$ and $h$
are constants, and the (complex) $\gamma$ is called the increment;
in terms of its real and imaginary parts, $\gamma=\gamma_{R}+i\gamma_{I}$,
where the real $\gamma_{R}$ is the growth rate of the normal mode.
\ The stability of the flow depends on the sign of $\gamma_{R}$:
if $\gamma_{R}>0$ for some normal modes then the system is unstable;
and if $\gamma_{R}<0$ for all normal modes, then the system is stable.
The complex amplitude $h$ is arbitrary; we choose it to be real and
positive. We substitute (\ref{eq:ModesNormal}) into the linearized
equations (\ref{eq:weak-eta-evolution}) and (\ref{eq:weak-surfactant-evolution})
to obtain the following equations for $\gamma$ and $G$: 
\begin{eqnarray}
\nonumber \\
 & \gamma h & =i\alpha\frac{2(m-1)n^{2}(n+1)s}{\psi}h-\alpha^{2}\frac{(m+n)n^{3}\Bo}{3\psi}h\label{eq:linear-eta-evolution}\\
 &  & +\alpha^{2}\frac{n^{2}(m-n^{2})\Ma}{2\psi}G,\nonumber 
\end{eqnarray}
and 
\begin{eqnarray}
\nonumber \\
 & \gamma G= & -i\alpha\frac{(n+1)\phi s}{\psi}h+\alpha^{2}\frac{(m-n^{2})n^{2}\Bo}{2\psi}h-\alpha^{2}\frac{(m+n^{3})n\Ma}{\psi}G.\label{eq:linear-surfactant-evolution}\\
\nonumber 
\end{eqnarray}
(Note that we have retained terms up to $\alpha^{2}$, but discarded
the (capillary) $\alpha^{4}$ terms corresponding to the fourth-derivative
terms in the weakly nonlinear system (\ref{eq:weak-eta-evolution})
- (\ref{eq:weak-surfactant-evolution}), since we are interested in
the threshold of instability only, and the latter is determined in
the long-wave limit, $\alpha\to0$. Also, note that, as is clear from
the linearization of equation (\ref{dKineBC}), in which the last
term vanishes since the base velocity is zero at the interface, $z=0$,
the right hand side of equation (\ref{eq:linear-eta-evolution}) is
$w_{1}(0)$. Hence, 
\begin{equation}
\gamma=\frac{w_{1}(0)}{h},\label{eq:gamma is w/h}
\end{equation}
which is used below.) This system of linear homogeneous equations
for the amplitudes $h$ and $G$ (\ref{eq:weak-eta-evolution}) -
(\ref{eq:weak-surfactant-evolution}), which has the matrix form 
\begin{equation}
\left[\begin{array}{cc}
\gamma-2i\frac{(m-1)n^{2}(n+1)s}{\psi}\alpha+\frac{n^{3}(m+n)\Bo}{3\psi}\alpha^{2} & -\frac{n^{2}(m-n^{2})\Ma}{2\psi}\alpha^{2}\\
i\frac{(n+1)\phi s}{\psi}\alpha-\frac{n^{2}(m-n^{2})\Bo}{2\psi}\alpha^{2} & \gamma+\frac{n(m+n^{3})\Ma}{\psi}\alpha^{2}
\end{array}\right]\left[\begin{array}{c}
h\\
G
\end{array}\right]=\left[\begin{array}{c}
0\\
0
\end{array}\right]\label{eq:matrix system}
\end{equation}
has non-trivial solutions only if its determinant is equal to zero.
This requirement yields a quadratic equation for $\gamma$: 
\begin{equation}
\psi\gamma^{2}+c_{1}\gamma+c_{0}=0,\label{eq:quadratic longwave}
\end{equation}
where 
\begin{equation}
c_{1}=n(m+n^{3})\Ma\alpha^{2}+\frac{1}{3}n^{3}(m+n)\Bo\alpha^{2}-2in^{2}(n+1)(m-1)s\alpha\label{eq:c1 lubrication}
\end{equation}
and$ $ 
\begin{equation}
c_{0}=\frac{1}{12}n^{4}\Ma\Bo\alpha^{4}-\frac{1}{2}in^{2}(n^{2}-1)\Ma s\alpha^{3}.\label{eq:c0 lubrication}
\end{equation}
The two solutions of (\ref{eq:quadratic longwave}) are 
\begin{equation}
\gamma=\frac{1}{2\psi}\left(-c_{1}+\left[c_{1}^{2}-4\psi c_{0}\right]^{1/2}\right)\label{eq:QuadEqnGamma}
\end{equation}
where the square root here has two (complex) values. Thus, for given
parameters and the wavenumber, there are, in general, two distinct
complex values of the increment $\gamma$. (We note that keeping the
capillary terms which are proportional to $\alpha^{4}$ in equations
(\ref{eq:linear-eta-evolution}) and (\ref{eq:linear-surfactant-evolution})
amounts to changing $\text{Bo}$ into $\text{Bo}+\alpha^{2}$. Then
the same change occurs in $c_{1}$, equation (\ref{eq:c1 lubrication}),
and $c_{0}$, equation (\ref{eq:c0 lubrication}). This will lead
to a higher order correction to the leading-order growth rate given
by equation (\ref{gamCsmallAlphaApprox}) below. This correction does
not affect the long-wave instability threshold (see equation (\ref{eq:BoCritical}).
It will determine the stabilization at shorter waves which we do not
consider in the present paper. Also, we note that the quadratic equation
of the form (\ref{eq:quadratic longwave}) holds for arbitrary wavelengths
with the expressions for $c_{0}$ and $c_{1}$ given in \citet{schweiger2013gravity}
without using the lubrication approximation, and their small-$\alpha$
asymptotics reproduce our coefficients $c_{0}$ and $c_{1}$, even
with the higher order capillary corrections.)

One can see that the dispersion function $\gamma_{R}$ (as well as
the increment function $\gamma$) has the following ``symmetry''
property 
\begin{equation}
\gamma_{R}(-n\alpha;\;ns,\;m^{-1},\;n^{-1},\;\text{Ma},\;n^{2}\text{Bo})=nm\gamma_{R}(\alpha;\;s,\;m,\;n,\;\text{Ma},\;\text{Bo}).\label{eq:gammasymmetrycondition-1}
\end{equation}
It is verified by changing $\gamma\to nm\;\gamma$ in the dispersion
equation (\ref{eq:quadratic longwave}), and $\alpha^{2}\to n^{2}\alpha^{2}$,
$s\to ns$, $m\to m^{-1}$, $n\to n^{-1}$, $\text{Ma}\to\text{Ma}$
and $\text{Bo}\to n^{2}\text{Bo}$ in the coefficients $c_{0}$, $c_{1}$
and $\psi$. In fact, this transformation is inferred by looking at
the flow from the ``upside down'' point of view, as was mentioned
in the preceding section. This implies the relations (with the left
superscript indicating the quantity in the new coordinate system)
$^{n}d_{1}=d_{2}$, $^{n}d_{2}=d_{1}$, $^{n}\rho_{1}=\rho_{2}$,
$^{n}\rho_{2}=\rho_{1}$, $^{n}\mu_{1}=\mu_{2}$, and $^{n}\mu_{2}=\mu_{1}$,
so that $^{n}m=1/m$, and $^{n}n=1/n$. Also, $^{n}x^{*}=-x^{*}$,
$^{n}z^{*}=-z^{*}$ (note that $^{n}\mathbf{\hat{z}=\mathbf{\hat{z}}}=<0,0,1>)$
and $^{n}t^{*}=t^{*}$ so that $^{n}x=-x/n$ , $^{n}z=-z/n$, $^{n}t=t/(nm)$
and $^{n}\alpha^{2}=n^{2}\alpha^{2}$. (Note that we used non-dimensionalization
(\ref{scales}) based on the bottom layer, so that the units of measurement
used there change since $^{n}d_{1}=d_{2}$, etc..) Furthermore, we
have $^{n}U_{1}^{*}=U_{2}^{*}$, $^{n}U_{2}^{*}=U_{1}^{*}$, $^{n}\eta^{*}=-\eta^{*}$,
and $^{n}g=-g$ so that $^{n}\text{Bo}=n^{2}\text{Bo}$. We find that
$^{n}s=ns$ and $^{n}\gamma=mn\gamma$. With the appropriate transformations
of the velocities and pressures, ($\boldsymbol{^{n}v}_{1}^{\ast}=-\boldsymbol{v}_{2}^{\ast}$,
etc.; $^{n}p_{1}^{*}=p_{2}^{*}$, etc.), the governing equations are
invariant under the ``upside down'' transformation, and we recover
the same dispersion relation. This implies the symmetry property given
by equation (\ref{eq:gammasymmetrycondition-1}). In view of this
symmetry of the growth rate function, it is sufficient to consider
linear stability for $n\ge1$. This range of $n$ is also sufficient
for nonlinear disturbances (see section \ref{sec:nonlinear}), for
the same reason.

Considering the limit of vanishing $\textrm{Ma},$ one observes that
the product of the increments of the two modes, $c_{0}/\psi$, vanishes.
So, at least one of the increments vanishes. However, the other increment
cannot vanish, because the sum $-c_{1}/\psi$ of the two increments,
the roots of the quadratic equation, does not vanish. We call the
non-vanishing continuous branch of the increment (and of the growth
rate function) the robust branch, and the other one the surfactant
branch of the increment (or of the growth rate). Correspondingly,
we sometimes speak of the robust and surfactant branches (sets) of
normal modes. (The robust mode is similar to the ``interface mode''
of two-layer surfactant-free flows down an inclined plane (see \citet{gao2007,samanta2014}
and references therein) in that both do not vanish in the limit of
surfactantless flows. \citet{Wei2007}, considering some single-fluid
surfactant-laden flows, calls the mode corresponding to our robust
mode the ``interface mode''. We, however, prefer the term ``robust
mode'', in order to avoid confusion due to the different meanings
of the term ``interface mode'' as used in the aforementioned references.)
Thus, there is just a single robust normal mode and a single surfactant
mode for each wavenumber.

\section{Increments, growth rates, and thresholds of instability}

\label{sec:Increments, growth rates}

\subsection{Leading-order long-wave increments and growth rates\label{subsec:Leading-order-long-wave-incremen}}

In this section we find the power series expansions in $\alpha$ of
the increment $\gamma$ for the case $s\ne0$, as well as for the
case $s=0$ and $\text{Bo}=0$, a solution to the quadratic equation
(\ref{eq:quadratic longwave}), in the form

\begin{equation}
\gamma=iI_{1}\alpha+R_{2}\alpha^{2}+iI_{3}\alpha^{3}+...,\label{eq:gamma series-1}
\end{equation}
where $I_{1}$, $R_{2}$ and $I_{3}$ are real and depend on the coefficients
of (\ref{eq:quadratic longwave}). In $c_{1}$, we denote $c_{11}$
the coefficient of $i\alpha$ and $c_{12}$ the coefficient of $\alpha^{2}$.
Similarly, in $c_{0}$, the coefficient of $i\alpha^{3}$ is $c_{03}$
and the coefficient of $\alpha^{4}$ is $c_{04}$. Note that these
coefficients are all real. For the first branch, by substituting (\ref{eq:gamma series-1})
into the quadratic equation (\ref{eq:quadratic longwave}), and balancing
the terms proportional to $\alpha^{2}$, 
\begin{equation}
I_{1}=-c_{11}/\psi.\label{eq:I1-1}
\end{equation}
Then, balancing the $\alpha^{3}$ terms, we obtain $R_{2}$, the leading
order of the growth rate, 
\begin{equation}
\gamma_{R}\approx\left({\frac{\,\varphi\left(m-{n}^{2}\right)}{4\left(1-m\right)\psi}}\text{Ma}-\frac{n^{3}(n+m)}{3\psi}\text{Bo}\right)\alpha^{2},\label{gamCsmallAlphaApprox}
\end{equation}
where we will confine ourselves to the case $m\ne1$. Also, the leading
order phase velocity is $c=-(\textrm{Im}\gamma)/\alpha=-I_{1}.$ (We
note that this $c$ is independent of wavenumber, and thus can be
made zero for all $\alpha$ at once by the Galilean transformation
to the reference frame moving with velocity $c$. To find the leading
non-constant phase velocity, we need 
\begin{equation}
I_{3}=-\psi\frac{c_{03}^{2}}{c_{11}^{3}}+\frac{c_{12}c_{03}}{c_{11}^{2}}-\frac{c_{04}}{c_{11}}-\frac{c_{13}}{\psi}.\label{eq:increment I3 unstable-1}
\end{equation}
However, determining $c_{13}$ requires the next correction in $\alpha^{2}$
to the lubrication approximation. This is done in Appendix \ref{sec:Augmented-lubrication-theory}.

In this connection, it is notable that the lubrication approximation
(\ref{eq:uj momentum})-(\ref{eq:continuityofw}) (as well as many
other similar lubrication approximation formulations such as those
in \citet{Babchin1983a,Blyth2004b,Charru2000,Oron1997,Wei2005c})
corresponds to the leading-order of expansions in the powers of the
small quantity $\alpha^{2}$, in which the coefficients are two-terms
expansions in powers of $i\alpha$ (with coefficients that are real
except for some special cases such as when $m=1$, see section \ref{subsec:The-finite-aspect}
where the coefficient expansions may be in powers of $\alpha^{1/2}$).
Thus, two leading orders in the small wavenumber $\alpha$ are captured
by the lubrication approximation.)

For the other mode, we have 
\begin{equation}
\gamma=S_{2}\alpha^{2}+iJ_{3}\alpha^{3}+S_{4}\alpha^{4}+\dots\label{eq:gamma series surfactant}
\end{equation}
and find $S_{2}=-c_{03}/c_{11}$. This gives the growth rate 
\begin{equation}
\gamma_{R}\approx\frac{(n-1)\text{Ma}}{4(1-m)}\alpha^{2}+k_{s}\alpha^{4},\label{gamSsmallAlphaApprox}
\end{equation}
where the expression for $k_{s}$ is given in Appendix \ref{sec:coefficient}.
(We have included the term with $\alpha^{4}$ because it becomes the
leading order term when $n=1$.) We also find the coefficient $J_{3}$
to be 
\begin{equation}
J_{3}=\psi\frac{c_{03}^{2}}{c_{11}^{3}}-\frac{c_{12}c_{03}}{c_{11}^{2}}+\frac{c_{04}}{c_{11}^{1}}.\label{eq:J_3}
\end{equation}
The leading term of $\textrm{Im}(\gamma)$ is $J_{3}\alpha^{3},$
and hence the leading phase velocity is $c=-\textrm{Im}(\gamma)/\alpha=-J_{3}\alpha^{2}.$
Thus, in contrast to the other branch, all the modes cannot be eliminated
at once by applying an appropriate Galilean transformation. We note
the relation $I_{3}=-J_{3}-c_{13}/\psi.$

The growth rate (\ref{gamCsmallAlphaApprox}) is a continuous function
of $\alpha$ which is identified as the robust branch of the growth
rate since it is nonzero even at $\text{Ma}=0$. The other continuous
branch of the growth rate, equation (\ref{gamSsmallAlphaApprox}),
that vanishes as $\text{Ma}\rightarrow0$, is the surfactant branch.

We note that there is another way to obtain the consecutive terms
of the power series for the increment $\gamma$ - by shuttling between
the thickness and surfactant equations (\ref{eq:linear-eta-evolution})
and (\ref{eq:linear-surfactant-evolution}), with increasing powers
in $\alpha$, to find the consecutive terms of the power series for
the quantities $G/h$ and $\gamma$ in turn. (This is equivalent to
the method of undetermined coefficients.) It works slightly differently
for the two modes, as follows:

For the surfactant mode, start with the thickness equation (\ref{eq:linear-eta-evolution})
at order $\alpha^{1}$ to find $G/h$ to its leading order $\alpha^{-1}$.
Use this in the surfactant equation taken at order $\alpha^{1}$ to
find $S_{2}$. Apply the latter in the thickness equation at order
$\alpha^{2}$ to find the $\alpha^{0}$ correction to the $G/h$.
Return with the latter to the surfactant equation taken at order $\alpha^{2}$
to find $J_{3}$.

For the robust mode, start with the thickness equation (\ref{eq:linear-eta-evolution})
at order $\alpha^{1}$ and find $I_{1}$. Use it in the surfactant
equation at order $\alpha^{1}$ to find $G/h$ to its leading order
$\alpha^{0}$. Then the thickness equation at order $\alpha^{2}$
yields $R_{2}$. Next, the surfactant equation taken at order $\alpha^{2}$
yields the $\alpha^{1}$ correction to $G/h$. The resulting expressions
for the eigenfunctions $G/h$ are as follows. For the robust branch,
\begin{equation}
\frac{G}{h}=-\dfrac{\varphi}{2n^{2}(m-1)}+i\alpha\frac{\psi}{4(m-1)^{2}n(n+1)s}\left(\frac{\text{Bo}}{3}-\phi\frac{(n-1)\text{Ma}}{4(m-1)n^{3}}\right),\label{eq:amplitude ratio for robust mode-1-1}
\end{equation}
and for the surfactant branch, 
\begin{equation}
\frac{G}{h}=i\dfrac{4(n+1)(m-1)s}{(n^{2}-m)\textrm{Ma}}\alpha^{-1}+\left(\dfrac{\psi(n-1)}{2n^{2}(n^{2}-m)(m-1)}-\dfrac{2\textrm{Bo}n(n+m)}{3\textrm{Ma}(n^{2}-m)}\right).\label{eq:G/h with correction surfactant branch-1-1}
\end{equation}
Note that for $m=n^{2}$ equations (\ref{eq:eta evolution equation})
and (\ref{eq:Gamma evolution equation}) imply that for the surfactant
branch $h=0$ and $G$ is arbitrary, and $\gamma_{R}$ is consistent
with equation (\ref{gamSsmallAlphaApprox}). The latter two equations
will be used in section \ref{sec:phase differences}.

For the case $s=0$ and $\text{Bo}=0$, consider the robust branch
first. Finding the leading-order growth rates requires the inclusion
of capillary, fourth derivative, terms which are found in the weakly
nonlinear equations (\ref{eq:weak-eta-evolution}) and (\ref{eq:weak-surfactant-evolution})
into the linear equations (\ref{eq:linear-eta-evolution}) and (\ref{eq:linear-surfactant-evolution}),
respectively. The leading-order balance of the capillary and the Marangoni
terms (the last two terms in equation (\ref{eq:weak-surfactant-evolution}))
yields in terms of the normal mode amplitudes 
\begin{equation}
\frac{G}{h}=\frac{(m-n^{2})n}{2\text{Ma}(m+n^{3})}\alpha^{2}.\label{eq:Gh s=00003D00003D0 Bo=00003D00003D0 Robust}
\end{equation}
The leading-order, $\alpha^{4}$, balance in the kinematic equation
involves three terms: the time derivative term; the capillary term;
and the Marangoni term. Substituting in there $G$ in terms of $h$
from the preceding equation yields the growth rate of the robust mode:
\begin{equation}
\gamma_{R}=-\frac{n^{3}}{12(m+n^{3})}\alpha^{4}.\label{eq:gammaR s=00003D00003D0 Bo=00003D00003D0 Robust}
\end{equation}
For the surfactant mode the capillary effects are negligible. So,
the leading-order balance in the surfactant transport equation involves
just the two terms with $G$, and immediately yields the growth rate:
\begin{equation}
\gamma_{R}=-\frac{n(m+n^{3})\text{Ma}}{\psi}\alpha^{2}.\label{eq:gammaR s=00003D00003D0 Bo=00003D00003D0 Surfactant}
\end{equation}
Substituting this into the kinematic condition yields 
\begin{equation}
\frac{G}{h}=-2\frac{(m+n^{3})}{n(m-n^{2})}.\label{eq:Gh s=00003D00003D0 Bo=00003D00003D0 Surfactant}
\end{equation}
These results are in agreement with FH.

Finally, for the case $s=0$ and $\text{Bo}\ne0$ there is no universal
expansion of the increment in powers of $\alpha$. However, it is
easy to see that both modes are stable if $\text{Bo}>0$ but there
is instability if $\text{Bo}<0$. Indeed, if $\text{Bo}<0$ then $c_{0}=\frac{1}{12}n^{4}\alpha^{4}\text{Ma}\text{Bo}<0$
(see (\ref{eq:c0 lubrication})). Therefore in equation (\ref{eq:QuadEqnGamma}),
the discriminant $c_{1}^{2}-4\psi c_{0}>c_{1}^{2}$ (note that for
this case equation (\ref{eq:c1 lubrication}) yields $c_{1}=n\alpha^{2}((m+n^{3})\text{Ma}+n(m+n)\text{Bo}/3)$),
and equation (\ref{eq:quadratic longwave}) yields one positive growth
rate value, so we have an instability. This is essentially the Rayleigh-Taylor
instability of a stagnant two-layer arrangement modified by the surfactant.
On the other hand, if $\text{Bo}>0$, then $c_{0}>0$ and $c_{1}>0$,
but the discriminant can be either positive or negative. If it is
negative, then the square roots in the solution (\ref{eq:QuadEqnGamma})
are purely imaginary and therefore both values of $\gamma_{R}$ are
negative. If the discriminant is positive, then $|\sqrt{c_{1}^{2}-4\psi c_{0}}|<c_{1}$,
so that both values of $\gamma$ given by equation (\ref{eq:QuadEqnGamma})
are negative again. It is clear that in all these cases $\gamma\propto\alpha^{2}$.
After the increment $\gamma$ is determined from equation (\ref{eq:QuadEqnGamma}),
(where the two possible values correspond to the two different modes),
the eigenfunction $G/h$ is found from the kinematic condition (\ref{eq:linear-eta-evolution})
as 
\begin{equation}
\frac{G}{h}=\frac{2}{(m-n^{2})\text{Ma}}\left(\frac{\psi}{n^{2}}\alpha^{-2}\gamma+\frac{(m+n)n\text{Bo}}{3}\right).\label{eq:G/h for s=00003D00003D0 nonzero Bo}
\end{equation}

We note that a growth-rate ``superposition principle'', $\gamma_{R}\left(\text{Ma, Bo}\right)=\ \gamma_{R}\left(\text{Ma, }0\right)+\gamma_{R}\left(0\text{, Bo}\right)$,
holds for the robust branch in the leading-order of equation (\ref{gamCsmallAlphaApprox}).
(The purely Marangoni growth rate $\gamma_{R}\left(\text{Ma, }0\right)$
is the one found in FH, and the other, purely Bond, term gives the
well-known growth rate of the (surfactantless) Rayleigh-Taylor instability.)
In contrast, the leading-order growth rate of the surfactant branch,
equation (\ref{gamSsmallAlphaApprox}), is independent of the Bond
number; the latter appears at higher orders, and is always multiplied
by some positive power of the Marangoni number (see Appendix \ref{sec:coefficient}).

\subsection{Instability thresholds in the three ($n,m$)-sectors}

\begin{figure}
\centering{}\includegraphics[clip,scale=0.6]{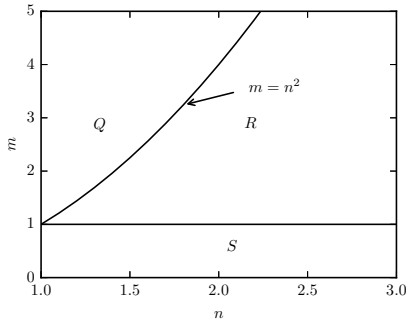}\protect\caption{The $(n,m)$-plane (for $n\ge1$) consists of three sectors ($Q$,
$R$, and $S$) which differ as regards the flow stability properties.
\label{fig:FigRegions}}
\end{figure}

From the long-wave approximation of FH, for $\Bo=0$, three sectors
were identified in the $n\ge1$ part of the $(n,m)$-plane as regards
the stability of the flow. The same three sectors turn out to be relevant
even when $\text{Bo}\ne0$ (as in our present case): the $Q$ sector
($1<n^{2}<m$); the $R$ sector ($1<m<n^{2}$); and the $S$ sector
($1<n<\infty$ and $0<m<1$). (The boundaries between the sectors
$m=n^{2}$ and $m=1$ correspond to the numerator and denominator
respectively of the coefficient of the Marangoni number in equation
(\ref{gamCsmallAlphaApprox}). Therefore it is clear why they appear
for the inertialess case of FH, with zero gravity and non-zero Marangoni
number. We note that the same curves, $m=n^{2}$ and $m=1$ appear
as neutral stability curves in the corrected figure 2 of \citet{Yiantsios1988}
(see the correction \citet{yiantsios1989erratum}) for the case with
neither surfactant nor gravity effects, where the instability hinges
on inertia - despite the fact that Poiseuille flow, and not Couette
flow, was the focus of attention in \citet{Yiantsios1988}. Part of
the reason for this is that the leading-order disturbance flow of
the robust mode (obtained in section 5.4; see equation (\ref{eq:re u1 robust}))
is a propagating wave which does not depend on the factors responsible
for the growth or decay of the disturbances \textendash{} such as
inertia, gravity or surfactant effects. (Also, it does not depend
on the details of the base velocity profile other than its interfacial
slope.) Therefore, this leading-order disturbance flow, that contains
the expressions $m-n^{2}$ and $m-1$, is essentially the same (up
to a scaling factor) for the Yih instability, the surfactant instability,
or the Rayleigh-Taylor instability of the basic flow, whether Couette
or Poiseuille one. Note, however, that for the inertial instability
of the Couette flow in \citet{Yih1967}, despite the fact that the
leading-order disturbance flow is still the same, the curve $m=n^{2}$
is not a curve of neutral stability. This is related to the following
difference between the Yih instability and the inertialess instability.
For the latter, the momentum equations for the correction of the disturbances
are homogeneous, not depending on the base flow. Thus, they are the
same for the Poiseuille flow $\bar{u}_{j}=(sz+qz^{2})/m_{j}$, with
$q(m-n^{2})=s(n+m)$, as for the Couette flow, equations (\ref{u1w1p1BSprofiles})-(\ref{u2w2p2BSprofiles}).
In contrast, for the cases with non-zero inertia, the momentum equations
for the disturbances in the correction order are non-homogeneous with
their sources depending on the base flow and the leading-order disturbance
of the flow. This is why the linear inertial instability results for
the Poiseuille flow of \citet{Yiantsios1988} differ from those for
the Couette flow of \citet{Yih1967} and of \citet{Charru2000} while
the inertialess surfactant instability results would be the same for
the Poiseuille flow as our results for the Couette flow.) Figure \ref{fig:FigRegions}
shows the three sectors and their borders. Stability properties of
the robust and surfactant modes can change significantly as one moves
from sector to sector. 

In both the $R$ sector and the $Q$ sector, according to (\ref{gamSsmallAlphaApprox}),
the surfactant branch is stable for all $\text{Bo}$. From equation
(\ref{gamCsmallAlphaApprox}) we can infer that the robust branch,
is unstable if $\Bo<\Bo_{cL}$, where the threshold value is 
\begin{equation}
\text{Bo}_{cL}=-\frac{3\varphi(m-n^{2})}{4n^{3}(m-1)(n+m)}\text{Ma}.\label{eq:BoCritical}
\end{equation}
This condition holds for all three sectors, as does also the fact
that gravity is stabilizing for $\text{Bo}>0$ and destabilizing for
$\text{Bo}<0$. In the $R$ sector, the Marangoni effect is destabilizing
and, from equation (\ref{eq:BoCritical}) with $m<n^{2}$, we have
$\text{Bo}_{cL}>0$. Gravity renders the flow stable for $\text{Bo}>\text{Bo}_{cL}$,
but for positive $\text{Bo }$ below $\text{Bo}_{cL}$, the flow is
still unstable (and it is unstable for all negative Bond numbers).
In the $Q$ sector, the Marangoni effect is stabilizing, $\text{Bo}_{cL}<0$,
and gravity renders the flow unstable only for the negative Bond numbers
below $\text{Bo}_{cL}$. In another interpretation of the same relation,
we say that surfactants with a given Marangoni number can stabilize
the Rayleigh-Taylor instability, provided that $\text{Bo}$ is above
the threshold (\ref{eq:BoCritical}).

\begin{figure}
\includegraphics[bb=9bp 200bp 603bp 600bp,clip,width=0.95\textwidth]{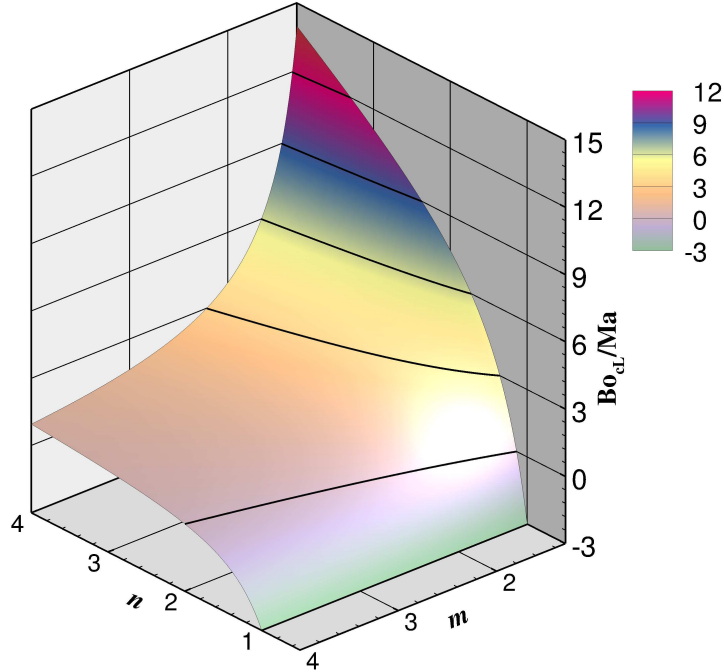}

\protect\protect\caption{The ratio $\Bo_{cL}/\text{Ma}$ as a function of the aspect ratio
and viscosity ratio for $n\ge1$ and $m>1$ (in the $R$ and $Q$
sectors). Here $\Ma=0.1$ and $s=1$.}

\label{fig:fig3} 
\end{figure}

From equation (\ref{eq:BoCritical}) the ratio $\text{Bo}_{cL}/\text{Ma}$
is a function of $m$ and $n$ only, and its graph is a surface in
the $(n,m,\text{Bo}_{cL}/\text{Ma})$-space. Figure \ref{fig:fig3}
represents the surface of the threshold ratio $\Bo_{cL}/\text{Ma}$
for the $R$ and $Q$ sectors combined, that is for the region $n>1$
and $m>1$. In particular, figure \ref{fig:fig3} reflects the fact
(which is clear from equation (\ref{eq:BoCritical}), in view of the
factor $m-1$ in the denominator) that $\Bo_{cL}\uparrow\infty$ as
$m\downarrow1$ at a fixed $n$ (which implies the $R$ sector). This
growth of $\text{Bo}_{cL}$ as $m\downarrow1$ is especially pronounced
for larger aspect ratios, as we see in the figure for the largest
value included, $n=4$; while for $n=1$ the critical ratio is constant,
$\text{Bo}_{cL}/\text{Ma}=-3$. (Because of the threshold ratio being
infinite at $m=1$, it is impossible to include in the figure all
the values of $m$ down to $m=1$; the cutoff in the figure \ref{fig:fig3}
is at $m=1.3$. ) Also, from equation (\ref{eq:BoCritical}), $\Bo_{cL}\downarrow0$
as $m\rightarrow n^{2}$ (in both $R$ and $Q$ sectors). In figure
\ref{fig:fig3}, the corresponding zero-level horizontal cross-section
is highlighted to appear different from the other horizontal cross-sections;
it is, clearly, the curve $m=n^{2}$ in the coordinate $(n,m)$-plane.
In contrast to the $R$ sector, in the $Q$ sector the (negative)
$\text{Bo}_{cL}$, is bounded: equation (\ref{eq:BoCritical}) has
a finite value at $n=1$, and a finite limit as $m\uparrow\infty,$
since the expression $\varphi$ (see equation (\ref{phi})) is linear
in $m$. 

In the $S$ sector ($1<n<\infty$ and $0<m<1$), the robust branch
(\ref{gamCsmallAlphaApprox}) is unstable when the Bond number is
below the threshold value given by equation (\ref{eq:BoCritical})
(which is negative in this sector since the Marangoni action is stabilizing,
in contrast to the $R$ sector) and stable otherwise. As to the surfactant
branch in this sector, equation (\ref{gamSsmallAlphaApprox}), which
does not contain the Bond number, indicates instability. Thus the
surfactant mode is unstable for any $\text{Bo}$ provided $\alpha$
is sufficiently small. The conclusion that no amount of gravity can
completely stabilize the flow may come across as somewhat counterintuitive.

We have omitted the stabilizing effects of capillary pressure since,
as was explained before, we are concerned with the instability threshold
and this is determined in the long-wave limit, in which these effects
are negligible. More detailed linear stability results, including
the asymptotic properties of the growth rate near the marginal wavenumber
and also on the borders between the $Q$, $R$ and $S$ sectors, are
found (in a different way) in \citet{schweiger2013gravity}, and will
be further investigated elsewhere.

\section{Phase shifts and their limitations as regards criteria of stability}

\label{sec:phase differences}

In the earlier literature, dealing with the case of zero gravity,
the authors have attempted to show that there must be correlations
of stability/instability of normal modes with the phase shift between
the surfactant and the interface displacement (FH) and/or, on the
other hand, between the interfacial vorticity and the interface displacement
(\citet{Wei2005c}, referred to as W below) being in certain two non-intersecting
subintervals splitting the total (-$\pi$, $\pi$) range of the phase.
Considering the more general case, of nonzero gravity, enables us
to clarify such statements and show their limitations, which is one
of the topic of the present section. We use the eigenfunction $G/h$
of the system (\ref{eq:matrix system}), a complex number whose argument
gives the phase shift considered in FH. We also find the amplitude
ratio of the bottom-layer interfacial vorticity to the interface displacement,
a complex number whose argument gives the phase shift considered in
W.

As in W, we exclude the case of $s=0$, but, in contrast to W, we
include gravity, thus allowing for $\text{Bo}\ne0$. As in W, we neglect
the capillary pressure.

\subsection{Robust branch}

We first consider the robust mode. In the first equation of the system
(\ref{eq:matrix system}) we note that the leading order imaginary
part of $\gamma$ determined by (\ref{eq:I1-1}) exactly cancels the
imaginary term next to $\gamma$. In the next order, the term $iI_{3}\alpha^{3}$
is added in $\gamma$. We mentioned before that $I_{3}$ includes
the term $-c_{13}/\psi$, which can be calculated only after the lubrication
approximation is corrected to the next order in $\alpha^{2}.$ However,
one realizes that the so corrected quadratic equation for the increment
means that the left upper entry of the matrix in (\ref{eq:matrix system})
must acquire an additional term $ic_{13}\alpha^{3}/\psi$. The latter
cancels the term with $c_{13}$ in the term $iI_{3}\alpha^{3}$ of
$\gamma$. Also, the $\textrm{Bo}$ part of $\gamma$ in the first
equation of the system (\ref{eq:matrix system}) exactly cancels the
$\textrm{Bo}$ term external to $\gamma$ in the left upper entry
of the system coefficient matrix. Thus, this entry consists of just
the real $\text{Ma}$ part of $\gamma$, i.e. the first term of (\ref{gamCsmallAlphaApprox}),
plus the imaginary term $iI_{3}^{-}\alpha^{3}$ (\ref{eq:gamma series-1}),
where 
\begin{equation}
I_{3}^{-}=-\psi\frac{c_{03}^{2}}{c_{11}^{3}}+\frac{c_{12}c_{03}}{c_{11}^{2}}-\frac{c_{04}}{c_{11}}.\label{eq:increment I3 unstable-1-1}
\end{equation}
So the coefficient of $h$ in the first equation of the system (\ref{eq:matrix system})
is $\varphi(n^{2}-m)\textrm{Ma}\alpha^{2}/[4(m-1)\psi]+iI_{3}^{-}\alpha^{3}$
, and the coefficient of $G$ is $\frac{n^{2}(n^{2}-m)\Ma}{2\psi}\alpha^{2}$.
The latter is non-zero (in fact positive) in the $S$ sector, $m<1,$
as well as in the $R$ sector, $n^{2}>m>1$. (It is convenient to
confine our considerations to the $R$ sector for now). This leads
to (a different form of (\ref{eq:amplitude ratio for robust mode-1-1}))
\begin{equation}
\frac{G}{h}=-\dfrac{\varphi}{2n^{2}(m-1)}-2i\alpha I_{3}^{-}\frac{\psi}{n^{2}(n^{2}-m)\text{Ma}},\label{eq:amplitude ratio for robust mode}
\end{equation}
and so the phase of $G/h$ is approximately $\pi+I_{3}^{-}\alpha4(m-1)\psi/[\varphi(n^{2}-m)\textrm{Ma}]$
(assuming that the second term is small as compared to unity). Note
that while the phase of $G/h$ is approximately $\pi$ in the $R$
sector where the robust mode is unstable provided $\text{Bo}<\text{Bo}_{cL}$,
and thus, in particular, for $\text{Bo}=0$, one can see from equation
(\ref{eq:amplitude ratio for robust mode}) that this phase is close
to zero ($\approx$ $-I_{3}^{-}\alpha4(1-m)\psi/[\varphi(n^{2}-m)\textrm{Ma}]$)
in the $S$ sector, where the robust branch is stable for $\text{Bo}=0$.

Let us consider now the situation near the onset of instability (in
the $R$ sector). For a fixed value of $\text{Bo}$, let $\textrm{Ma}_{0}$
denote the critical value such that $R_{2}=0$ (so that the growth
rate is zero) at $\textrm{Ma}=\textrm{Ma}_{0}$. For the limit of
$n\gg1$ and (a positive) $m-1\sim1$, we have approximately $\textrm{Bo}\sim3n\textrm{Ma}_{0}/[4(m-1)],$
$c_{11}\sim-2n^{3}(m-1)s$, $c_{12}\sim n^{5}\textrm{Ma}_{0}/[4(m-1)]$,
$c_{03}\sim-n^{4}s\textrm{Ma}/2$, and $c_{04}\sim n^{5}\textrm{Ma}\textrm{Ma}_{0}/[16(m-1)]$
. (Also note that $\phi\sim n^{3}$ and $\psi\sim n^{4}$.) This leads
to the phase $\theta_{G}$ of $G/h$ being 
\begin{equation}
\theta_{G}=\pi+\frac{\alpha n\textrm{Ma}_{0}(m-1+n\varDelta_{M})}{8s(m-1)^{2}},\label{eq:phase G/h}
\end{equation}
where $\varDelta_{M}\text{\textmd{:}}=\textrm{Ma}/\textrm{Ma}_{0}-1$
can be as large as $O(1)$. Also, the expression for the growth rate
(\ref{gamCsmallAlphaApprox}) used here (as well as in FH and W for
the case $\textrm{Bo}=0$) comes from the expansion (\ref{eq:gamma series-1}),
with certain assumptions concerning the smallness of $\alpha$, namely
$|R_{2}|\alpha^{2}\ll|I_{1}|\alpha$. Using (\ref{eq:I1-1}) and (\ref{gamCsmallAlphaApprox})
for $n\gg1$, one can see that this requires 
\begin{equation}
\dfrac{\alpha n^{2}\textrm{Ma}}{s}\ll1.\label{eq:validity}
\end{equation}
In particular, (\ref{eq:validity}) implies that the correction to
$\pi$ in the phase expression (\ref{eq:phase G/h}) is small compared
to unity. This is discussed later in connection with the numerical
results.

We turn now to the phase shift between the vorticity and interface
amplitudes, which we denote by $\theta_{\omega}$. The $y$-component
of the interfacial vorticity in the lower layer, denoted by $\omega_{1}$,
is, to the leading order in $\alpha$, 
\begin{equation}
\omega_{1}=u_{1z}(z=0)=A_{1}\label{eq:vorticity is A_1}
\end{equation}
from equation (\ref{eq:uj vel}). (Note that the sign in (\ref{eq:vorticity is A_1})
is opposite to that in W. This is due to the fact that our coordinate
axes are related to those of W by a rotation about the $x$-axis,
so that our spanwise axis (our $y$-axis) is directed opposite to
that of W (his $z$-axis). This is the root of the opposite signs
difference between the two spanwise components of the vorticity vector.)
To find the phase $\theta_{\omega}$ of $\omega_{1}/h$, we substitute
the normal modes expressions for $\eta$ and $\Gamma$ into the linear
form of equation (\ref{eq:A1}) to obtain

\begin{align}
A_{1} & =\left(-\frac{4(m-1)(m+n^{3})s}{\psi}h+i\alpha\frac{2mn^{2}(n+1)\Bo}{\psi}h\right.\nonumber \\
 & \left.-i\alpha\frac{n(n^{3}+4m+3mn)\Ma}{\psi}G\right)e^{i\alpha x+\gamma t}\label{eq:A1 in terms of h and G}
\end{align}
We then substitute into (\ref{eq:A1 in terms of h and G}) the expression
for $G$ in terms of $h$ from equation (\ref{eq:amplitude ratio for robust mode}),
where we need just the leading order:, 
\begin{equation}
\frac{G}{h}=-\dfrac{\varphi}{2n^{2}(m-1)},\label{eq:G over h leading order robust}
\end{equation}
which simplifies to 
\begin{equation}
G=-\frac{hn}{2(m-1)}\label{eq:G in terms of h large n}
\end{equation}
in the limit of large $n$. (One may note that this signifies the
surfactant and interface are approximately in anti-phase. Clearly,
this result holds, in particular, at $\textrm{Bo}=0$.) As a result
we find 
\[
\frac{\omega_{1}}{h}=-4\frac{(n^{3}+m)(m-1)s}{\psi}+i\frac{\alpha}{\psi}\left(2n^{2}(n+1)m\text{Bo}+\frac{\varphi(n^{3}+3nm+4m)\text{Ma}}{2n(m-1)}\right).
\]
Hence, for large $n$ we have 
\begin{equation}
\theta_{\omega}=\pi-\frac{\alpha\textrm{Ma}n^{2}}{8(m-1)^{2}s}.\label{eq:vorticity phase, growing}
\end{equation}
(The last term here is small in view of (\ref{eq:validity})). This
phase expression, being independent of $\text{Bo}$, is valid in particular
for $\textrm{Bo}=0$.

For the zero-gravity case, using the same considerations with $\textrm{Bo}=0$
(in place of $\textrm{Bo}\sim3n\textrm{Ma}_{0}/[4(m-1)]$ used above),
one finds the results for the case considered in W. In particular,
the growth rate for this mode (see equation (\ref{gamCsmallAlphaApprox})),
\[
\gamma_{R}=\dfrac{n\textrm{Ma}}{4(m-1)}\alpha^{2},
\]
is positive for all Marangoni numbers, in contrast to crossing from
negative values to positive as the Marangoni number increases through
the nonzero $\textrm{Ma}_{0}$ with a fixed nonzero Bond number.

\subsection{Surfactant branch}

For the surfactant mode, there is no cancellation of the (imaginary)
order $\alpha$ terms in the first equation of the system (\ref{eq:matrix system}).
From equation (\ref{gamSsmallAlphaApprox}), to the leading order
in $\alpha$ (with $n\ne1$), the increment is real and, for $m>1$,
negative : 
\begin{equation}
\gamma=-\dfrac{(n-1)\textrm{Ma}}{4(m-1)}\alpha^{2}.\label{eq:increment of decaying lubr mode}
\end{equation}
(This is independent of the Bond number and for $\textrm{Bo}=0$ appeared
already in FH (see also W).) Equation (\ref{eq:linear-eta-evolution})
yields now the leading order term of (\ref{eq:G/h with correction surfactant branch-1-1})
\begin{equation}
\frac{G}{h}=i\dfrac{4(n+1)(m-1)s}{(n^{2}-m)\textrm{Ma}\alpha}.\label{eq:G decaying}
\end{equation}
Thus, to the leading order, 
\[
\theta_{G}=\frac{\pi}{2}.
\]
Using $G$ in terms of $h$ in equation (\ref{eq:A1 in terms of h and G}),
one obtains that the leading-order terms (the first and last terms
on the r.h.s.) cancel to zero. Therefore, we need to use the full,
two-order, expression $G/h$, equation (\ref{eq:G/h with correction surfactant branch-1-1}).This
leads to 
\[
\frac{\omega_{1}}{h}=\dfrac{4(m-1)s}{n^{2}-m}+i\alpha\left(\dfrac{2n^{2}\textrm{Bo}}{3(n^{2}-m)}-\dfrac{(n^{3}+3nm+4m)(n-1)\textrm{Ma}}{2n(n^{2}-m)(m-1)}\right).
\]
Thus, for large $n$, the phase shift, to the leading order, is 
\begin{equation}
\theta_{\omega}=\arctan\left[\left(\dfrac{2\textrm{Bo}}{3}-\dfrac{n\textrm{Ma}}{2(m-1)}\right)\alpha\dfrac{n^{2}}{4(m-1)s}\right].\label{eq:vorticity phase for zero Bo, decaying}
\end{equation}
For the zero gravity case, we have 
\[
\theta_{\omega}=\arctan\left[-\dfrac{(n^{3}+3nm+4m)(n-1)\textrm{Ma}}{8n(m-1)^{2}s}\alpha\right].
\]
Note that for large $n$, the argument of the arctan here is of the
order of the product of the small parameter (\ref{eq:validity}) and
the large factor $n$, so it can range from small to large. Only when
it is large (e.g., for $n=1000$, $\textrm{Ma}=1$, $\alpha=10^{-6},$
$m=2$, and $s=10$), we have the result, $\theta_{\omega}=-\pi/2$,
which was (erroneously) stated in W as a general one. When $\textrm{Bo}$
is nonzero, for $\text{Ma}$ near $\textrm{Ma}_{0}$ (that is, for
the fixed $\textrm{Bo}$, the threshold value for the robust mode,
used just formally here for the surfactant mode, which is decaying
in the $R$ sector for all $\text{Ma}$ and $\text{Bo}$), we have
\begin{equation}
\theta_{\omega}=\arctan\left[-\dfrac{n^{3}\textrm{Ma}_{0}\Delta_{M}\alpha}{8(m-1)^{2}s}\right],\label{eq:vorticity phase for nonzero Bo, decying}
\end{equation}
and so, as the Marangoni number is increased, the phase shift crosses
at $\textrm{Ma}_{0}$ from positive to negative values. This can be
re-stated in terms of a fixed Marangoni number and increasing Bond
number: 
\[
\theta_{\omega}=\arctan\left[-\dfrac{n^{2}\textrm{Bo}_{0}\Delta_{B}\alpha}{6(m-1)^{2}s}\right],
\]
where $\textrm{Bo}_{0}$ is the ``critical'' value of $\textrm{Bo}$
and $\Delta_{B}=(\textrm{Bo}-\textrm{Bo}_{0})/\textrm{Bo}_{0}$. Therefore,
as the growing $\textrm{Bo}$ passes through $\textrm{Bo}_{0}$, the
vorticity-interface phase shift also grows and crosses zero, changing
its sign from negative to positive.

\subsection{Disconnect of phase differences from stability/instability at nonzero
gravity}

\begin{figure}
\includegraphics[clip,width=0.95\textwidth]{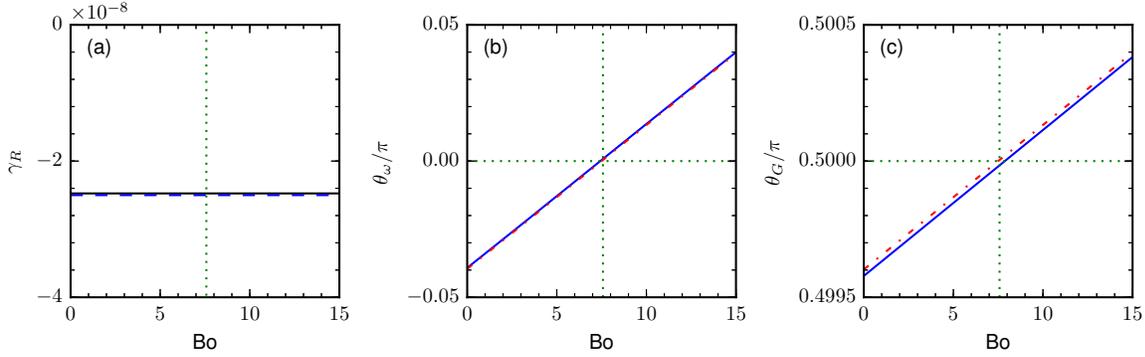}\caption{Dependencies on the Bond number for the surfactant mode in the $R$
sector, with $m=2$ and $n=100$: (a) growth rate, (b) phase difference
between disturbances of the interfacial vorticity and the interface
, and (c) phase difference between surfactant and interface disturbances
. The solid lines are full numeric results and the dashed lines are
large $n$ asymptotic approximations. The other parameters are $\text{Ma=0.1},$
$s=10$ and $\alpha=10^{-4}$. \label{fig:figD1}}
\end{figure}

The W paper, next to the statement (equivalent to (\ref{eq:vorticity phase for zero Bo, decaying}))
concerning the decaying mode, only tells about the growing mode that
it ``does the opposite'' of the decaying one. This may suggest to
a reader the (wrong) phase shift of $\pi/2$ for the unstable mode;
whereas in fact, according to (\ref{eq:vorticity phase, growing}),
the vorticity-interface phase shift is close to $\pi$ (which would
correspond to neutral stability according to W), and only the correction
puts the phase shift into the interval $(0,\pi)$, which corresponds
to instability according to W. (Note that the value of $\theta_{\omega}$
given by (\ref{eq:vorticity phase for zero Bo, decaying}) is within
the interval $(-\pi,0)$ corresponding, according to W, to stability.)
However, this correspondence between the intervals $(-\pi,0)$/$(0,\pi)$
and stability/instability evidently does not necessarily hold for
nonzero Bond numbers: first, as we established above, on passing the
Bond number value $\textrm{Bo}_{0}$, the vorticity-interface phase
shift goes from negative to positive values, although the mode remains
stable all along. In Figure (\ref{fig:figD1}), the full numerical
results (based on the numerical solution of the linear system (\ref{eq:matrix system}))
for the decaying mode are plotted along with the asymptotic approximate
dependencies obtained above for large $n$. Note that the first correction
to $\theta_{G}/\pi$ has been included in Figure \ref{fig:figD1}(c).
One can see excellent agreement, and in some cases the difference
is even hardly discernible.

\begin{figure}
\includegraphics[bb=0bp 0bp 576bp 180bp,clip,width=0.95\textwidth]{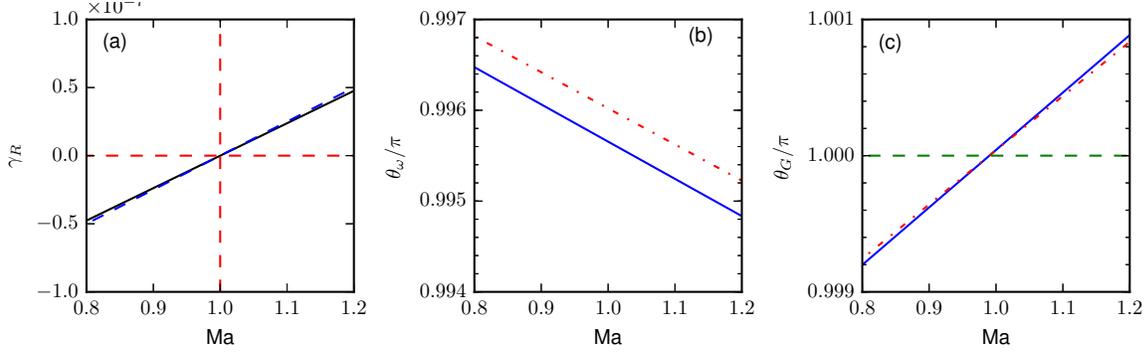}\caption{Dependencies on the Marangoni number for the robust mode in the $R$
sector, with $m=2$, $n=100$, $s=10$, and $\alpha=10^{-4}$. Here
$\text{Bo=Bo}_{cL}(\text{Ma=1)}$, so that, as (a) shows, the growth
rate $\gamma_{r}$ crosses 0 at $\text{Ma}=1$. The other panels show
(b) the phase difference between disturbances of the interfacial vorticity
and the interface, and (c) the phase difference between surfactant
and interface disturbances. The solid lines are full numeric results
and the dashed lines are large $n$ asymptotic approximations. \label{fig:figD2}}
\end{figure}

Also, Figure (\ref{fig:figD2}) shows that at a fixed Bo, as the Marangoni
number passes through the threshold value at which the instability
sets in for the robust mode, the vorticity-interface phase shift remains
within the same subinterval $(0,\pi)$ (of the full phase range $(-\pi,\pi)$)
for both the stability and instability subintervals of the Marangoni
number. Clearly, the phase differences $\theta_{\omega}$ and $\theta_{G}$,
plotted respectively in parts (b) and (c), are close to $\pi$ over
the displayed range of $\text{Ma}$, and $\theta_{\omega}$ remains
below $\pi$. In this figure as well, the full numerical results (solid
lines) are plotted along with the asymptotic approximate dependencies
(dashed lines), and we see very good agreement between them.

To understand why the 'vorticity argument' of W does not work for
the nonzero gravity case, we retrace the consideration in W. It is
based on the kinematic relation (at the bottom of p. 185 of W) between
the rate of growth (in time) of the interface displacement and the
disturbance flow rate of the bottom layer. In terms of amplitudes
it is given by 
\[
\gamma h=-i\alpha Q_{1},
\]
where $Q_{1}=\intop_{-1}^{0}u_{1}dz$ is the flow rate in the bottom
layer. Then a positive coefficient of proportionality is shown between
this flow rate and the (disturbance) vorticity of the bottom layer
at the basic interface, $\omega_{1}$, whenever the vorticity of the
top layer, $\omega_{2}$, may be neglected; for example, when $n$
is large and $m-1$ is not large as compared to unity. It follows
that 
\[
\theta(\gamma)=-\dfrac{\pi}{2}+\theta_{\omega},
\]
where $\theta(\gamma)$ denotes the argument of the increment. Since
instability means positive growth rate, it clearly implies that $-\pi/2<\theta(\gamma)<\pi/2$,
i.e. $0<\theta_{\omega}<\pi$. Thus, the instability corresponds to
the interval $(0,\pi)$ of the vorticity-interface phase difference.
Similarly, one finds that stability corresponds to this phase shift
being between $0$ and $-\pi$. (We call these correspondences ``the
rule of vorticity phase intervals''.) However, we find that the proportionality
between the flow rate and vorticity breaks down when $\textrm{Bo}$
is nonzero, because an extra term proportional to $\textrm{Bo}$ appears
in the relation: the W equation (25) modified by gravity is (in our
notation and in terms of amplitudes) 
\[
Q_{1}=\dfrac{n^{2}m}{2(m+n^{3})}\left(\dfrac{n}{m}\omega_{1}+\omega_{2}\right)-\textrm{Bo}\dfrac{n^{3}}{3(m+n^{3})}i\alpha h.
\]
Thus, ``the rule of vorticity phase intervals' for stability/instability
suggested by W, in general, may fail even when the $\omega_{2}$ term
is negligible. It will hold only when the $\textrm{Bo}$ term is also
negligible, along with the $\omega_{2}$ one.

In general, it is clear that, dynamically, only the surfactant is
responsible for instability in the absence of gravity. The nonzero
vorticity component is just one of the kinematic fields that are present
even in the absence of surfactants. It is interesting to consider,
instead of vorticity, the upward component of the disturbance velocity
$w_{1}$ in the bottom layer. From equation (\ref{eq:gamma is w/h}),
it is clear the growth rate $\gamma_{R}=Re(w/h)$, where for simplicity
we use the notation $w=w_{1}(0)$, and the wave velocity $c=-\alpha^{-1}Im(w/h)$.
Hence one obtains the universal ``rule of velocity phase intervals''
for stability/instability: we have instability when $\theta_{w},$
the upward-velocity phase shift relative to the interface, is in the
interval $(-\pi/2,\pi/2)$, and stability corresponds to the rest
of the interval $(-\pi,\pi)$, that is the open region. Consider first
the robust branch by using the shuttling method described above. We
have $Im(\gamma)\sim\alpha\gg\alpha^{2}\sim Re(\gamma)$., From (\ref{eq:linear-eta-evolution}),
$Im(\gamma)$ is positive in the $Q$ and $R$ sectors, and hence
$\theta_{w}$ is close to $\pi/2$. Similarly, in the $S$ sector,
$\theta_{w}\approx-\pi/2$. The magnitude of the small corrections
to these values is $\left|Re(\gamma)/Im(\gamma)\right|=\alpha\left|R_{2}/I_{1}\right|$,
and it is easy to see that for the cases of instability, $R_{2}>0,$
the phase shift $\theta_{w}$ is indeed in the interval $(-\pi/2,\pi/2)$,
while stability cases correspond to $\theta_{w}$ being outside this
interval (and close to its endpoints).

For the surfactant branch, $Re(\gamma)\sim\alpha^{2}\gg\alpha^{3}\sim Im(\gamma)$,
so the phase shift $\theta_{w}$ is close to $0$ for the unstable
modes in the $S$ sector, and close to $\pi$, or $-\pi$, for the
stable modes in the $R$ and $Q$ sectors. Again, we see that instability
corresponds to the interval $(-\pi/2,\pi/2)$ of the phase shift $\theta_{w}$
while stability corresponds to the complementary region $(-\pi,-\pi/2)\cup(\pi/2,\pi)$.
This ''rule of the upward velocity phase intervals'' is universally
true, even in the presence of gravity. However, it does not give any
advantages in determining the stability properties over simply considering
the system of the thickness and surfactant equations by the shuttling
method. The same is true for the rule of vorticity phase intervals
(besides, as we have seen above, the latter may fail altogether in
the presence of gravity,- and in many other situations as well.)

The surfactant-interface phase shift was considered in FH for a particular
case, with $m=1$ and the semi-bounded geometry, only in order to
make a plausibility argument about how the instability was possible
- since there seemed to be in the literature the (erroneous) idea,
based on systems with no base flow, that the surfactant is always
stabilizing. The surfactant concentration, in addition to being dynamically
natural, has the advantage over the vorticity (or velocity) that it
is a scalar, and not a component of a vector. The plausible ``rule
of surfactant phase interval'' seemed to be that stability corresponds
to the phase shift being closer to the in-phase case, that is to being
in the interval $(-\pi/2,\pi/2)$ (excluding its (corresponding to
neutral stability) midpoint, $\theta_{G}=0$). The above considerations
show that in the absence of gravity this rule works also for the case
(not considered in this regard in FH) with $m\neq1$, and (bounded)
channel flows. However, in the presence of gravity this phase interval
rule is also fallible: figure \ref{fig:figD2} shows that this phase
shift remains in the same interval, close to $\pi$ (or $-\pi$, in
other words), as Ma is increased through the threshold value and stability
gives way to instability.

We conclude that, in general, any of the phase shifts in the normal
modes appear to be hardly suitable for explaining the mechanisms of
the instability. Instead, as we show in the next section, the horizontal
velocity constituents which have a quarter-circle phase shift relative
to the interface play a key role in the instability mechanism.

\section{Instability mechanisms: Marangoni stresses and out-of-phase velocities}

\label{sec:mechanical}

In this section, we endeavor to elucidate the mechanism of instability
for the two branches of normal modes, somewhat in the spirit of \citet{Charru2000}
(CH for short) Like them, we will sometimes use dimensional quantities
. For simplicity, we omit the stars in their notations for the latter,
and also omit tildes denoting disturbances; the context indications
are sufficient for avoiding confusion. Like in CH, we first consider
the case with large aspect ratio $n$ and no gravity; these restrictions
will be relaxed in the last subsection of this section. (However,
unlike CH, interfacial surfactant is present, and may cause instability
despite the absence of inertia. In contrast, inertia is necessary
for Yih's instability treated in CH.) Similar to CH, all parameters
other than $n$ are tentatively assumed to be of order one.

\subsection{Robust branch}

Considering a small-amplitude, normal-mode disturbance of the interface,
$\eta=h\exp(\Sigma t)$ $\cos\alpha(x-ct)$ (where $\Sigma:=\Real(\gamma)$
is by definition the growth rate, the real part of the increment,
and $c:=-\alpha^{-1}\Imag(\gamma)$ is the wave velocity), and first
neglecting the surfactant disturbance, we can repeat the considerations
of CH to find the same leading order flows. Namely, the flow in the
thick layer is a pressure-gradient driven one with the zero net flow
rate, that is (omitting hats for the amplitudes) 
\begin{equation}
u_{2}=u_{0}\left(1-\dfrac{z}{d_{2}}\right)\left(1-3\dfrac{z}{d_{2}}\right).\label{eq:u_2}
\end{equation}
Here 
\begin{equation}
u_{0}:=u_{2}(0)=\frac{\mu_{2}-\mu_{1}}{\mu_{2}}sh,\label{eq:u_2 at 0}
\end{equation}
where $s$ is the (dimensional) base shear rate. Clearly, the pressure
gradient is $6\mu_{2}u_{0}/d_{2}^{2}$. Its interfacial shear stress
drives a Couette flow in the thin layer, 
\begin{equation}
u_{1}=-4u_{0}\frac{\mu_{2}/\mu_{1}}{d_{2}/d_{1}}\left(1+\frac{z}{d_{1}}\right).\label{eq:u_1}
\end{equation}
(The two tangential stresses, of (\ref{eq:u_2}) and (\ref{eq:u_1}),
are equal as they should be since for this branch the surfactant Marangoni
stress is negligible in the leading order.) The balance of mass in
a control volume {[}$0\leq x\leq\lambda/2,$ $z<0${]} over a short
time $0<t<\delta t$ gives, exactly as in CH, the wave velocity 
\begin{equation}
\frac{c}{sd_{1}}=-2\frac{(m-1)}{n}.\label{eq:wave velocity}
\end{equation}
(This result corresponds to the large $n$ limit of (\ref{eq:I1-1}).
Notably, the sign of $c$ is opposite to that of $m-1$. We note also
that while $u_{2}(0)\sim(1/n)^{0}$, the transverse differentiation
reduces the order: $Du_{2}(0)\sim(1/n)^{1}$, which is then the order
of the tangential stress; and $D^{2}u_{2}(0)\sim(1/n)^{2}$, which
is clearly the order of the pressure gradient $6\mu_{2}u_{0}/d_{2}^{2}$.)

However, instead of originating from inertia, in our case the velocity
correction to this flow, denoted $u_{1}^{M}$, comes by the Marangoni
action of the surfactant disturbance $\varGamma=|G|\exp(\Sigma t)$
$\cos(\alpha(x-ct)+\theta_{\varGamma})$, where $\theta_{\varGamma}$
is a phase shift (undetermined for the moment), that is the argument
of the complex amplitude of the surfactant eigenfunction, $G=|G|\exp i\theta_{\varGamma}$.
The Marangoni tangential stress should balance the viscous stress
of the thin film, as the thick film contribution is much smaller in
the interfacial condition (5); indeed, the interfacial (correction)
velocities of the layers must be equal, and then the shear rate in
the thick layer is $d_{1}/d_{2}$ times smaller than in the thin one.
Thus, we find the linear velocity profile (satisfying also the no
slip condition at the bottom plate): 
\begin{equation}
u_{1}^{M}=-\frac{\sigma_{0}}{\mu_{1}\varGamma_{0}}\textrm{Ma}i\alpha G(z+d_{1}).\label{eq:correction velocity in terms of G}
\end{equation}
(This corresponds to the large $n$ limit of the dimensionless equation
(\ref{eq:im u1 robust}). Similarly, $u_{2}^{M}$ would correspond
to equation (\ref{eq:im u2 robust}); however, like in CH, it is not
needed in the argument.) Note that the velocity phase here is $90\si{\degree}$
less than that of the surfactant. The surfactant transport equation
(\ref{dSurfConc}), at the leading order, yields 
\begin{equation}
\frac{\partial\Gamma}{\partial t}+\varGamma_{0}\frac{\partial}{\partial x}\left(\overline{u}_{1}(\eta)\right)=0,\label{eq:surfactant transport}
\end{equation}
where $\varGamma_{0}$ is the (uniform) base surfactant concentration
and $\overline{u}_{1}(\eta)=s\eta$ is the base velocity of the thin
layer (\ref{u1w1p1BSprofiles}). Hence, in terms of the amplitudes,
we get, at the leading order, 
\[
-ic\alpha G+\varGamma_{0}si\alpha h=0,
\]
or 
\begin{equation}
G=h\frac{\varGamma_{0}s}{c}=-h\frac{\varGamma_{0}n}{2d_{1}(m-1)},\label{eq:G in terms of H}
\end{equation}
where the second equality follows from equation (\ref{eq:wave velocity})
(cf. equation (\ref{eq:G in terms of h large n})). (It follows that
for $m>1$ the phase shift of the surfactant wave relative to the
interface displacement wave is $\theta_{\varGamma}=\pi$, i.e. they
are in anti-phase, and for $m<1$ the phase shift is zero, so the
surfactant disturbance is in phase with the interface displacement.)
Substituting (\ref{eq:G in terms of H}) into the velocity expression
(\ref{eq:correction velocity in terms of G}), we obtain the flow
in terms of the interface displacement amplitude $h$: 
\begin{equation}
u_{1}^{M}=h\frac{\sigma_{0}}{\mu_{1}}\textrm{Ma}i\alpha\frac{n}{2(m-1)}\left(\frac{z}{d_{1}}+1\right),\label{eq:velocity in terms of h}
\end{equation}
Since this velocity amplitude is purely imaginary, the velocity is
out-of-phase with the interface, either by $90\si{\degree}$ for $m>1$,
or $-90\si{\degree}$ for $m<1$. (We note that there is another correction
velocity, of order $(1/n)^{2}$, needed to satisfy the normal stress
condition, the equality of pressures. However, its amplitude is real
and hence it is either in phase or anti-phase with the interface,
and not $\pm90\si{\degree}$ out-of-phase as in (\ref{eq:velocity in terms of h}).
Therefore, it leads to a small correction to the wave velocity, but
is irrelevant to the growth rate.) The growth rate is found, as in
CH, from the mass conservation law, by equating the change of volume
of the thin layer over the interval $\lambda/4\leq x\leq3\lambda/4$
(where $\lambda:=2\pi/\alpha$ is the wavelength), over a small time
interval $\delta t$, to the sum of inflow volumes through the two
boundaries, at $x=\lambda/4$ and $x=3\lambda/4$ (at which point
the flux magnitude attains its maximum, since the amplitude of the
fluid flux is purely imaginary along with the velocity amplitude).
Here, the inflow through the left boundary is given by the mid-layer
correction velocity, that is half the interfacial velocity at the
boundary location, times the layer thickness $d_{1}$, and that at
the right boundary is similarly equal to minus the interfacial velocity
at that boundary location times $d_{1}/2$. The velocity at $x=\lambda/4$
is $u_{1}^{M}(0)\exp i\alpha\lambda/4=iu_{1}^{M}(0)$, where from
equation (\ref{eq:velocity in terms of h}), 
\begin{equation}
iu_{1}^{M}(0)=-h\frac{\sigma_{0}}{\mu_{1}}\textrm{Ma}\alpha\frac{n}{2(m-1)}.\label{eq:velocity at left boundary}
\end{equation}
The expression on the right-hand side here is similarly found to also
give the negative of the velocity at the right boundary, so the two
boundary inflows are exactly equal. As a result, the mass conservation
equation is 
\[
iu_{1}^{M}(0)d_{1}(\delta t)=\intop_{\lambda/4}^{3\lambda/4}h(e^{\Sigma\delta t}-1)(\cos\alpha x)dx=-\frac{2h\Sigma}{\alpha}(\delta t).
\]
Substituting here for $iu_{1}^{M}(0)$ given by (\ref{eq:velocity at left boundary}),
we solve for the growth rate and obtain 
\begin{equation}
s\Sigma=\alpha^{2}\frac{\sigma_{0}n\textrm{Ma}d_{1}}{4\mu_{1}(m-1)}.\label{eq:Sigma_plus}
\end{equation}
(This result corresponds to the large $n$ limit of equation (\ref{gamCsmallAlphaApprox})
with $\text{Bo}=0$, and allows one to identify this mode as the robust
one.) Clearly, this corresponds to instability for $m>1$. Recall
that for $m>1$, the surfactant and the interface displacement are
in anti-phase. For $m<1$, we have in-phase propagation of the surfactant
and interface-displacement waves, hence the reversed velocities, and
consequently the stability of the normal mode. (A (different) link
between the surfactant-interface phase shift and the stability of
the normal mode was first noted in FH for a case with $m=1$.)

A clearly equivalent way to this integral mass (of the fluid) conservation
method of finding the growth rate is as follows. Use the divergenceless
relation, with the horizontal velocity correction $u_{1}^{M}$ given
by equation (\ref{eq:velocity in terms of h}), to determine the vertical
velocity correction $w_{1}^{M}(z=0)$. Then the corresponding (real)
correction to the increment is found from equation (\ref{eq:gamma is w/h})
to be $\varSigma=$$w_{1}^{M}(z=0)/h$.

In summary, the growth/decay mechanism for the robust branch is described
as follows. The leading order flow is the same as in \citet{Yih1967}
and leads to the same, imaginary, increment. Then, the surfactant
wave of the normal mode must propagate either in anti-phase for $m>1$
or in phase for $m<1$ with the interface. The Marangoni tangential
stress exerted by the surfactant drives a (linear-profile) correction
flow whose velocity $u_{1}^{M}$ is $-90\si{\degree}$ out-of-phase
with the surfactant. Thus, this velocity is either $90\si{\degree}$,
for $m>1$, or $-90\si{\degree}$, for $m<1$, out of phase with the
interface. For $m>1$, this leads to a net outflow for the half-period
part of the thin layer with the thickness minimum at the middle point,
that is instability for the normal mode, and for $m<1$, the velocity
is reversed, which yields stability. (Note that in difference with
the Yih instability induced by inertia which was explained in \citet{Charru2000}
in terms of vorticity, the latter does not play any natural role in
the mechanism of the surfactant driven instability.)

\subsection{Surfactant branch\label{subsec:Surfactant-branch}}

In this subsection we consider the other normal mode, the surfactant
mode. It turns out that here the surfactant effect appears at the
leading order of disturbances. At the interface, the shear stress
exerted on the thin layer by the thick layer (whose velocity is still
given by (\ref{eq:u_2})) and the Marangoni stress cancel each other
to the leading order in $1/n$. Therefore, the velocity is zero to
this order, and hence the wave velocity to the order $\alpha$, in
contrast to the robust mode, is zero. (There is a weak, pressure-gradient
related, flow in the thin layer, of order $(1/n)^{2}$ (see (\ref{eq:re u1 surfactant}));
however, as was discussed regarding the robust mode, it is irrelevant
to the growth rate, and actually gives a zero contribution to the
increment.) The condition of such cancellation of the stresses is
\[
-\frac{4u_{0}\mu_{2}}{d_{2}}=\frac{\sigma_{0}}{\varGamma_{0}}\textrm{Ma}i\alpha G.
\]
Using the expression for $u_{0}$, equation (\ref{eq:u_2 at 0}),
we find the following relation between $G$ and $h$, 
\begin{equation}
G=ih\frac{4\varGamma_{0}(\mu_{2}-\mu_{1})s}{\alpha\textrm{Ma}\sigma_{0}d_{2}}.\label{eq:h/G}
\end{equation}
(Note that this is equivalent to the leading order in $\alpha^{1}$
of $G/h$ found from the film thickness equation (\ref{eq:linear-eta-evolution})
, as in the first step in the 'shuttling' method discussed above.
Also, note that this $G$ is purely imaginary (unlike being real for
the robust mode), whereas the surfactant flux is always real at leading
order. Hence, the growth rate of the surfactant mode is found (immediately
below) by using the surfactant conservation law in integral form.)
One finds the growth rate of this branch by equating the change over
a small time $\delta t$ of the total quantity of the surfactant over
the interval $0\leq x\leq\lambda/2$, with the surfactant disturbance
inflow through the interface boundaries, at $x=0$ and $x=\lambda/2$.
The inflow rate is, at the leading order, $\Gamma_{0}\bar{u}_{1}(\mbox{\ensuremath{\eta})}=\Gamma_{0}sh\cos\alpha x$,
(note that this surfactant flux is in phase with the interface, and
thus is positive (and maximum) at $x=0$ and negative at $x=\lambda/2$,
corresponding to a positive net influx of the surfactant into the
control part of the interface), which gives the sums of the inflows
through the two boundaries to be $2\Gamma_{0}sh\delta t$. The surfactant
wave is $\varGamma=G_{I}\cos(\alpha x+\pi/2)=-G_{I}\sin\alpha x,$
where, in view of equation (\ref{eq:h/G}), 
\begin{equation}
G_{I}:=h\frac{4\varGamma_{0}(\mu_{2}-\mu_{1})s}{\alpha\textrm{Ma}\sigma_{0}d_{2}}\label{eq:G_I}
\end{equation}
is real. We see that the surfactant concentration reaches its minimum
value at the middle point of the interval $0\leq x\leq\lambda/2$
for $m>1$, since then the phase shift of the surfactant (from the
interface) is $90\si{\degree}$; but reaches its maximum value for
$m<1$ since then the phase shift of the surfactant is $-90\si{\degree}$.
Together with the aforementioned positive net influx of surfactant
through the interval endpoints, this yields stability for $m>1$ and
instability for $m<1$. Quantitatively, the integral form of the mass
conservation law for the surfactant implies 
\[
2\Gamma_{0}sh(\delta t)=\intop_{0}^{\lambda/2}-G_{I}(e^{\Sigma\delta t}-1)(\sin\alpha x)dx,=-\frac{2\Sigma(\delta t)G_{I}}{\alpha}.
\]
(Note that this equation is equivalent to using the differential surfactant
equation (\ref{eq:linear-surfactant-evolution}), which is the second
step in the 'shuttling' method.) Substituting the expression (\ref{eq:G_I})
for $G_{I}$, we arrive at the growth rate 
\begin{equation}
\Sigma=-\alpha^{2}\textrm{Ma}\frac{\sigma_{0}d_{2}}{4\mu_{1}(m-1)},\label{eq:Sigma}
\end{equation}
in agreement with the large $n$ limit of (\ref{eq:increment of decaying lubr mode}).
Clearly, this formula shows stability for $m>1$, which corresponds
to the phase shift $\theta_{G}=$ $\pi/2$ (see (\ref{eq:h/G})),
and instability for $m<1$, corresponding to $\theta_{G}=$ $-\pi/2$
.

This growth/decay mechanism is summarized as follows. The leading
order flow in the thick layer is the same as that for the robust branch,
but vanishes in the thin layer because of the cancellation of the
tangential stress of the thick layer by the Marangoni stress. This
cancellation requires that the surfactant phase shift with respect
to the interface is $90\si{\degree}$ for $m>1$ and $-90\si{\degree}$
for $m<1$; whereas the surfactant flux (the product of the base concentration
and the base velocity at the disturbed interface) is always in phase
with the interface. Hence, the surfactant flux is out-of-phase with
the surfactant wave, $90\si{\degree}$ for $m>1$ and $-90\si{\degree}$
for $m<1$. Thus, considering the surfactant for $0\leq x\leq\lambda/2$,
the net influx through the endpoints is always positive. For $m>1$,
the surfactant concentration is a minimum at the midpoint, and the
positive influx implies stability. For $m<1$, the surfactant concentration
is a maximum at the midpoint, and the positive influx implies instability.

(If the assumed cancellation of the tangential stresses is relaxed,
the two stresses in question are still of the same order, and this
implies that $h\sim i\alpha G$. Since $u_{1}\sim h,$ we get $w_{1}(0)\sim i\alpha h,$
and then (\ref{eq:gamma is w/h}) yields $\gamma\sim i\alpha$. Hence,
in the surfactant equation, the left-hand side term is $\gamma G\sim\alpha G$,
while no term on the right-hand side is of a lower order than $i\alpha h\sim\alpha^{2}G$.
Thus, the leading-order term $\gamma G$ cannot be balanced. This
contradiction can be resolved only by returning to the cancellation
of the tangential stresses.)

As a consistency check, this growth rate, as well as the wave velocity
for this branch, can be also recovered in the manner that was used
for the other branch, by considering the volume balance of the bottom
liquid film over the intervals of length $\pi/\alpha$, starting at
$x=\pi/2\alpha$ and $x=0$, correspondingly. (Here, we use the dimensionless
form of all quantities for the rest of this subsection). For this,
the leading non-vanishing approximations of both the real and the
imaginary parts of the velocity $u_{1}(z)$, equation (\ref{eq:uj vel}),
are needed. To find them, the pressure gradient (\ref{eq:p1x}) and
the coefficient $A_{1}$ (\ref{eq:A1}) of the linear part of the
velocity are linearized in $\eta$ and used in (\ref{eq:uj vel})
so that the velocity is expressed in terms of $h$ and $G$ (using
the amplitudes of the normal mode): 
\begin{align}
i\alpha p_{1} & =\psi^{-1}[-6(m-1)(m-n^{2})sh-6i\alpha mn(n+1)\textrm{Ma}G\nonumber \\
 & +i\alpha(3m+4mn+n^{2})n^{2}(\textrm{Bo+}\alpha^{2})h]\label{eq:p_1x linear}
\end{align}
and 
\begin{align}
A_{1} & =\psi^{-1}[-4(m-1)(m+n^{3})sh-i\alpha(4m+3mn+n^{3})n\textrm{Ma}G\nonumber \\
 & +2i\alpha m(n+1)n^{2}(\textrm{Bo+}\alpha^{2})h].\label{eq:A_1 linear}
\end{align}
For the present case only, the $sh$ and $\textrm{Ma}G$ terms are
retained in the above expressions, and the last terms containing gravity
and capillarity are neglected. We express $G$ in terms of $h$ by
using equation (\ref{eq:G/h with correction surfactant branch-1-1}),
retaining only the necessary leading orders in the limit of large
$n$ 
\[
i\alpha\textrm{Ma}G=-\dfrac{4(n+1)(m-1)s}{n^{2}}h+i\alpha\dfrac{n\textrm{Ma}}{2(m-1)}h.
\]
Note the factor $(n+1)$ in the first term has been retained because
the contribution corresponding to $n$ in ($n+1)$ clearly cancels
with the order $n^{3}$ coefficient of $sh$ in (\ref{eq:A_1 linear}).
The real part of $A_{1}$ is therefore found to be of order $n^{-4}n^{2}=n^{-2}$
since $\psi$ is of order $n^{4}$. On the other hand, we can see
that the first term in $i\alpha p_{1}$ is of order $n^{2}$, much
greater than the real part of the second term of $i\alpha p_{1}$.
As a result, we find 
\begin{equation}
u_{1}(z)=\dfrac{(m-1)sh}{n^{2}}(z+1)(3z+1)-i\alpha\dfrac{n\textrm{Ma}h}{2(m-1)}(z+1),\label{eq:u1surfactant}
\end{equation}

\[
u_{2}(z)=(z-n)(3z-n)\left(\frac{(m-1)sh}{mn^{2}}-i\alpha\frac{\text{Ma}h}{2(m-1)n}\right).
\]
(Alternatively, we can look for the correction velocities $u_{j}^{c}(z)$
in the form of quadratic polynomials with undetermined coefficients.
The interfacial and wall conditions lead to an expression of the coefficients
in terms of the surfactant correction $G^{c}$, since the only nonhomogeneous
condition is the tangential stress one, $mDu_{2}-Du_{1}=i\alpha\textrm{Ma}G^{c}.$
As a result, we find $u_{1}^{c}=-i\alpha\textrm{Ma}G^{c}(z+1).$ Hence,
the kinematic equation is $\gamma h=-\alpha^{2}\textrm{Ma}G^{c}/2$,
and thus, using the leading order expression for $\gamma,$ we find
that $G^{c}=nh/[2(m-1)]$. Substituting this into the expression for
the velocity recovers the formula $u_{1}^{c}=-i\alpha n\textrm{Ma}h(z+1)/[2(m-1)]$,
the imaginary part of (\ref{eq:u1surfactant})).

By using the imaginary term of $u_{1}(z)$, similar to the liquid
mass conservation equation used for the robust branch, we write 
\[
i\left(-i\alpha\dfrac{n\textrm{Ma}h}{2(m-1)}\right)(\delta t)=\intop_{\pi/2\alpha}^{3\pi/2\alpha}h(e^{\Sigma_{1}\delta t}-1)(\cos\alpha x)dx=-\frac{2h\Sigma_{1}}{\alpha}(\delta t).
\]
Hence, the growth rate is 
\begin{equation}
\Sigma_{1}=-\alpha^{2}\textrm{Ma}\frac{n}{4(m-1)},\label{eq:Sigma-1}
\end{equation}
so that the large $n$ approximation of the growth rate (\ref{eq:increment of decaying lubr mode})
of the surfactant branch is reproduced. However, this calculation
should be considered to be merely a consistency check, since equation
(\ref{eq:G/h with correction surfactant branch-1-1}) used here already
utilized the resulting expression for $\Sigma_{1}$. This is in contrast
with the derivation using the integral form of the mass conservation
law for the surfactant as given above.

In the integral balance for finding the wave velocity, the fluxes
at $x=0$ and $x=\pi/\alpha$ are determined by integrating the real
part of $u_{1}(z)cos(\alpha x)$ over the interval of $z$ from $z=-1$
to $z=0$. These fluxes are equal to zero for the quadratic velocity
profile given by $(z+1)(3z+1)$, which is consistent with the fact
that, in contrast to the robust branch, there is no term proportional
to $\alpha$ in the increment (\ref{eq:gamma series surfactant}).
Instead, the leading term in the imaginary part of the increment is
$\alpha^{3}J_{3}$. Hence, the wave velocity varies as $\alpha^{2}$,
to the leading approximation, and cannot be eliminated for all normal
modes simultaneously by a single Galilean transformation, as can be
done for the robust branch whose leading-order phase velocity is independent
of $\alpha$. To recover the phase velocity 
\[
c=-J_{3}\alpha^{2}
\]
by integral balance considerations, one must augment $G/h$ with the
term $-2\psi i\alpha J_{3}/[n^{2}(n^{2}-m)\textrm{Ma}]$, which for
large $n$ simplifies to $2i\alpha J_{3}/\textrm{Ma}$.

\subsection{Comparison of the two modes and intermediate asymptotics }

In general, we may start the analysis for either mode with the disturbance
flow in the thick layer. Remarkably, it is decoupled from the thin
layer and is completely determined by the base flow and the condition
of zero net flow in the thick layer. Returning to dimensionless quantities,
this flow is 
\[
u_{2}=\frac{(m-1)}{m}sh\left(1-\frac{z}{n}\right)\left(1-3\frac{z}{n}\right).
\]
It exerts a viscous tangential stress $\tau_{2}=-4(m-1)sh/n$ on the
thin layer at the interface. There is an additional tangential stress
due to the Marangoni effect of the surfactant, which is $\tau_{M}=-\text{Ma}i\alpha G$.
The flow in the thin layer has a linear velocity profile driven by
the sum of these two interfacial tangential stresses, $\tau_{1}=\tau_{2}+\tau_{M}$,
\[
u_{1}=-\frac{4(m-1)sh}{n}(z+1)-\text{Ma}i\alpha G(z+1),
\]
where the first term is ultimately due to the base flow (via interfacial
friction), and the second one to the surfactant. Using the continuity
equation (1), we find the vertical velocity 
\[
w_{1}(z)=\frac{2(m-1)si\alpha h}{n}\left(z+1\right)^{2}-\frac{1}{2}\text{Ma}\alpha^{2}G\left(z+1\right)^{2}.
\]
We use the interfacial value of this velocity component to write the
kinematic boundary condition (\ref{dKineBC}) in the form 
\begin{equation}
\gamma h-\frac{2s(m-1)}{n}i\alpha h+\frac{\text{Ma}}{2}\alpha^{2}G=0\label{eq:kinematic dimensionless}
\end{equation}
(cf. equation (\ref{eq:linear-eta-evolution})). The second term here
originates from the base flow and the third one is due to the surfactant.
The surfactant transport equation (\ref{dSurfConc}) takes the form
\begin{equation}
\gamma G+i\alpha sh+\text{Ma}\alpha^{2}G=0\label{eq:dimensionless transport}
\end{equation}
(cf. equation (\ref{eq:linear-surfactant-evolution})). Note that
the term coming from the non-surfactant part of the disturbance velocity,
$4(m-1)shi\alpha/n$, has been neglected by comparison with the second
term of (\ref{eq:dimensionless transport}).

There are three possibilities regarding the relative size of the Marangoni
term (containing $\text{Ma}G$ and corresponding to the Marangoni
tangential stress at the interface) and the base flow term (containing
$sh$ and corresponding to the interfacial tangential stress induced
by the base flow) of the kinematic equation (\ref{eq:kinematic dimensionless}):
(1) the Marangoni term is much smaller than the base flow term; (2)
both terms are of the same order; and (3) the Marangoni term is much
larger than the base flow term. We consider them in turn.

In the first case, when the Marangoni term is negligible, the kinematic
equation gives $\gamma=2i\alpha s(m-1)/n$ to the leading order. Also,
in this leading-order flow $G=-\frac{n}{2(m-1)}h$ from the surfactant
equation (cf. equation (\ref{eq:G in terms of h large n})). The surfactant
driven flow is a correction to this leading order, with the correction
to increment $\gamma_{c}$ satisfying the correction to the kinematic
equation 
\[
\gamma_{c}h=-\frac{\text{Ma}}{2}\alpha^{2}G=\frac{\text{Ma}n}{4(m-1)}\alpha^{2}h.
\]
So the growth rate is 
\[
\gamma_{c}=\frac{\text{Ma}n}{4(m-1)}\alpha^{2},
\]
which is the dimensionless form of the result (\ref{eq:Sigma_plus}),
the robust branch.

Assuming now the second case, the last two terms of the kinematic
equation being of the same order $h/n\sim\alpha G$, it is clear that
the Marangoni term in the transport equation is negligible. At a fixed
$n,$ a solution can be found if the first term in the kinematic equation
(\ref{eq:kinematic dimensionless}) is negligible. Then we have 
\[
h=-Gi\frac{\alpha\textmd{Ma}n}{4s(m-1)},
\]
which agrees with the large $n$ limit of equation (\ref{eq:G decaying}).
With this, the transport equation (\ref{eq:dimensionless transport})
yields 
\[
\gamma=-\frac{\alpha^{2}\textmd{Ma}n}{4(m-1)},
\]
which is the dimensionless form of the previous result (\ref{eq:Sigma}),
the surfactant mode. Thus the two normal modes are characterized in
terms of the relative strengths of the two tangential stresses at
the interface.

Turning now to the last case, when the term with $\textmd{Ma}$ dominates
the term with $s$ in the kinematic equation, we must have $\alpha h/n\ll\alpha^{2}G$
and hence $h\ll\alpha nG\ll G$ (since $\alpha n\ll1)$, and the transport
and kinematic equations simplify to 
\begin{equation}
\gamma G=-i\alpha sh\label{eq:simple transport}
\end{equation}
and 
\begin{equation}
\gamma h=-\alpha^{2}\textmd{Ma}G/2,\label{eq:simple kinematic}
\end{equation}
respectively. From these two equations, we obtain $\gamma^{2}=i\alpha^{3}\textmd{Ma}s/2$,
so 
\[
\gamma=\pm(1+i)\textmd{Ma}^{1/2}s^{1/2}\alpha^{3/2}/2.
\]
Thus, the growth rates for the two modes are 
\begin{equation}
\Sigma=\pm\textmd{Ma}^{1/2}s^{1/2}\alpha^{3/2}/2,\label{eq:Sigma_3}
\end{equation}
so one of the modes is stable and the other one unstable. Writing
the fact that the second term in the kinematic equation is negligible
in comparison with the third term, $h\ll\alpha nG$, and taking into
account that, from the simplified transport equation with $\gamma\sim\alpha^{3/2}$,
we have $G\sim\alpha^{-1/2}h$, it follows that $1\ll\alpha^{1/2}n$.
Together with $\alpha n\ll1$, this means that the modes (\ref{eq:Sigma_3})
exist in the interval 
\[
\frac{1}{n^{2}}\ll\alpha\ll\frac{1}{n},
\]
which is bounded away from zero. Thus, this case is generic, but the
asymptotics (\ref{eq:Sigma_3}) is merely intermediate since it does
not persist in the limit $\alpha\downarrow0$. One should note that
the condition of validity for $\alpha$ is more accurately given by
\[
\alpha\ll\frac{1}{n}\ll\alpha^{1/2}.
\]

\subsection{The finite aspect ratio case in the presence of nonzero gravity\label{subsec:The-finite-aspect}}

Turning next to the less simple situation of the layer thicknesses
being comparable, allowing for gravity effects, the flows in the two
layers are fully coupled. Both governing equations, the kinematic
equation (\ref{eq:linear-eta-evolution}) and the transport equation
(\ref{eq:linear-surfactant-evolution}), have three different terms
in their right-hand sides: one term due to the base shear (containing
$sh$), one due to the surfactant (containing $\textmd{Ma}G$), and
one due to gravity (containing $\textmd{Bo}h$). The two modes found
previously still have the following physical characterization, similar
to the simpler case of large aspect ratio and no gravity: the robust
mode has the gravity and surfactant effects absent at the leading-order
in $i\alpha$, so that only the first right-hand side term is retained
in both governing equations. The simplified kinematic equation at
once yields the leading-order increment (see (\ref{eq:I1-1})) 
\begin{equation}
\gamma=\frac{2i\alpha(m-1)n^{2}(n+1)s}{\psi},\label{eq:increment leading robust-1}
\end{equation}
(and thus the wave velocity $c=2(m-1)(n+1)n^{2}s/\psi$). Therefore,
the simplified surfactant transport equation determines (cf. equation
(\ref{eq:amplitude ratio for robust mode-1-1})) 
\begin{equation}
G=-h\frac{\varphi}{2n^{2}(m-1)}.\label{eq:G in terms of h leading-1-1}
\end{equation}
(Note that this $G$ is real and hence the surfactant is either in
phase (for $m<1$) or in anti-phase (for $m>1$) with the interface.)
Thus, the leading-order yields just the wave velocity found in \citet{Yih1967}
(and, for large $n$, reproduces the previously obtained expression
(\ref{eq:wave velocity})). The growth rate due to surfactant and
gravity effects appears in the correction to the leading-order disturbance
flow. Substituting the leading-order $G$ (\ref{eq:G in terms of h leading-1-1})
into the kinematic equation (\ref{eq:linear-eta-evolution}) with
the left-hand side $\gamma_{c}h$, where $\gamma_{c}$ is the correction
to the increment, and the first term on the right-hand side absent
in the correction equation, we reproduce the growth rate (\ref{gamCsmallAlphaApprox})
of the robust mode.

The surfactant mode hinges on the Marangoni effect, the leading-order
flow being determined, just like in the previous case of large $n$,
by the dominant balance of only the base-shear and surfactant terms
in the kinematic equation. This reproduces the leading-order of the
result (\ref{eq:G/h with correction surfactant branch-1-1}) for the
relation between $G$ and $h$, which we write now in the form 
\[
h=-i\dfrac{(n^{2}-m)\textrm{Ma}\alpha}{4(n+1)(m-1)s}G.
\]
(Hence, the surfactant shift from the interface is $-90\si{\degree}$
in the $S$ ($m<1$) and $Q$ ($m>n^{2}$) sectors and $90\si{\degree}$
in the $R$ ($1<m<n^{2}$) sector.) Substituting this into the surfactant
transport equation (\ref{eq:linear-surfactant-evolution}), the gravity
term is of a higher order in $\alpha$, and hence the growth rate
is found to be independent of $\textmd{Bo}$, reproducing equation
(\ref{gamSsmallAlphaApprox}).

One way to find the velocity profile $u_{1}(z)$ for the surfactant
branch is to apply the same procedure as was used in Section \ref{subsec:Surfactant-branch}
for the case of large $n$ and no gravity. Namely, we use expression
(\ref{eq:uj vel}) for normal modes along with the linearized pressure
gradient (\ref{eq:p1x}) and the interfacial vorticity (\ref{eq:A1})
to obtain the velocity amplitude in terms of $G$ and $h,$ and then
substitute $G$ in terms of $h$ from equation (\ref{eq:G/h with correction surfactant branch-1-1}).
The result is 
\begin{align}
u_{1}(z) & =-\frac{(m-1)s}{m-n^{2}}h(z+1)(3z+1)\nonumber \\
 & +\frac{i\alpha h}{m-n^{2}}\left\{ \frac{(n-1)\text{Ma}}{2(m-1)n}\left[3m(n+1)(z^{2}-1)+(4m+3mn+n^{3})(z+1)\right]\right.\nonumber \\
 & \left.-\frac{n^{2}\text{Bo}}{6}(z+1)(3z+1)\right\} .\label{eq:u1expression}
\end{align}
To reproduce the leading nonzero phase velocity, we should augment
$G/h$ with the term $2i\alpha J_{3}\psi/[(m-n^{2})n^{2}\textrm{Ma}]$
(as is found from the order $\alpha^{3}$ of the kinematic condition
(\ref{eq:linear-eta-evolution})). This adds the term $2\alpha^{2}J_{3}(m-n^{2})^{-1}n^{-1}[3m(n+1)(z^{2}-1)+(4m+3mn+n^{3})(z+1)]$
to (\ref{eq:u1expression}), so that the integral form of the liquid
conservation equation yields 
\[
c=-J_{3}\alpha^{2}.
\]
(All the other velocities, for both branches and both layers, can
be found similarly, and are listed in Appendix \ref{sec:Eigenfunctions}.)
As was noted in CH, any flow of this type, with a complex-valued horizontal-velocity
amplitude as in (\ref{eq:u1expression}), can be considered as a superposition
of two flows, one of which is in phase or anti-phase with the interface
and the other is $\pm90^{o}$ out of phase with the interface. The
in-phase (or anti-phase) and out-of-phase flows correspond, respectively,
to the real and purely imaginary addends in the amplitude of the horizontal
velocity. The real part of this velocity component generates the imaginary
part of the vertical velocity, whose interfacial value divided by
$h$ equals the imaginary part of the increment, which determines
the wave velocity; while the imaginary part generates the real part
of the vertical velocity, whose interfacial value divided by $h$
is the growth rate (see equation (\ref{eq:gamma is w/h})). Thus,
the stability/instability is due solely to the out-of-phase flow.
(From the alternative point of view based on the integral form of
the mass conservation law, this is so because the fluid flux wave
is $\pm90^{o}$ out of phase with the thickness wave.) Note that,
in difference with the robust branch, whose in-phase flow is due to
the base shear only, the in-phase flow of (\ref{eq:u1expression}),
a surfactant mode, has a contribution from the surfactant. This is
due to the fact that, for the surfactant branch, to the leading-order
$\alpha^{-1}$, the surfactant amplitude $G$ is purely imaginary,
i.e. $\pm90^{o}$ out of phase with $h$, and so expressing $G$ in
terms of $h$ converts the surfactant terms in the pressure, vorticity,
velocity, and the kinematic equations into the form of the base-shear
terms there. The wave velocity, which corresponds to the real part
of (\ref{eq:u1expression}), vanishes since $\int_{-1}^{0}(z+1)(3z+1)dz=0.$
(For the same reason, there is no term proportional to $\textrm{Bo}$
in the growth rate (\ref{gamSsmallAlphaApprox}) determined by the
out-of-phase flow, by integrating the imaginary part of $u_{1}$ (\ref{eq:u1expression}).)
Also, this integral being zero is interpreted as the annihilation
of the flux of the in-phase flow.

From these considerations, it transpires that the robust modes can
be regarded as a modification of the (single) mode of \citet{Charru2000},
in which the leading-order flow is the same in-phase, non-dissipative
Yih wave, but the next order, out-of-phase, dissipative correction
is determined, instead of inertia, by the Marangoni tangential stress
and/or the pressure difference generated by the gravitational normal
stress. Thus, the robust mode corresponds to the Marangoni tangential
stress being of higher order than that of the tangential viscous stresses
of the liquid layers (whose thicknesses are comparable) at their interface.
The robust branch is also characterized by the leading-order surfactant
concentration being either in phase or totally, $180^{o}$ out of
phase with the interface, while for the surfactant branch the leading-order
surfactant concentration is $\pm90^{o}$ out of phase with the interface.
The surfactant mode can be recovered, in this more physical way, by
starting with the only other possible assumption about the Marangoni
tangential stress: that the latter is of the same, order as the viscous
stresses of the liquid layers and thus participates in the leading-order
interfacial balance of the tangential stresses (and not only in the
correction order of the tangential stress condition, as in the robust
mode).

In more detail, in this alternative way of finding the complete normal
modes 
\[
[u_{j}(z),w_{j}(z),p_{j},h,G]e^{i\alpha x+\gamma t},
\]
the corresponding algebra-differential eigenvalue problem is given
by the normal-form version of equations (\ref{eq:uj momentum}), (\ref{eq:incompress}),
(\ref{eq:Pi}), (\ref{eq:tangential stress lubrication}), (\ref{eq:continuityofu}),
(\ref{eq:continuityofw}), the kinematic equation $\gamma h=w_{1}(0)$
(see equation (\ref{eq:gamma is w/h})), and the amplitude form of
the linearized surfactant evolution equation (\ref{dSurfConc}) (with
the diffusion term discarded): 
\begin{equation}
\gamma G=-i\alpha(sh+u_{1}(0)).\label{eq:surf transp amplitude form-1}
\end{equation}
There are also the zero conditions for the both velocity components
at the plates. The solutions of the eigenvalue problems corresponding
to the two branches of normal modes are obtained using the appropriate
assumptions about the tangential stresses and the flow fluxes in the
two layers (as was mentioned above).

We deal with the surfactant branch first. From the momentum equations,
we know that the horizontal velocities are quadratic functions of
$z$ that can be written, without loss of generality, in the form
satisfying the no-slip wall conditions, as 
\begin{equation}
u_{1}=(z+1)[A_{1}(z-1)+A_{0}]\textrm{ and }u_{2}=(z-n)[B_{1}(z+n)+B_{0}].\label{eq:horiz velocities}
\end{equation}
(cf. (\ref{eq:uj vel}). The same considerations hold for the robust
branch, and below we will use for this velocity the same form with
four undetermined coefficients.) Integrating the incompressibility
equation yields the vertical velocities given by equation (\ref{eq:w in terms of u-1-1}):
\begin{equation}
w_{1}(z)=-i\alpha\int_{-1}^{z}u_{1}(\xi)d\xi\textrm{ and }w_{2}(z)=-i\alpha\int_{n}^{z}u_{2}(\xi)d\xi.\label{eq:vert velocities-1}
\end{equation}
For the surfactant branch, as discussed above, the Marangoni stresses
act already in the leading-order, and in such a way that, to the leading-order,
(but not necessarily for the correction, as will be seen later), the
fluxes vanish through each layer separately. This is equivalent to
requiring $w_{1}(0)=w_{2}(0)=0$ (which, from the kinematic condition,
implies that the leading-order $\alpha^{1}$ increment is zero). Integrating
(\ref{eq:vert velocities-1}) with the upper limit $z=0$ yields the
following two relations for the coefficients: 
\begin{equation}
-\frac{2}{3}A_{1}+\frac{1}{2}A_{0}=0\textrm{ and }\frac{2}{3}n^{3}B_{1}+\frac{1}{2}n^{2}B_{0}=0.\label{eq:vert veloc cond for a and b-1}
\end{equation}
This allows eliminating two of the constants and thus writing each
velocity with just one undetermined coefficient: $u_{1}=A(z+1)(3z+1)sh$
and $u_{2}=B(z-n)(3z-n)sh$, (where the factor $sh$ has been introduced
for future convenience). The pressures are equal since the gravity
effects are of a higher order; this requires $mB=A.$ Hence, we eliminate
$A$ from the interfacial condition for the horizontal velocities,
at $z=0$: $n^{2}Bsh-mBsh=sh(m-1)/m.$ This implies the solution 
\[
B=\frac{(m-1)}{m(n^{2}-m)},\textrm{ }A=\frac{(m-1)}{(n^{2}-m)}.
\]
With this, we obtain exactly the leading-order velocities (\ref{eq:re u1 surfactant})
and (\ref{eq:re u2 surf}). Next, the tangential stress condition
(\ref{eq:tangential stress lubrication}), written in the amplitude
form, gives a relation between $h$ and $G$, $4sh(n+1)(m-1)/(m-n^{2})=i\alpha\textrm{Ma}G$
(thus reproducing the leading-order of (\ref{eq:G/h with correction surfactant branch-1-1})),
which we use in equation (\ref{eq:surf transp amplitude form-1}),
$\gamma G=-i\alpha(sh+u_{1}(0))$, where $u_{1}(0)=-sh(m-1)/(m-n^{2}).$
Substituting the latter into the surfactant equation (\ref{eq:surf transp amplitude form-1}),
followed by expressing $h$ in terms $G$ from the tangential stress
condition, yields, after cancelling out $G$, exactly the explicit
expression (\ref{gamSsmallAlphaApprox}) for the leading-order increment
$\gamma,$ (which is real and thus the leading-order nonzero growth
rate for the surfactant mode).

The (purely imaginary, out-of-phase with $h$) corrections to these
leading-order disturbances of the horizontal velocities are written
in the same quadratic form (\ref{eq:horiz velocities}), but with
the four coefficients having the superscript ``c'' and, for anticipated
convenience, a factor $i\alpha h$. However, in contrast to the leading-order,
each correction flux is not required to be zero; instead, the kinematic
condition in order $\alpha^{2}$ requires $w_{1}^{c}(0)=\gamma=w_{2}^{c}(0)$
where $\gamma$ is the leading-order growth rate given by (\ref{gamSsmallAlphaApprox}).
This yields the following two relations for the coefficients: 
\begin{equation}
-\frac{2}{3}A_{1}^{c}+\frac{1}{2}A_{0}^{c}=-\frac{(n-1)\textrm{Ma}}{4(m-1)}\label{eq:vert vel a_j^c cond-1}
\end{equation}
and 
\begin{equation}
n^{3}B_{1}^{c}+\frac{1}{2}n^{2}B_{0}^{c}=-\frac{(n-1)\textrm{Ma}}{4(m-1)}.\label{eq:vert vel b_j^c cond-1}
\end{equation}
We use the latter equation to express $B_{0}^{c}$ in terms of $B_{1}^{c}$
and $\textrm{Ma}$. From the normal stress condition, we obtain $A_{1}^{c}$
in terms of $B_{1}^{c}$ and $\textrm{Bo}$. Then, using the continuity
of the horizontal velocities at $z=0$, $A_{0}^{c}$ is obtained in
terms of $B_{1}^{c}$, $\textrm{Bo}$ and $\textrm{Ma}$. Substituting
the latter expressions into (\ref{eq:vert vel a_j^c cond-1}) yields
an equation for $B_{1}^{c}$, whose solution is 
\[
B_{1}^{c}=\frac{1}{2(m-n^{2})}\left(\textrm{Ma}\frac{3(n^{2}-1)}{(m-1)n}-\textrm{Bo}\right).
\]
Using this, all the other coefficients are written in terms of the
system parameters, which reproduces the imaginary parts of the horizontal
velocity eigenfunctions, equations (\ref{eq:im u1 surf}) and (\ref{eq:im u2 surf})
of Appendix \ref{sec:Eigenfunctions}. The tangential stress condition
(\ref{eq:tangential stress lubrication}), in this order, leads to
the relation $i\alpha h(mB_{0}^{c}-A_{0}^{c})=i\alpha\textrm{Ma}G^{c}$,
from which we obtain the $G^{c}/h$ in terms of the parameters. It
is exactly the second term of (\ref{eq:G/h with correction surfactant branch-1-1}).
Finally, the surfactant equation in the order $\alpha^{2}$ is used
to obtain the increment correction term. The latter is purely imaginary,
of the form $i\alpha^{3}J_{3}$, where the $J_{3}$ is found in terms
of the parameters to be the same as given by equation (\ref{eq:J_3})
(which leads to the leading nonzero term of the wave velocity proportional
to $\alpha^{2}$). Thus, we have determined completely the eigenfunctions
and eigenvalues of the surfactant branch.

We turn now to the robust branch. The leading-order disturbances $u_{j}$
take the same form as (\ref{eq:horiz velocities}) but with a relabelling
of the coefficients: $C$ in place of $A,$ and $D$ instead of $B$.
The relation following from the continuity of vertical velocities
is 
\begin{equation}
-\frac{2}{3}C_{1}+\frac{1}{2}C_{0}=\frac{2}{3}n^{3}D_{1}+\frac{1}{2}n^{2}D_{0}.\label{eq:vert vel relation robust-1}
\end{equation}
Since the pressures are equal, $C_{1}=mD_{1}$. The Marangoni term
is absent in the tangential stress condition, so that $mDu_{2}-Du_{1}=0$;
hence, $C_{0}=mD_{0}$. Then the horizontal velocity condition (\ref{eq:continuityofu})
yields 
\begin{equation}
n^{2}D_{1}+nD_{0}-m(D_{1}-D_{0})=-\frac{sh(m-1)}{m}.\label{eq:hor vel relation for d_k-1}
\end{equation}
Also, equation (\ref{eq:vert vel relation robust-1}) becomes a relation
between $D_{1}$ and $D_{0}$ only, which can be written as 
\[
D_{1}=\frac{3(m-n^{2})}{4(m+n^{3})}D_{0}.
\]
Substituting this into (\ref{eq:hor vel relation for d_k-1}) yields
the following expression for $D_{0}$, 
\[
D_{0}=-\frac{4(m-1)(m+n^{3})}{m\psi}sh.
\]
Using this, the other coefficients can be written in terms of the
system parameters, and we thus obtain the leading-order horizontal
velocities (\ref{eq:re u1 robust}) and (\ref{eq:re u2 robust}) given
(as the real parts) in Appendix \ref{sec:Eigenfunctions}. (From the
above derivation, it is clear that these horizontal velocities must
coincide with those obtained by \citet{Yih1967} and given as the
leading-order eigenfunctions in \citet{Charru2000}, and with those
obtained by \citet{Yiantsios1988} for the two-layer Poiseuille base
flow when written in terms of the base shear parameter.) The kinematic
condition $\gamma h=w_{1}(0)$ yields the purely imaginary leading-order
increment (\ref{eq:increment leading robust-1}). Using the expressions
in terms of the system parameters for $\gamma$ and $u_{1}(0)$ in
the surfactant equation (\ref{eq:surf transp amplitude form-1}) yields
the robust branch relation for $G$ given by the leading-order of
(\ref{eq:amplitude ratio for robust mode-1-1}), which will be used
to find the corrections to the horizontal velocities. Namely, the
tangential stress condition in the order $\alpha^{1}$, which now
includes the Marangoni term $i\alpha\textrm{Ma}G$, yields 
\begin{equation}
C_{0}^{c}=mD_{0}^{c}+i\frac{\alpha h\textrm{Ma}\varphi}{2(m-1)n^{2}}.\label{eq:robust correction tangential}
\end{equation}
The normal stress condition yields 
\begin{equation}
C_{1}^{c}=mD_{1}^{c}+i\frac{\alpha h\textrm{Bo}}{2}.\label{eq:robust correction normal}
\end{equation}
We use these relations to eliminate $C_{0}^{c}$ and $C_{1}^{c}$
in the continuity conditions for the vertical and horizontal velocity
corrections in the order $\alpha^{1}$ given correspondingly by equation
(\ref{eq:vert vel relation robust-1}) (in which all the unknowns
should be endowed with the superscript 'c') and the equation (cf.
equation (\ref{eq:hor vel relation for d_k-1})) 
\begin{equation}
n^{2}D_{1}^{c}+nD_{0}^{c}-(C_{1}^{c}-C_{0}^{c})=0.\label{eq:robust correction horizontal}
\end{equation}
Solving this system of two equations for the unknowns $D_{0}^{c}$
and $D_{1}^{c}$ leads to the expressions for the (higher-order correction)
imaginary parts of horizontal velocities for the robust branch given
in Appendix \ref{sec:Eigenfunctions}. Finally, using the order $\alpha^{2}$
kinematic condition $\gamma^{c}h=w_{1}^{c}(0)$ yields the growth
rate (\ref{gamCsmallAlphaApprox}). The correction to the leading-order
$G/h$ can be found from the surfactant equation taken in the order
$\alpha^{2}$.

Returning to the surfactant branch, note that the requirement we have
used, that each vertical velocity is zero at the interface, can be
relaxed. If we just impose equality of these velocities, along with
requiring that the surfactant Marangoni term is not negligible in
the leading-order tangential stress condition, then it turns out that
these velocities must automatically vanish. Thus, the surfactant mode
is recovered from the sole assumption that the surfactant Marangoni
tangential stress is present in the leading-order balance, while the
robust mode is characterized, to the contrary, by the Marangoni tangential
stress being neglected in the leading-order and first appearing in
the next order correction.

Finally, for $s\ne0$, consider the special case of $m=1$, the $R-S$
boundary. (Note that then $\psi=(n+1)^{4}$ and $\varphi=(n+1)^{3}$).
The leading-order disturbances $u_{j}$ are given in standard form
by (\ref{eq:horiz velocities}), but with the coefficients labeled,
say, $F_{k}$ and $G_{k}$ instead of $A_{k}$ and $B_{k}$, respectively.
These four coefficients are determined from the velocity and stress
conditions at the interface. The horizontal velocity relation is homogeneous
since the right-hand side (see equation (\ref{eq:hor vel relation for d_k-1}))
vanishes for $m=1$. To have non-trivial results, the Marangoni forcing
term must be present in the tangential stress relation. As a result,
\begin{equation}
u_{1}=-i\alpha\textrm{Ma}G\frac{n}{(n+1)^{3}}(z+1)[3(z-1)+(n^{2}-n+4)]\label{eq:u_1 m=00003D1}
\end{equation}
and 
\begin{equation}
u_{2}=i\alpha\textrm{Ma}G\frac{1}{(n+1)^{3}}(z-n)[-3n(z+n)+(4n^{2}-n+1)].\label{eq:u_2 m=00003D1}
\end{equation}
Hence we find $w_{1}(0)$, and thus the kinematic condition in the
leading-order yields 
\begin{equation}
\gamma h=-\alpha^{2}\textrm{Ma}G\frac{n^{2}(n-1)}{2(n+1)^{3}}.\label{eq:kinem for m is 1-1}
\end{equation}
The surfactant conservation equation is found in the leading-order
$\alpha^{1}$ to be 
\begin{equation}
\gamma G=-i\alpha sh.\label{eq:surf for m is 1-1}
\end{equation}
Multiplying these two equations, 
\begin{equation}
\gamma^{2}=i\alpha^{3}s\textrm{Ma}\frac{n^{2}(n-1)}{2(n+1)^{3}},\label{eq:equal viscosities increment}
\end{equation}
while dividing them yields 
\[
\left(\frac{G}{h}\right)^{2}=2i\alpha^{-1}\frac{s(n+1)^{3}}{\textrm{Ma}n^{2}(n-1)}.
\]
We see that there are two solutions with $\gamma\propto\alpha^{3/2}$
and $G/h\propto\alpha^{-1/2}$. Thus, the leading-order of the kinematic
condition (\ref{eq:kinem for m is 1-1}) is $\alpha^{3/2}$. It is
clear that for the growing mode with $\textrm{Arg}(\gamma)=\pi/4$
we have $\textrm{Arg(}G)=-3\pi/4$ and for the decaying mode, with
$\textrm{Arg(}\gamma)=-3\pi/4$, we have $\textrm{Arg}(G)=\pi/4$.
We can rewrite the velocities in terms of $h$ (see equations (\ref{eq:u1m=00003D00003D1})
and (\ref{eq:u2m=00003D00003D1}) in Appendix \ref{sec:Eigenfunctions}).
The effect of gravity comes in the next order correction. The normal
stress relation is $G_{1}^{c}-F_{1}^{c}=-i\alpha\textrm{Bo}h/2$ and
the tangential stress relation is $G_{0}^{c}-F_{0}^{c}=i\alpha\textrm{Ma}G^{c}$
(where the superscript indicates a correction). Correspondingly, we
find the coefficients, and thus the velocity corrections, in terms
of $h$ and the surfactant correction $G^{c}.$ Then, the kinematic
condition in order $\alpha^{2}$ yields 
\[
\gamma^{c}h=-\alpha^{2}\textrm{Bo}h\frac{n^{3}}{3(n+1)^{3}}-\alpha^{2}\textrm{Ma}G^{c}\frac{n^{2}(n-1)}{2(n+1)^{3}},
\]
and the surfactant equation of order $\alpha^{3/2}$ is $\gamma^{c}G+\gamma G^{c}=0$.
It follows that 
\begin{equation}
\gamma^{c}=-\alpha^{2}\textrm{Bo}\frac{n^{3}}{6(n+1)^{3}}\label{eq:equal visc increm Bo correction}
\end{equation}
and 
\[
\frac{G^{c}}{h}=-\frac{\textrm{Bo}n}{3\textrm{Ma}(n-1)}.
\]
With this, the velocity corrections are written in terms of $h$,
and are given in Appendix \ref{sec:Eigenfunctions}. We note that
the requirement that the gravity term is negligible in the kinematic
condition is satisfied when 
\[
\frac{\alpha^{1/2}\textmd{Bo}s^{1/2}}{\textmd{Ma}^{1/2}(n+1)^{3/2}(n-1)^{1/2}}\ll1,
\]
which is the case, e.g., even if $\textmd{Bo}\gg1$, but at the same
time $n$ is sufficiently large.

Clearly when in addition to $m=1$, also $n=1$, equation (\ref{eq:equal viscosities increment})
does not yield a non-zero leading order result. We can obtain the
leading order growth rates by using the quadratic equation (\ref{eq:quadratic longwave})
and specifying $m=n=1$ in the coefficients $c_{1}$ and $c_{0}$.
Then it is straightforward to obtain from the solution (\ref{eq:QuadEqnGamma})
that $\gamma_{R}=-\text{Bo}\alpha^{2}/24$ for the robust branch and
$\gamma_{R}=-\text{Ma}\alpha^{2}/8$ for the surfactant branch.

The consideration for the special case $s=0$, which implies that
both the base flow and the leading-order disturbance flow are absent,
proceeds in the same manner as before, starting with the same quadratic
ansatz (\ref{eq:horiz velocities}) for the correction-order horizontal
velocity (which in this case is actually the leading nonzero order
flow), but has certain differences between the cases of $\text{Bo}=0$
and $\text{Bo}\ne0$. In all these cases, whether $\text{Bo}=0$ or
$\text{Bo}\ne0$, the surfactant mode or the robust one, the four
coefficients of the two horizontal-velocity expressions for the fluid
layers are found in terms of $G$ and $h$ by solving the system of
four linear non-homogeneous equations, which consists of the two interfacial
velocity conditions, the normal stress condition and the tangential
stress condition. Substituting these velocity expressions into the
kinematic equation (\ref{eq:gamma is w/h}) and the surfactant equation
(\ref{eq:surf transp amplitude form-1}), which simplifies to $\gamma G=-i\alpha u_{1}(0)$,
we obtain the same two equations for the eigenvalue $\gamma$ and
eigenfunction $G/h$ as in section \ref{subsec:Leading-order-long-wave-incremen},
whose solution reproduces the eigenvalues and $G/h$ found there.
Then the velocities are written in term of $h$ only as $G$ is eliminated
from their expressions by using the appropriate ratios $G/h$. These
expressions for the velocities in terms of $h$ are given in Appendix
\ref{sec:Eigenfunctions}. (We note that the special cases $m=1$,
$s\ne0$ and $s=0$, $\text{Bo}\ne0$ are more complicated than the
other cases in that we arrive at a quadratic equation for $\gamma$
(or for $G/h$) rather than a linear one. For the former case, it
is the incomplete quadratic equation (\ref{eq:equal viscosities increment})
for the leading-order $\gamma$ (while the equation for the correction
of the increment $\gamma^{c}$ is linear again). For the case $s=0$,
$\text{Bo}\ne0$, the equation for $\gamma$ (which coincides with
$\gamma^{c}$) is a full quadratic equation. The equations for the
four undetermined coefficients of $u_{j}$ are the same as those for
the velocity corrections of the robust branch with $s\neq0$, equations
(\ref{eq:robust correction tangential})-(\ref{eq:robust correction horizontal})
and (\ref{eq:vert vel relation robust-1}), and therefore the corresponding
velocity expressions written in terms of $G$ and $h$ are the same.
However, they differ when written in terms of $h$ only, because the
corresponding eigenfunctions $G/h$ are different.)

It is worth noting that in the conditions of the flows considered
by \citet{Charru2000}, which included the effects of inertia but
assumed constant surface tension, the advection of the leading, in-phase,
vorticity by the base flow, (clearly, an inertial term), acts as a
source for the out-of-phase corrections to the vorticity and the horizontal
velocity, and therefore to the in-phase vertical velocity, whose interfacial
value is identical to the growth rate. Thus, the leading-order vorticity
is solely responsible for the dissipative effects of the growth or
damping of the infinitesimal disturbances. In contrast, for our (and
W) case of inertialess flow, but with surfactants and/or gravity,
it is clear that vorticity plays no such dynamical role at all. Instead,
the out-of-phase horizontal velocities (solely responsible for resolving
the stability/instability question) are produced by the Marangoni
forces due to the interfacial surfactant and/or gravity. Although
it is possible to formulate the criterion of stability/instability
in terms of the intervals for the phase (i.e. argument) of the complex-valued
interfacial vertical velocity, it is clearly more natural, and simpler,
to use for this purpose the real part of $w_{1}(0)$ since (see equation
(\ref{eq:gamma is w/h})) the latter divided by $h$ is identically
equal to the growth rate. This vertical velocity is closely related
to the out-of-phase horizontal velocity; as was mentioned before,
the spanwise integral of the horizontal velocity is proportional to
the interfacial value of the vertical velocity.

\section{Nonlinear stages of instability}

\label{sec:nonlinear}

\subsection{Small-amplitude saturation in the $R$ and $Q$ sectors with linearly
unstable robust modes}

Regimes in which the amplitudes of the deviations of the interface
thickness and the surfactant concentration remain small are described
by the weakly nonlinear equations (\ref{eq:weak-eta-evolution}) and
(\ref{eq:weak-surfactant-evolution}). As was mentioned above, by
changing $x$ to a new variable $x\rightarrow x+Vt$ where $V$ is
the coefficient of $\eta_{x}$ in (\ref{eq:weak-eta-evolution}),
we eliminate the $\eta_{x}$ in that equation. However, performing
this change of variable, an additional term $-V\Gamma_{x}$ appears
in the surfactant equation (\ref{eq:weak surfactant evolution 2-1})
below: 
\begin{equation}
\eta_{t}+sN_{1}\eta\eta_{x}-\frac{n^{3}(m+n)\Bo}{3\psi}\eta_{xx}+\frac{n^{3}(m+n)}{3\psi}\eta_{xxxx}-\frac{n^{2}(n^{2}-m)\Ma}{2\psi}\varGamma_{xx}=0\label{eq:eta-evolution-1-2-1-1}
\end{equation}
and 
\begin{eqnarray}
\nonumber \\
 & \varGamma_{t} & +\frac{2(m-1)n^{2}(n+1)s}{\psi}\Gamma_{x}-\frac{n(m+n^{3})\Ma}{\psi}\varGamma_{xx}+\frac{(n+1)\phi s}{\psi}\eta_{x}\label{eq:weak surfactant evolution 2-1}\\
 &  & -\frac{n^{2}(n^{2}-m)\Bo}{2\psi}\eta_{xx}+\frac{n^{2}(n^{2}-m)}{2\psi}\eta_{xxxx}=0.\nonumber \\
\nonumber 
\end{eqnarray}
Note that the transport equation is now linear to this leading-order;
we have neglected the nonlinear term $sN_{2}\eta\eta_{x}$ by comparison
with the retained term proportional to $\eta_{x}$. Some examples
of such weakly nonlinear regimes follow.

If $m=n^{2}$ (the border between the $R$ and $Q$ sectors), the
surfactant term in the kinematic equation vanishes, so it decouples
(also, in the transport equation, two terms vanish). Note that $\phi=4n^{2}(n+1)$,
$\psi=4n^{3}(n+1)^{2}$ and $N_{1}=1/n$. If, in addition, $n$ is
large (note that then $\varphi\sim4n^{3}$ and $\psi\sim4n^{5}$),
the weakly nonlinear system (\ref{eq:eta-evolution-1-2-1-1})-(\ref{eq:weak surfactant evolution 2-1})
simplifies to 
\begin{equation}
\eta_{t}+\frac{s}{n}\eta\eta_{x}-\frac{\Bo}{12}\eta_{xx}+\frac{1}{12}\eta_{xxxx}=0\label{eq:eta-evolution-1-1-1}
\end{equation}
and 
\begin{equation}
\varGamma_{t}+\frac{s}{2}\Gamma_{x}-\frac{\Ma}{4n}\varGamma_{xx}+\frac{s}{n}\eta_{x}=0.\label{eq:surfactant-evolution-1-1-1-1-1}
\end{equation}
The first equation here is a KS equation for $\eta$ (provided the
Bond number is negative). It gives a saturated chaotic state with
the characteristic length scale, say, $L$, the time scale $T$, and
the amplitude of undulations $N$, which can be estimated (in terms
of the Bond number, the thickness ratio and the shear parameter) from
the pairwise balance of the four terms as $L\sim(-\Bo)^{-1/2}$, $N\sim n/(12L^{3}s)$,
and $T\sim12L^{4}$. (For example, choosing $\textmd{Bo}=-10^{-2}$,
$n=100$, and $s=1$, we get $L\sim10$, $N\sim10^{-2}$, $T\sim10^{5}$.)
The transport equation has the form of a diffusion equation for the
surfactant, with the $\eta_{x}$ term acting as a source. The ratio
of the third to the second terms of equation (\ref{eq:surfactant-evolution-1-1-1-1-1})
is of order $\text{Ma}/(snL)$ and since $n$ and $L$ are large,
assuming $\Ma/s=O(1)$ or less, the second derivative term in equation
(\ref{eq:surfactant-evolution-1-1-1-1-1}) is neglected. Thus (\ref{eq:surfactant-evolution-1-1-1-1-1})
simplifies to the form 
\begin{equation}
\Gamma_{t}+\frac{s}{2}\Gamma_{x}=-\frac{s}{n}\eta_{x}\label{eq:linear-first-order-gamma-equation-1}
\end{equation}
(where $s=1$ for this example). Note that the dominant balance is
between the two terms with the first order $x$ derivatives. Hence
we see that 
\begin{equation}
\Gamma\approx-\frac{2}{n}\eta.\label{eq:gammaproptoeta-1}
\end{equation}
This means that $\Gamma$ and $\eta$ are in anti-phase. The right
hand side of equation (\ref{eq:linear-first-order-gamma-equation-1})
is a known function, a solution of the Kuramoto-Sivashinsky equation
(\ref{eq:eta-evolution-1-2-1-1}). It is well known and also it can
be easily checked that the solution of the equation of the form 
\[
u_{t}+au_{x}=f(t,x)
\]
with the initial condition $u(0,x)=u_{0}(x)$ is 
\begin{equation}
u(t,x)=u_{0}(x-at)+\int_{0}^{t}\;f(\tau,x-at+a\tau)\;d\tau,\label{eq:general solution to  1st order u equation-1}
\end{equation}
where, in our case $u(t,x)=\Gamma(t,x)$, $a=s/2$, and $f(t,x)=-\frac{s}{n}\eta_{x}(t,x)$.
We change the variable $\tau$ to $y$ where $y=x-at+a\tau$ so that
$\tau(y)=\frac{y-x}{a}+t$. Then 
\[
\eta_{x}(t,x)=\eta_{y}(\frac{y-x}{a}+t,y),
\]
where the partial derivative is with the first variable being fixed
at the value $\frac{y-x}{a}+t$. As a result, the integral in (\ref{eq:general solution to  1st order u equation-1})
takes the form 
\[
-\frac{s}{n}\int_{x-at}^{x}\frac{\partial\eta}{\partial y}\left(\frac{y-x}{a}+t,y\right)\;\frac{dy}{a}.
\]
Note that the partial derivative under the integral is related - and
briefly will be seen to be approximately equal - to the ordinary derivative
as 
\begin{equation}
\frac{d\eta}{dy}=\frac{\partial\eta}{\partial\tau}(\tau(y),y)\frac{d\tau}{dy}+\frac{\partial\eta}{\partial y}(\tau,y).\label{eq:full deta dy-1}
\end{equation}
Here, the partial derivatives of the Kuramoto-Sivashinsky solution
have the following estimates: 
\[
\frac{\partial\eta}{\partial y}\sim\frac{N}{L},\quad\frac{\partial\eta}{\partial\tau}\sim\frac{N}{T}=\frac{N}{12L^{4}}.
\]
So the first term in (\ref{eq:full deta dy-1}) can be neglected.
Thus the integral in question is approximately 
\[
-\frac{s}{n}\int_{x-at}^{x}\frac{d\eta}{dy}(\tau(y),y)\;\frac{dy}{a}=-\frac{s}{an}\left(\eta(t,x)-\eta(0,x-at)\right).
\]
Therefore the solution is 
\begin{equation}
\Gamma(t,x)=-\frac{2}{n}\eta(t,x)+\left(\Gamma(0,x-at)-\frac{2}{n}\eta(0,x-at)\right).\label{eq:gamma solution large time large n-1}
\end{equation}
Hence, when the initial conditions can be neglected as compared to
the saturated solutions, we return to (\ref{eq:gammaproptoeta-1}).

\begin{figure}
\includegraphics[bb=0bp 0bp 576bp 288bp,clip,width=0.95\textwidth]{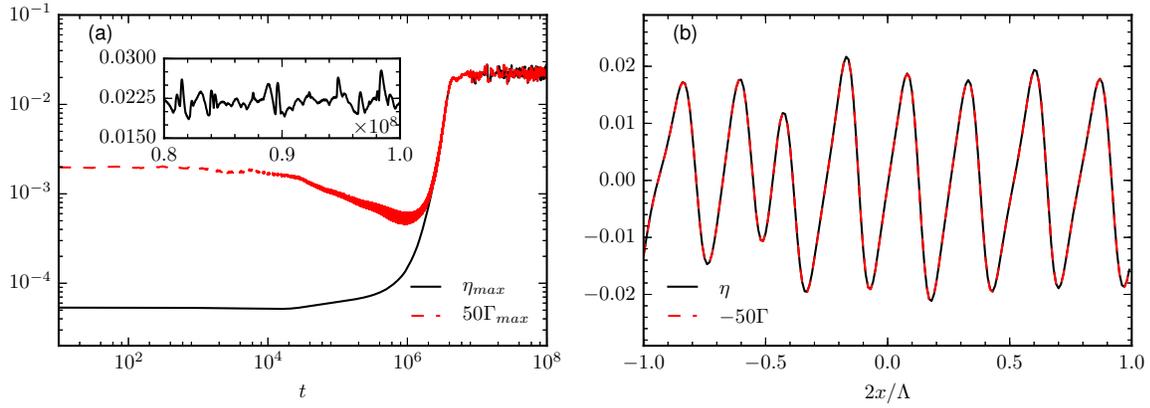}\caption{(a) Time dependence of the maximum values of $\eta$ and $\Gamma$
over the spatial domain $-\Lambda/2\le x\le\Lambda/2$, with $m=n^{2}=100^{2}$,
$s=1$, $\text{Ma}=1$, $\text{Bo}=-0.01$ and $\Lambda=200\pi$,
obtained by solving the coupled equations (\ref{eq:eta-evolution-1-1-1})
and (\ref{eq:surfactant-evolution-1-1-1-1-1}) for $0<t<10^{8}$.
The factor $-50$ multiplying the surfactant concentration $\Gamma$
corresponds to (\ref{eq:gammaproptoeta-1}). The (linear-scales) inset
zooms in on $\eta_{\textrm{max}}$ for a later part of the numerical
run, $8\times10^{7}<t<10^{8}$. (b) Snapshot of small-amplitude spatial
profiles typical of the ultimate, post-saturation, stage of evolution.
It shows $\eta$ and scaled $\Gamma$ at $t=2\times10^{8}$ for the
evolution pertaining to part (a). \label{fig:etagamma_weakn}}
\end{figure}

With no constraints on $m$ and $n$ (so that $m$ is not necessarily
equal to $n^{2}$ and $n$ is not necessarily large), we solved the
strongly nonlinear system of equations, (\ref{eq:eta evolution equation})
and (\ref{eq:Gamma evolution equation}), (except for the figure \ref{fig:etagamma_weakn}(a)
obtained with the weakly nonlinear equations) numerically on the interval
$-\Lambda/2\le x\le\Lambda/2$ with periodic boundary conditions using
the method of lines, where the spatial derivatives were approximated
using fourth-order finite differences. A variable time-stepping scheme
was used from the software package SUNDIALS (\citet{hindmarsh2005sundials}).
The length of the computation domain, $\Lambda$, was chosen to be
large enough so that the choice of initial conditions did not significantly
influence the large time solutions of the system of equations.

Figure \ref{fig:etagamma_weakn}(a) shows the time evolution of $\eta_{\max}=\max_{-\Lambda/2\le x\le\Lambda/2}(\eta(t,x))$
and $50\Gamma_{\max}=50\max_{-\Lambda/2\le x\le\Lambda/2}(\Gamma(t,x))$
for the following set of parameters: $\Lambda=200\pi$, $m=n^{2}=100^{2}$,
$s=1$, and $\text{Bo}=-0.01$. It bears out that eventually there
is small-amplitude saturation of the instability and that in this
ultimate regime the solution to (\ref{eq:eta-evolution-1-1-1}) and
(\ref{eq:surfactant-evolution-1-1-1-1-1}) satisfies the proportionality
property, equation (\ref{eq:gammaproptoeta-1}). Figure \ref{fig:etagamma_weakn}(b)
also shows that the large-time prediction, equation (\ref{eq:gammaproptoeta-1}),
is corroborated, this time for the spatial profiles obtained in the
numerical simulation of the strongly nonlinear equations. (This also
provides an additional testimony to the veracity of the weakly-nonlinear
numerical solutions.) The inset in part (a) of this figure zooms in
on a part of the ultimate evolution of $\eta_{\textrm{max}}$ resolving
its fluctuations and revealing their characteristic time scale. 

\begin{figure}
\includegraphics[bb=0bp 0bp 576bp 288bp,clip,width=0.95\textwidth]{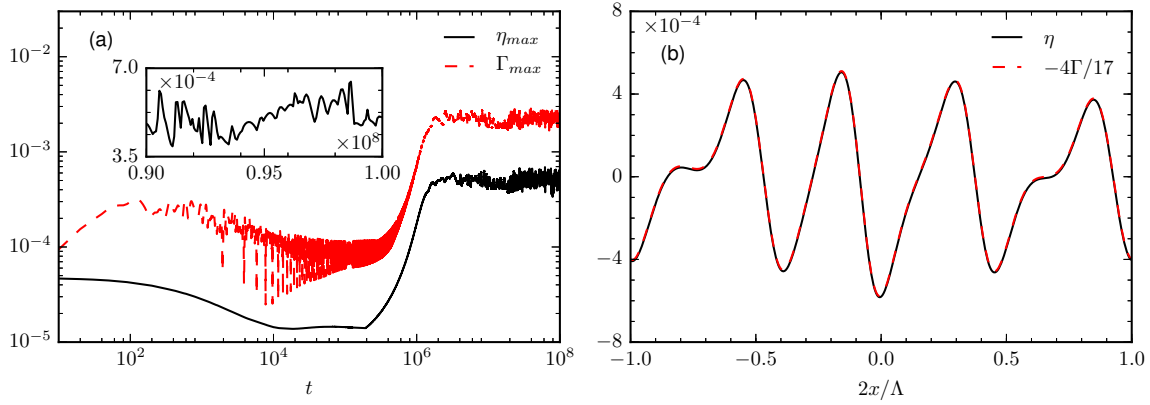}\caption{(a) The maximum values of $\eta$ and $\Gamma$ as functions of time
in the $R$ sector. Here $n=m=2$, $s=1$, $\text{Ma}=0.01$, $\text{Bo}=0.001$,
and $\Lambda=100\pi$. The (linear-scales) inset zooms in on $\eta_{\textrm{max}}$
for a later part of the run, $9\times10^{7}<t<10^{8}$. (b) Spatial
profiles of $\eta$ and $\Gamma$ at the end time, $t=10^{8}$, of
the evolution pertaining to part (a). \label{fig: etagammaRsector-1}}
\end{figure}

Such small-amplitude saturation solutions are found even with zero
Bond number in the $R$ sector. The Bond number terms disappear from
equations (\ref{eq:eta-evolution-1-2-1-1}) and (\ref{eq:weak surfactant evolution 2-1}).
In the latter, if the Marangoni number is of order one or less, for
the long waves, the term with the second derivative of $\Gamma$ is
much smaller than the term with the first derivative of $\Gamma$,
and the dominant balance is between the $\Gamma_{x}$ and the $\eta_{x}$
terms in equation (\ref{eq:weak surfactant evolution 2-1}) provided
that the time scale is sufficiently large. This, similar to equation
(\ref{eq:gammaproptoeta-1}), implies the relation 
\begin{equation}
\Gamma\approx-\frac{\phi}{2n^{2}(m-1)}\eta.\label{eq:gammapropto eta R sector-1}
\end{equation}
In the $R$ and $Q$ sectors this clearly implies that $\Gamma$ and
$\eta$ are in anti-phase. (Note that the same relation is found for
the normal modes of the linear theory given by equation (\ref{eq:matrix system}).
Also, in the limit of $m=n^{2}$ and $n\to\infty$ we recover the
relation (\ref{eq:gammaproptoeta-1}).) Substituting (\ref{eq:gammapropto eta R sector-1})
into the kinematic equation (\ref{eq:eta-evolution-1-2-1-1}) we obtain
the Kuramoto-Sivashinsky equation 
\begin{align}
\eta_{t}+\frac{n^{3}(m+n)}{3\psi}\eta_{xxxx}+\frac{\phi(n^{2}-m)\Ma}{4\psi(m-1)}\eta_{xx}+sN_{1}\eta\eta_{x} & =0.\label{eq:kuramoto-sivashinky large time}\\
\nonumber 
\end{align}
The characteristic scales (assuming that all other parameters except
for $\text{Ma}$ are of order one including $m-1$) become $L\sim\text{Ma}^{-1/2}$,
$N\sim\text{Ma}/L\sim\text{Ma}^{3/2}$ and $T\sim L^{4}\sim\text{Ma}^{-2}$.
Hence for $\text{Ma}\ll1$, the length scale is large, the time scale
is even much larger, the amplitudes are small, and the previously
assumed dominant balance in the surfactant equation is justified.
With these scales, neglecting the term with $\text{Bo}$ in equation
(\ref{eq:eta-evolution-1-2-1-1}) in comparison with the fourth derivative
term is consistent if $\text{Bo}\ll L^{-2},$ that is $\text{Bo}\ll\text{Ma}$.
(Neglecting the term with $\text{Bo}$ in equation (\ref{eq:weak surfactant evolution 2-1})
as compared to the term with $\eta_{x}$ leads to a weaker requirement,
$|\text{Bo}|\ll(n+1)\phi sL/(n^{2}(m-n^{2}))$. By considerations
similar to those which led to equation (\ref{eq:gamma solution large time large n-1}),
we obtain a correction to (\ref{eq:gammapropto eta R sector-1}) due
to the initial conditions, 
\begin{equation}
\Gamma(t,x)=-\frac{\phi}{2n^{2}(m-1)}\eta(t,x)+\left(\Gamma(0,x-at)-\frac{\phi}{2n^{2}(m-1)}\eta(0,x-at)\right),\label{eq:Gamma driven by KS equation}
\end{equation}
where $a=\frac{2(m-1)n^{2}(n+1)s}{\psi}$. For example, for $m=n=2$,
the relation (\ref{eq:gamma solution large time large n-1}) gives
$\Gamma=-\frac{17}{4}\eta$ for the large-time, permanent, saturated
state. Thus we have two chaotic functions, $\eta$ and $\Gamma$,
which differ by just a constant factor. As an illustration, a numerical
simulation (of the strongly nonlinear system) yields the time dependencies
of $\eta_{max}$ and $\Gamma_{\max}$, figure \ref{fig: etagammaRsector-1}(a),
which show the saturation of instability, and the spatial profiles,
figure \ref{fig: etagammaRsector-1}(b), which are all in excellent
agreement with the predictions. The present result of small-amplitude
saturation is in marked contrast with our earlier findings (see \citet{Frenkel2006})
that for the semi-infinite system, also with no gravity, no small
amplitude saturation is possible (which was also confirmed in the
numerical simulations of \citet{Bassom2010,kalogirou2016} ). 

The above results have been obtained for the $R$ sector where the
robust mode is unstable and the surfactant one is stable. To the contrary,
in the $S$ sector, where the surfactant mode is unstable and the
robust mode is stable for zero Bond number, there appears to be no
small-amplitude saturation. Moreover, as is discussed in the next
section, even the long-wave assumption may get violated after some
time so that no long-wave solutions exist at large time.

\begin{figure}
\includegraphics[bb=0bp 0bp 576bp 288bp,clip,width=0.95\textwidth]{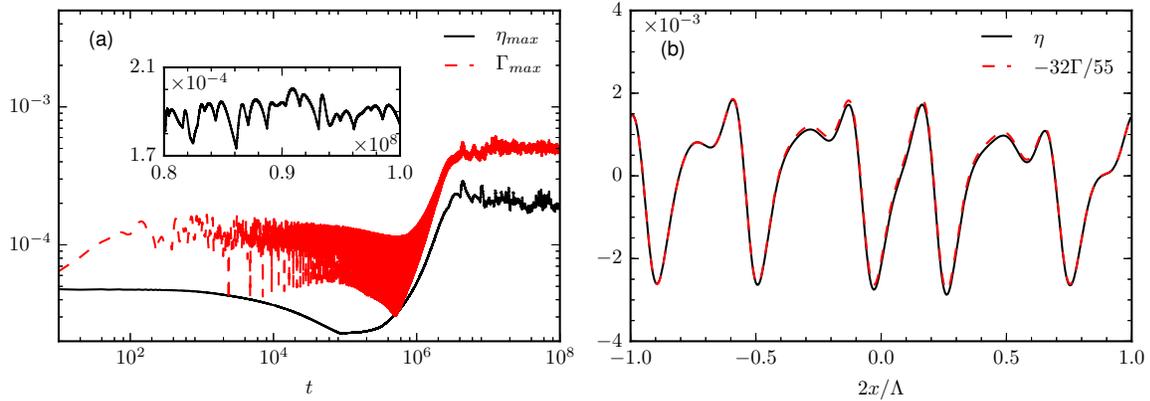}
\protect\caption{(a) Time dependence of the maximum values of $\eta$ and $\Gamma$
in the $Q$ sector. The parameter values are $n=2$, $m=5$, $s=1$,
$\text{Ma}=0.001$, $\text{Bo}=-0.01$, and $\Lambda=200\pi$. The
(linear-scales) inset zooms in on $\eta_{\textrm{max}}$ for a later
part of the run, $8\times10^{7}<t<10^{8}$. (b) Spatial profiles of
$\eta$ and $\Gamma$ at the end time, $t=10^{8}$, of the evolution
pertaining to part (a). \label{fig:etagammaQsector-1}}
\end{figure}

In the $Q$ sector, for finite $m$ and $n$, small negative $\text{Bo}$,
and assuming that $\text{Ma}$ is so small that the terms containing
it can be discarded, we obtain again the decoupled Kuramoto-Sivashinsky
equation, leading to $L\sim(-\Bo)^{-1/2}$, $N\sim n^{3}(m+n)/(3\psi L^{3}s)$,
and $T\sim3\psi L^{4}n^{-3}/(m+n)$. We can see that the relation
(\ref{eq:gammapropto eta R sector-1}) holds here as well as in the
$R$ sector. Such solutions belonging to the $Q$ sector are illustrated
in figure \ref{fig:etagammaQsector-1}. Note that $\Gamma\approx-\frac{55}{32}\eta$,
exactly as equation (\ref{eq:gammapropto eta R sector-1}) predicts
for $n=2$ and $m=5$. 

As we know in the $S$ sector the robust modes are unstable provided
that the Bond number is negative and below the threshold given by
(\ref{eq:BoCritical}). They still saturate with small amplitude like
in the $R$ and $Q$ sectors. The difference is that $\Gamma$ and
$\eta$ are in phase as opposed to anti-phase. For example, in the
cases where the Marangoni number is essentially zero, we have the
Kuramoto-Sivashinsky equation (\ref{eq:eta-evolution-1-2-1-1}) with
a destabilizing gravity term. This leads to small-amplitude saturation
of the Rayleigh-Taylor instability. Similar saturation was found e.g.
in \citet{Babchin1983a} but for $n=\infty$. (Note, however, that
the saturation of the Rayleigh-Taylor instability in the finite channels
has not been demonstrated before the present study.) The surfactant
in this case plays no dynamical role, and is just advected passively
by the flow.

\subsection{Nonlinear saturation in the $S$ sector with linearly unstable surfactant
modes}

In the previous subsection we established that unstable robust modes
saturate with the amplitudes of both $\eta$ and $\Gamma$ being small.
To the contrary, in the $S$ sector, there appears to be no small-amplitude
saturation of the linearly unstable surfactant mode. (Recall that
the surfactant modes are linearly stable in the $R$ and $Q$ sectors.)
However, it is possible that the saturated $\eta$ amplitude is still
small while the saturated $\Gamma$ is not small. For such regimes,
as was noted above, the linear transport equation (\ref{eq:weak surfactant evolution 2-1})
acquires a nonlinear term and thus takes the form 
\begin{align*}
\varGamma_{t}+\frac{(n+1)\phi}{\psi}s\left[\eta(1+\Gamma)\right]_{x}+\frac{2(m-1)n^{2}(n+1)s}{\psi}\Gamma_{x}-\frac{n(m+n^{3})\Ma}{\psi}\left[\Gamma_{x}(1+\Gamma)\right]_{x}\\
-\frac{n^{2}(n^{2}-m)\Bo}{2\psi}\eta_{xx}+\frac{n^{2}(n^{2}-m)}{2\psi}\eta_{xxxx} & =0.
\end{align*}
Note that a nonlinear term containing Marangoni number has been included,
as it may be comparable with the $s$ term, since the extra differentiation
in the former can be balanced by the smallness of $\eta$ in the latter.

\begin{figure}
\includegraphics[bb=0bp 0bp 576bp 193bp,clip,width=0.95\textwidth]{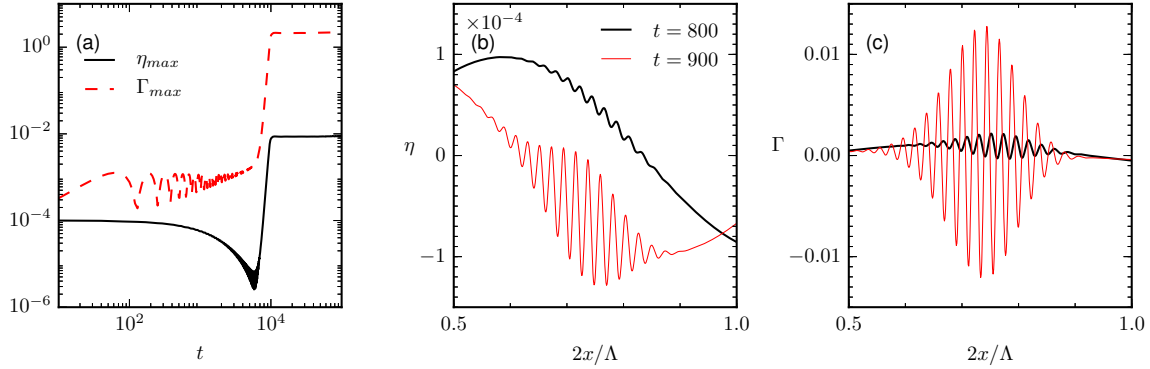}
\caption{(a) The maximum values of $\eta$ and $\Gamma$ as functions of time
in the $S$ sector, with $n=2$, $m=1/2$, $s=1$, $\text{Ma}=10^{-2}$,
$\text{Bo}=0$ and $\Lambda=20\pi$. (b) $\eta$ and (c) $\Gamma$
profiles near the moment when small scale disturbances appear if large
scale but small-amplitude initial conditions are used. Note that only
the right half of the spatial domain is shown, where the small scales
first appear. \label{fig:etagammamax S Sector-1}}
\end{figure}

\begin{figure}
\includegraphics[bb=0bp 0bp 576bp 288bp,clip,width=0.95\textwidth]{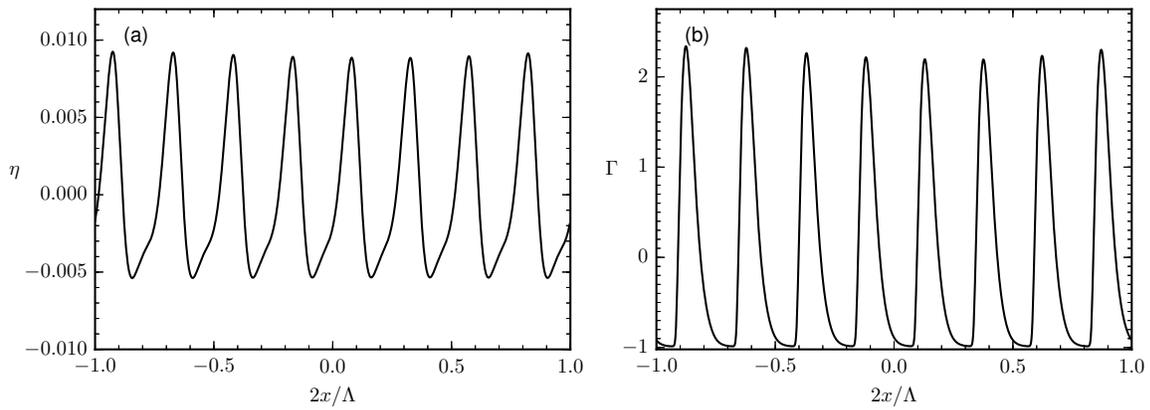}
\caption{Small amplitude $\eta$ and large amplitude $\Gamma$ profiles at
the end time, $t=10^{5}$, of the evolution corresponding to figure
\ref{fig:etagammamax S Sector-1}(a). \label{fig:Saturated-eta-and-gamma-S-sector-1}}
\end{figure}

Moreover, even the long wave assumption may get violated after some
time so that no long wave solutions exist at large time. As an example,
the run corresponding to figure \ref{fig:etagammamax S Sector-1}(a)
starts with a very long-wave sinusoidal initial condition, but later,
as shown in part (b) of figure \ref{fig:etagammamax S Sector-1},
a short-wave disturbance appears on a limited part of the profiles.
As time goes on, the amplitude and the extent of the disturbance grow
(see part (c) of figure \ref{fig:etagammamax S Sector-1} ), and on
the post-saturation stage (see figure \ref{fig:Saturated-eta-and-gamma-S-sector-1}),
we have small-amplitude $\eta$ but $\Gamma$ of order one, and the
characteristic length of the pulses is not large. Then, even the lubrication
approximation assumptions are not satisfied. 

\begin{figure}
\includegraphics[bb=0bp 0bp 576bp 288bp,clip,width=0.95\textwidth]{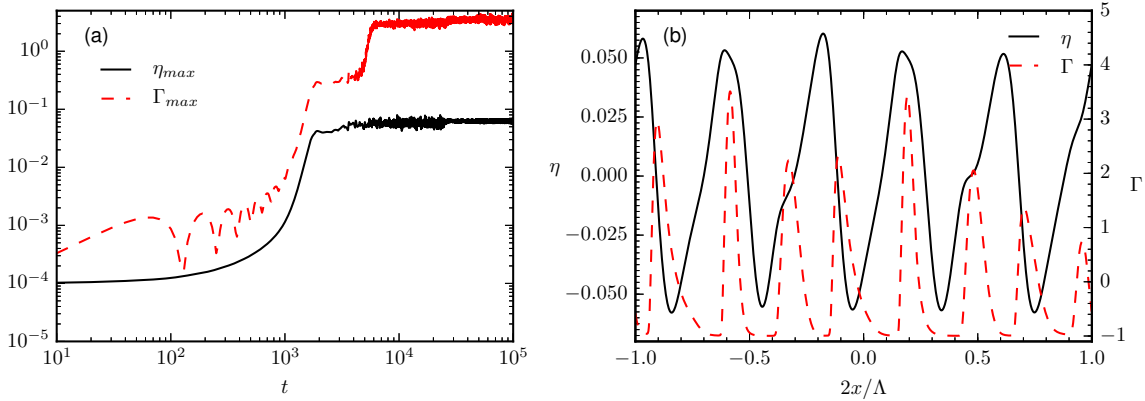}
\caption{(a) Evolution of the maximum amplitude of $\eta$ and $\Gamma$ for
the same parameter values as in the preceding two figures (thus in
the $S$ sector), with the exception that $\text{Bo}=-0.5$ here.
(b) Small-amplitude $\eta$ and large-amplitude $\Gamma$ at the end
time, $t=10^{5}$. \label{fig:etagamma_S_Sector_Bo-1}}
\end{figure}

This non-applicability of the lubrication approximation seems to be
a general feature for the $S$ sector. For example, if we take the
Bond number sufficiently negative so that there are unstable robust
modes along with the unstable surfactant ones, we get results similar
to the ones shown in figure \ref{fig:etagamma_S_Sector_Bo-1}. We
note that for these ``deeply-robust'' regimes the number of $\Gamma$
pulses appears to be different from that of $\eta$ pulses. This contrasts
with the purely surfactant-mode regimes, and the robust-surfactant
regimes with a smaller negative value of the Bond number. (We also
note that the amplitude of fluctuations of $\eta_{\textrm{max}}$
and $\varGamma_{\textrm{max}}$ in the post-saturation state may change
with the number of pulses on the computation interval. This occurs
due to coalescences of pulses and the emergence of new pulses, similar
to such phenomena observed for a different strongly nonlinear equation
in \citet{Kerchman1994}.) 

\section{Summary and discussion}

\label{sec:Conclusions}

This study concerned the linear and nonlinear stages of evolution
of initially small disturbances of the horizontal two-fluid Couette
flow (with top-to-bottom aspect ratio $n$ and viscosity ratio $m$
) in the presence of surfactants and gravity, and with negligible
inertia (figure \ref{fig:FigDefinitionSketch}). For any flow with
$n<1$, it is described as one with $n>1$ in a new coordinate system
obtained by reversing the (spanwise) $z$-axis direction. Therefore,
without loss of generality, we consider the flows with $n\geq1$.
The lubrication approximation yields two coupled strongly nonlinear
evolution equations for the interface thickness and the insoluble
surfactant concentration. These equations take a weakly nonlinear
form when the amplitudes of disturbances are small, but finite. The
onset of instability is investigated by linearizing (for the infinitesimal
disturbances) these evolution equations and applying, as usual, the
normal mode analysis.

The dispersion relation for the increment (a complex eigenvalue which
determines the real growth rate and the phase velocity) of the linear
instability is found to be a quadratic equation whose coefficients
depend on the interfacial shear-rate, the aspect and viscosity ratios,
the Marangoni number and the Bond number. As introduced in HF for
the case of no gravity, the subdivision of the $n\geq1$ part of the
$(n,m)$-plane into the three sectors - called here the $Q$ sector,
in which $m>n^{2}>1$; the $R$ sector, characterized by $1<m<n^{2}$;
and the $S$ sector, $m<1$ (figure \ref{fig:FigRegions}) - turns
out to be useful even in the presence of gravity.

The growth rate dependence on the wavenumber, being the real part
of the solution to the quadratic equation, has two single-valued continuous
branches, called the robust branch and the surfactant one. Correspondingly,
for each wavenumber, there is a single robust normal mode that exists
even when $\text{Ma}=0$ and $s=0$, and a single surfactant normal
mode that vanishes when $\text{Ma}\downarrow0$, and we speak of the
two branches (sets) of modes. The expressions for the growth rates
for the base flows with a nonzero shear rate $s$ differ from those
for a stagnant base two-layer system.

For $s\neq0$, the growth rate for the robust branch (see equation
(\ref{gamCsmallAlphaApprox})) is the sum of two terms, which are
both independent of $s$: a Marangoni term (which equals the growth
rate due to the surfactant in the absence of gravity, first found
in \citet{Frenkel2002}), and a Bond term (which gives the well-known
growth rate of the Rayleigh-Taylor instability of the flow with no
surfactants). The Marangoni term is negative in the $S$ and $Q$
sectors and positive in the $R$ sector. The Bond term (with its negative
sign included) clearly increases when the Bond number decreases. Therefore,
the instability sets in when the Bond number is less than some threshold
value denoted $\text{Bo}_{cL}$. In the $Q$ and $S$ sectors, where
the Marangoni term is negative, the surfactant acts on the robust
modes in a stabilizing way, so $\text{Bo}_{cL}<0$, while in the $R$
sector, the surfactant action has a destabilizing character and hence
$\text{Bo}_{cL}>0$. The ratio of the threshold Bond number to the
Marangoni number, being independent of the base shear-rate $s$, varies
with the two remaining variables, the aspect ratio $n$ and the viscosity
ratio $m$, only (figure \ref{fig:fig3}).

In contrast to the robust modes, the growth rate for the surfactant
branch with $s\neq0$ (see equation (\ref{gamSsmallAlphaApprox}))
has no purely Bond term; the leading term is purely Marangoni, not
containing the Bond number at all; and the higher order terms, if
they have the Bond number as a factor, always contain some Marangoni
number factor as well. These surfactant modes are stable in the $Q$
and $R$ sectors and unstable in the $S$ sector. Thus, in the $S$
sector, a finite band of the long-wave surfactant modes, somewhat
surprisingly, are unstable even for arbitrarily large Bond number
(albeit the band width is expected to decrease as gravity grows stronger).
Thus, no amount of gravity, however strong, can completely stabilize
the surfactant instability in the $S$ sector.

On the other hand, in the $Q$ sector, the surfactant can stabilize
the Rayleigh-Taylor instability. Only the robust branch needs to be
stabilized since the surfactant one is stable independent of the Bond
number (see equation \ref{gamSsmallAlphaApprox}). For example, the
flow with $n=2$, $m=5$, and $\text{Bo}=-0.015$ is Rayleigh-Taylor
unstable in the absence of surfactant. But in the presence of surfactant,
such that, say, $\text{Ma}=0.1$, the flow is stable according to
equation (\ref{gamCsmallAlphaApprox}). This value of the Bond number
corresponds to $(\rho_{2}-\rho_{1})gd_{1}^{2}=0.015\sigma_{0}$ in
view of equation (\ref{eq:Bond number definition}), that is, for
$\sigma_{0}=10$ (in cgs units), $(\rho_{2}-\rho_{1})gd_{1}^{2}=0.15$.
For the Earth's gravity, $g\approx10^{3}$, and $\rho_{2}-\rho_{1}\sim O(1)$,
this means the thickness $d_{1}\sim10^{-2}\text{cm}$, a rather thin
film. But, under the conditions of microgravity, with (say) $g\sim10^{-1}$,
the bottom layer is much thicker, $d_{1}\sim1\text{\;cm}$. (Even
with the Earth's gravity, the film thickness is $d_{1}\sim10^{-1}\text{cm}$
if the densities are almost equal, $\rho_{2}-\rho_{1}\sim10^{-2}$.)
It is remarkable that the interfacial surfactant can completely suppress
the Rayleigh-Taylor instability under quite realistic conditions.

The lubrication approximation is sufficient for finding the results,
including increments, in the leading-order and also in the next order
correction. For the robust mode, the leading-order of the increment
determines the wave velocity which turns out to be independent of
the wavenumber $\alpha$ and hence can be eliminated by using the
co-moving reverence frame. To find the first truly nonzero term of
the wave velocity, scaling as $\alpha^{2}$, one needs the first post-lubrication
corrections to the governing equations (as found in Appendix B), whereas
the lubrication-theory increment correction gives the leading-order
growth rate.

For the case of equal viscosities, $m=1$ (with $s\ne0$), i.e. on
the boundary between the $R$ and $S$ sectors, the gravitational
effects are absent in the leading-order, and so, as was found in \citet{Frenkel2002},
both growth rates scale as $\alpha^{3/2}$ (equation (\ref{eq:equal viscosities increment})).
The correction to them, proportional to $-\alpha^{2}\textrm{Bo}$,
is the same for both modes (equation (\ref{eq:equal visc increm Bo correction})).

For the case with no base flow, i.e. with $s=0$, both modes are stable
if the Bond number is positive, but one of the modes is unstable if
the Bond number is negative. This is essentially the Rayleigh-Taylor
instability of a stagnant system modified by the surfactant.

The eigenfunction amplitudes, including those of the surfactant concentration
$G$ (where the arbitrary interface deviation amplitude $h$ is taken
to be real and positive), the velocities, and pressures, are determined
as well. This can be done using the eigenvectors of the system for
$G$ and $h$. However, we have used also a different way, linearizing
the primitive governing equations (rather than the two evolution equations
derived from them) and using the method of undetermined coefficients,
which has advantages in uncovering the physical mechanisms of instability
for the two modes.

We suggested that in the inertialess settings, the vorticity lacks
the dynamic significance which was shown by \citet{Charru2000} for
a surfactantless case of the Yih instability, whose very existence
depends on inertia. Thus, in contrast to the case of Yih instability,
vorticity does not appear to be a suitable agent for the mechanism
of the surfactant instability. \citet{Wei2005c} showed that under
certain conditions, without gravitational effects, there is a correlation
of stability of normal modes with $\theta_{\omega}$, the phase shift
between the interfacial vorticity and the interfacial displacement.
Namely, $\theta_{\omega}$ being in the interval $(0,\pi)$ corresponds
to instability while $\theta_{\omega}$ within the interval $(-\pi,0)$
corresponds to stability. However, we showed that, under the same
conditions, except for the Bond number being nonzero, this correspondence
does not necessarily hold. For example, figure \ref{fig:figD2} shows
that the growth rate changes from negative to positive as the Marangoni
number grows, but the vorticity-interface phase shift remains in the
same interval $(0,\pi)$ all along, thus for both the stable and unstable
flows. This is related to the lack of any significant role of vorticity
for instability in the absence of inertial effects, with or without
gravitational effects.

To uncover the mechanisms of instability for the two modes, we considered
the case of large thickness ratio and used the mass conservation laws
in their integral forms (similar to \citet{Charru2000}). The growth/decay
mechanism for the robust branch is as follows: the leading-order disturbance
flow is the same as in \citet{Yih1967} and leads to the same, purely
imaginary, increment. This flow is found from physical considerations
as in \citet{Charru2000}, using the fact that the thick layer disturbances
uncouple in the case of the large aspect ratio. The surfactant transport
is determined by the base velocity at the perturbed interface. As
a result, the surfactant wave of the normal mode must propagate either
in anti-phase, for $m>1$, or in phase, for $m<1$, with the interface.
The Marangoni tangential stress exerted by the surfactant drives a
correction flow in the thin layer whose horizontal velocity is $-90\si{\degree}$
out-of-phase with the surfactant. (This holds for all elevations,
since the vertical profile of this velocity is linear, and thus it
has the same sign at all elevations.) Thus, this velocity is either
$90\si{\degree}$, for $m>1,$ or $-90\si{\degree}$, for $m<1$,
out-of-phase with the interface. For $m>1$, this leads to a net outflow
for the half-period part of the thin layer with the thickness minimum
at the interval midpoint, which means instability for the normal mode;
and for $m<1$, the velocity is reversed, which yields stability.

The surfactant branch corresponds to the Marangoni stresses playing
a role already in the leading-order of disturbances (in contrast to
their correction role for the robust branch). The leading-order flow
disturbance in the thick layer is still the same as that for the robust
mode. We find that the Marangoni stress must cancel the viscous tangential
stress of the thick layer at the interface. As a result, the surfactant
phase shift with respect to the interface is $90\si{\degree}$ for
$m>1$ and $-90\si{\degree}$ for $m<1$. On the other hand, the surfactant
flux, the product of the base concentration and the thin-layer base
velocity at the perturbed interface, is always in phase with the interface.
Hence, the surfactant flux is out-of-phase with the surfactant wave,
$90\si{\degree}$ for $m>1$ and $-90\si{\degree}$ for $m<1$. Thus,
considering the half-period interval of the wave with the maximum
positive net influx through its endpoints, the surfactant concentration
is minimum at the midpoint for $m>1$, but the magnitude of this minimum
gradually decreases, which means stability. In contrast, for $m<1,$
the surfactant concentration is maximum at the midpoint, which grows
because of the positive net influx of the surfactant, and this corresponds
to instability.

With no gravity, small-amplitude nonlinear saturation of the surfactant
instability is possible (figure \ref{fig: etagammaRsector-1}), in
contrast to the semi-infinite case studied by \citet{Frenkel2006}.
For non-zero Bond number, the small-amplitude saturation in the $Q$
sector is seen in figure \ref{fig:etagammaQsector-1}(a). It also
occurs along the border between the $R$ and $Q$ sectors, where $m=n^{2}$,
for $\text{Bo}<0$ (figure \ref{fig:etagamma_weakn}(a)).

For certain ranges of $(m,n)$, in the $R$ and $Q$ sectors, the
interface is governed by a decoupled Kuramoto-Sivashinsky equation,whose
solution provides a source term for the linear convection-diffusion
equation of the surfactant. When diffusion is negligible, the surfactant
equation has an analytic solution. As a result, the surfactant wave
is as chaotic as the interface; however, the ratio of the two waves
is constant at sufficiently large times such that the saturated state
has been reached (figures \ref{fig:etagamma_weakn}(b), \ref{fig: etagammaRsector-1}(b),
and \ref{fig:etagammaQsector-1}(b)). These analytical predictions
are confirmed by the full numerical solution of the nonlinear evolution
equations.

In contrast, we have never seen the small-amplitude saturation in
the $S$ sector, $m<1$. Instead, numerical results show that the
instability saturates with only the interface disturbances being small-amplitude
but the surfactant ones large (figures \ref{fig:etagammamax S Sector-1}
and \ref{fig:etagamma_S_Sector_Bo-1}). However, the final characteristic
length scale of these solutions is not as large as is required by
the lubrication approximation (figures \ref{fig:Saturated-eta-and-gamma-S-sector-1}(b)
and \ref{fig:etagamma_S_Sector_Bo-1}(b)). To the best of our knowledge,
the only other simulations for the case of finite aspect ratio, even
with zero gravity, were performed in \citet{Blyth2004b}. They were
limited to the $S$ sector and small computational intervals. The
saturation that they observed was not small-amplitude, and we checked
that if extended to sufficiently large intervals, the evolution leads
to a characteristic length scale being small, and thus not consistent
with the lubrication approximation. The question whether such partly
weakly and partly strongly nonlinear saturated regimes are real may
be decided by a future non-lubrication theory. Also, the inertial
effects could be included, for example, similar to \citet{Frenkel2005}.
The three-dimensional disturbances could be considered similar to
\citet{frenkel2000saturation}.

\appendix

\section{Coefficient $k_{s}$ \label{sec:coefficient}}

The coefficient of the $\alpha^{4}$ term that appears in equation
(\ref{gamSsmallAlphaApprox}) is

\begin{align}
k_{S}= & \frac{\text{Ma}\left(n^{3}-4n^{2}+4n-1\right)}{60(m-1)}\nonumber \\
 & +\frac{\text{Ma}^{3}}{128(m-1)^{5}n^{4}(n+1)s^{2}}(n-1)\left(m^{4}(3n+1)+2m^{3}\left(-3n^{3}-2n^{2}+4n+1\right)n\right.\nonumber \\
 & +\left.4m^{2}\left(n^{3}-2n^{2}-2n+1\right)n^{3}+2m\left(n^{3}+4n^{2}-2n-3\right)n^{5}+(n+3)n^{8}\right)\nonumber \\
 & +\frac{\text{Bo}\text{Ma}^{2}}{192(m-1)^{4}n(n+1)^{2}s^{2}}\left(m^{3}\left(3n^{2}-4n-3\right)\right.+m^{2}\left(2n^{3}+13n^{2}-6n-5\right)n\nonumber \\
 & +\left.m\left(-5n^{3}-6n^{2}+13n+2\right)n^{3}+\left(-3n^{2}-4n+3\right)n^{5}\right)\nonumber \\
 & +\text{Bo}^{2}\text{Ma}\frac{n^{2}\left(-m^{2}+m(n-1)n+n^{3}\right)}{144(m-1)^{3}(n+1)^{2}s^{2}}.\label{eq:ks}
\end{align}

\section{Augmented lubrication theory\label{sec:Augmented-lubrication-theory}}

Write 
\[
u_{fj}=u_{j}+\alpha^{2}u_{cj}+O(\alpha^{4}),\;w_{fj}=w_{j}+\alpha^{2}w_{cj}+O(\alpha^{4}),\;p_{fj}=p_{j}(x)+\alpha^{2}p_{cj}(x,z)+O(\alpha^{4}),
\]
where the subscript $f$ marks the full disturbances, the first term
in each right-hand side is the lubrication-approximation value, the
next term, with the subscript $c$, is the leading order correction
to the lubrication-approximation, and the last term indicates the
higher order error. The continuity equation yields 
\begin{equation}
Dw_{cj}=-i\alpha u_{cj}.\label{eq:wcj in terms of ucj}
\end{equation}
Note that from the lubrication approximation $w_{j}=O(\alpha u_{j})$,
$p_{j}=O(\alpha^{-1}u_{j})$, and thus $w_{j}=O(\alpha^{2}p_{j}).$
The (Stokes flow) horizontal momentum equation 
\[
(D^{2}-\alpha^{2})u_{fj}=\frac{1}{m_{j}}i\alpha p_{fj},
\]
with the corresponding lubrication approximation equation 
\[
D^{2}u_{j}=\frac{1}{m_{j}}i\alpha p_{j},
\]
yields the correction equation 
\begin{equation}
D^{2}u_{cj}=u_{j}+\frac{i\alpha}{m_{j}}p_{cj}.\label{eq:pc equation}
\end{equation}
(Having in mind the normal modes, we take the liberty of using the
same notation for a variable and its amplitude, and may use interchangeably
the operator $\partial/\partial x$ and the multiplication by $i\alpha$.)
The vertical-momentum lubrication-approximation equation is 
\[
Dp_{j}=0.
\]
Thus the correction equation 
\begin{equation}
\alpha^{2}Dp_{cj}=m_{j}\left(D^{2}w_{j}\right).\label{eq:pcj}
\end{equation}
Eliminating $p_{cj}$ from the momentum equations (\ref{eq:pc equation})
and (\ref{eq:pcj}) (and making use of the continuity equations) yields
the following nonhomogeneous equation for $u_{cj}$: 
\begin{equation}
D^{3}u_{cj}=2Du_{j}.\label{eq:3rd order ucj equation}
\end{equation}
The solution satisfying the boundary condition at the plates has the
form 
\[
u_{cj}=\frac{i\alpha}{12m_{j}}p_{j}(z^{4}-n_{j}^{4})+\frac{1}{3}A_{j}(z^{3}-n_{j}^{3})+\frac{B_{cj}}{2}\left(z^{2}-n_{j}^{2}\right)+A_{cj}\left(z-n_{j}\right),
\]
where $p_{j}$ and $A_{j}$ are determined by the linearization of
equations (\ref{eq:p1x}) and (\ref{eq:A1}): 
\[
\psi p_{1}=6i\alpha^{-1}(m-1)(m-n^{2})sh-6m(n+1)n\text{Ma}G+(3m+4mn+n^{2})n^{2}\Pi,
\]
\[
\psi A_{1}=-4(m-1)(m+n^{3})sh-(4m+3mn+n^{3})n\text{Ma}i\alpha G+2m(n+1)n^{2}i\alpha\Pi,
\]
and $p_{2}$ and $A_{2}$ are given in terms of $p_{1}$ and $A_{1}$
by equations (\ref{eq:p2 in terms of p1}) and (\ref{eq:A2 in terms of A1}).
The coefficients $B_{cj}$ and $A_{cj}$ are independent of $z$ and
are to be determined by the interfacial conditions. The continuity
equation (\ref{eq:wcj in terms of ucj}) yields $w_{cj}$ in the form
\begin{align*}
w_{cj} & =\frac{\alpha^{2}}{60m_{j}}p_{j}\left(z^{5}-5n_{j}^{4}z+4n_{j}^{5}\right)-\frac{i\alpha}{12}A_{j}\left(z^{4}-4n_{j}^{3}z+3n_{j}^{4}\right)\\
 & -\frac{i\alpha}{6}B_{cj}\left(z-n_{j}\right)^{2}\left(z+2n_{j}\right)-\frac{i\alpha}{2}A_{cj}\left(z-n_{j}\right)^{2}.
\end{align*}
The conditions of velocity continuity, $u_{c1}(0)=u_{c2}(0)$ and
$w_{c1}(0)=w_{c2}(0)$, yield 
\begin{equation}
-\frac{n^{2}}{2}B_{c2}-nA_{c2}+\frac{1}{2}B_{c1}-A_{c1}=\frac{i\alpha n^{4}}{12m}p_{2}+\frac{n^{3}}{3}A_{2}-\frac{i\alpha}{12}p_{1}+\frac{1}{3}A_{1},\label{eq:contin. u_c}
\end{equation}
\begin{equation}
-\frac{i\alpha n^{3}}{3}B_{c2}-\frac{i\alpha n^{2}}{2}A_{c2}-\frac{i\alpha}{3}B_{c1}+\frac{i\alpha}{2}A_{c1}=-\frac{\alpha^{2}n^{5}}{15m}p_{2}+\frac{i\alpha n^{4}}{4}A_{2}-\frac{\alpha^{2}}{15}p_{1}-\frac{i\alpha}{4}A_{1}.\label{eq:contin. w_c}
\end{equation}
The tangential and normal stress conditions at the interface $z=0$,
equations (\ref{TanStressBC}) and (\ref{NrmStressBC}), whose lubrication
approximations are equations (\ref{eq:Pi}) and (\ref{eq:tangential stress lubrication}),
yield the correction equations 
\begin{equation}
m(\alpha^{2}Du_{c2}+i\alpha w_{2})=\alpha^{2}Du_{c1}+i\alpha w_{1},\label{eq:tang.  stress correction}
\end{equation}
and 
\begin{equation}
\alpha^{2}p_{c2}-2mDw_{2}=\alpha^{2}p_{c1}-2Dw_{1},\label{eq:norm. stress correction}
\end{equation}
where, using (\ref{eq:pc equation}) (at $z=0$), $p_{cj}$ is substituted
as 
\[
p_{cj}=-i\alpha^{-1}m_{j}B_{cj}+\frac{n_{j}^{2}}{2}p_{j}-i\alpha m_{j}^{-1}n_{j}A_{j}.
\]
We use the normal stress equation (\ref{eq:norm. stress correction})
to eliminate $B_{c2}$, 
\[
mB_{c2}=B_{c1}-\frac{3}{2}n^{2}i\alpha p_{2}-3mnA_{2}+\frac{3}{2}i\alpha p_{1}-3A_{1}
\]
and the tangential stress equation, (\ref{eq:tang.  stress correction})
to eliminate $A_{c2},$ 
\[
mA_{c2}=A_{c1}-\frac{i\alpha n^{3}}{3}p_{2}-\frac{mn^{2}}{2}A_{2}-\frac{i\alpha}{3}p_{1}+\frac{A_{1}}{2}.
\]
With these substitutions for $A_{c2}$ and $B_{c2}$, equations (\ref{eq:contin. u_c})
and (\ref{eq:contin. w_c}) become a linear nonhomogeneous system
of two equations for the unknowns $A_{c1}$ and $B_{c1}.$ After obtaining
its unique solution in terms of $h$ and $G$, we determine the velocity
corrections at $z=0$ and use them to obtain the corrections to the
coefficient matrix of the system of transport equations (\ref{eq:matrix system})
and hence the corrections to the coefficients of the quadratic dispersion
equation (\ref{eq:quadratic longwave}). Of the latter corrections,
as was discussed in the text, only the coefficient $c_{13}$ plays
a role in determining the coefficient $I_{3}$ of the increment $\gamma$.
The expression for $I_{3}$ is given by 
\begin{align*}
I_{3} & =\frac{n(m-n^{2})}{24(n+1)(m-1)^{2}}\frac{\text{Bo}\text{Ma}}{s}-\frac{(n-1)(m-n^{2})\phi}{32n^{2}(n+1)(m-1)^{3}}\frac{\text{Ma}^{2}}{s}\\
 & +\frac{2n^{2}(n+1)(m-1)(n^{4}(27-(n-3)n)-2mn^{2}(3+n(3n-17))+m^{2}(3n(9n+1)-1))}{15\psi^{2}}s,
\end{align*}
while $c_{13}$ is found as 
\[
c_{13}=-(2/5)(m-1)n^{2}(n^{3}+1)s.
\]

\section{Eigenfunctions: velocities and pressure\label{sec:Eigenfunctions}}

Provided $s\ne0$ the velocities are as follows. For the surfactant
branch, the real part of the (bottom-layer) horizontal velocity component
$u_{1}$, to its leading order $\alpha^{0}$, is 
\begin{equation}
\Real(u_{1})=-hs\frac{(m-1)}{(m-n^{2})}(z+1)(3z+1);\label{eq:re u1 surfactant}
\end{equation}
the imaginary part of $u_{1},$ to its leading order $\alpha^{1},$
is 
\begin{align}
\Imag(u_{1}) & =\alpha h\frac{1}{(m-n^{2})}(z+1)\left\{ \text{Ma}\frac{(n-1)}{2(m-1)n}\left[3m(n+1)(z-1)\right.\right.\nonumber \\
 & +\left.\left.4m+3mn+n^{3}\right]-\text{Bo}\frac{n^{2}}{6}(3z+1)\right\} .\label{eq:im u1 surf}
\end{align}
The real part of the (top-layer) velocity component $u_{2}$, to its
leading order $\alpha^{0}$, is 
\begin{equation}
\Real(u_{2})=-hs\frac{(m-1)}{m(m-n^{2})}(z-n)(3z-n);\label{eq:re u2 surf}
\end{equation}
(the imaginary part of $u_{2},$ to its leading order $\alpha^{1},$
is 
\begin{align}
\Imag(u_{2}) & =-\alpha h\frac{1}{(m-n^{2})}(z-n)\left\{ \textrm{Ma}\frac{(n-1)}{2(m-1)n^{2}}\left[-3(z+n)n(n+1)+(m+3n^{2}+4n^{3})\right]\right.\nonumber \\
 & \left.+\frac{\text{Bo}}{6}(3z-n)\right\} .\label{eq:im u2 surf}
\end{align}
(Note that $h=$ for $m=n^{2}$. For this case, the velocities can
be expressed in terms of $G$ rather than $h$, in the same way as
equations (\ref{eq:u1s0Bo}) and (\ref{eq:u2s0Bo}) below.)

For the robust branch, the real part of the (bottom-layer horizontal
velocity component) $u_{1},$ to its leading order $\alpha^{0},$
is 
\begin{equation}
\Real(u_{1})=-hs\frac{(m-1)}{\psi}(z+1)[3(m-n^{2})(z-1)+4(m+n^{3})];\label{eq:re u1 robust}
\end{equation}
the imaginary part of $u_{1},$ to its leading order $\alpha^{1},$
is 
\begin{align}
\Imag(u_{1}) & =\alpha h\frac{1}{2\psi}(z+1)\left\{ \textrm{Ma}\frac{\varphi}{(m-1)n}\left[3m(n+1)(z-1)+(4m+3nm+n^{3})\right]\right.\nonumber \\
 & \left.\phantom{\frac{\phantom{a}}{\phantom{b}}}+\textrm{Bo}n^{2}\left[(3m+4nm+n^{2})(z-1)+4m(n+1)\right]\right\} .\label{eq:im u1 robust}
\end{align}
The real part of $u_{2}$, to its leading order $\alpha^{0}$, is
\begin{equation}
\Real(u_{2})=-hs\frac{(m-1)}{m\psi}(z-n)[3(m-n^{2})(z+n)+4(m+n^{3})];\label{eq:re u2 robust}
\end{equation}
the imaginary part of $u_{2}$ its leading order $\alpha^{1},$ is
\begin{align}
\Imag(u_{2}) & =\alpha h\frac{1}{2\psi}(z-n)\left\{ \textrm{Ma}\frac{\varphi}{(m-1)n^{2}}\left[3n(n+1)(z+n)-(m+3n^{2}+4n^{3})\right]\right.\nonumber \\
 & +\left.\textrm{Bo}\left[-(m+4n+3n^{2})(z+n)+4(n+1)n^{2}\right]\right\} .\label{eq:im u2 robust}
\end{align}

For the special case $m=1$, the leading-order velocities are 
\begin{equation}
u_{1}=\pm\alpha^{1/2}(1-i)h(2\textrm{Ma}s\frac{n^{2}}{(n+1)^{3}(n-1)})^{1/2}(z+1)[3(z-1)+(n^{2}-n+4)],\label{eq:u1m=00003D00003D1}
\end{equation}
\begin{equation}
u_{2}=\mp\alpha^{1/2}(1-i)h(2\textrm{Ma}s\frac{1}{(n+1)^{3}(n-1)})^{1/2}(z-n)[-3n(z+n)+(4n^{2}-n+1)],\label{eq:u2m=00003D00003D1}
\end{equation}
with the upper/lower signs for the growing/decaying modes, respectively.
The next order corrections are 
\begin{equation}
u_{1}^{c}=\alpha ih\textrm{Bo}\frac{n^{2}}{6(n+1)^{3}(n-1)}(z+1)[3(z-1)(n^{2}+2n-1)+2(n^{2}+5n-2)],\label{eq:u1cm1}
\end{equation}
\begin{equation}
u_{2}^{c}=\alpha ih\textrm{Bo}\frac{1}{6(n+1)^{3}(n-1)}(z-n)[3(z+n)(-n^{2}+2n+1)+2n(2n^{2}-5n-1)].\label{eq:u2cm1}
\end{equation}
(Note that the corrections for the growing mode are the same as for
the decaying one.) Considering the special case $s=0$ and $\text{Bo}=0$,
(that implies that the leading-order disturbances vanish, along with
the base flow), the horizontal velocities are as follows. For the
surfactant branch, the bottom-layer horizontal velocity is 
\begin{equation}
u_{1}=i2h\frac{(m+n^{3})}{n(m-n^{2})\psi}(z+1)\left[3mn(n+1)(z-1)+4m+3mn+n^{3}\right]
\end{equation}
while for the top-layer, it is 
\begin{equation}
u_{2}=-i2h\frac{(m+n^{3})}{n(m-n^{2})\psi}(z-n)\left[-3n(n+1)(z+n)+m+3n^{2}+4n^{3}\right].
\end{equation}
For the robust branch, the horizontal velocity in the bottom-layer
is 
\begin{equation}
u_{1}=-i\frac{2h(m+n)n^{2}\alpha^{2}}{3(m-n^{2})\psi}(z+1)\left[3(n+1)(z-1)+4m+3mn+n^{3}\right]
\end{equation}
and for the top-layer, it is 
\begin{equation}
u_{2}=i\frac{2h(m+n)n\alpha^{2}}{3(m-n^{2})\psi}(z-n)\left[-3n(n+1)(z+n)+m+3n^{2}+4n^{3}\right].
\end{equation}

For the special case $s=0$ and $\text{Bo}\ne0$, the horizontal velocities
in terms of $h$ and $G$ are 
\begin{align}
u_{1} & =i\alpha\frac{1}{2\psi}(z+1)\left\{ -2\textrm{Ma}Gn\left[3m(n+1)(z-1)+(4m+3nm+n^{3})\right]\right.\nonumber \\
 & \left.+\textrm{Bo}hn^{2}\left[(3m+4nm+n^{2})(z-1)+4m(n+1)\right]\right\} \label{eq:u1s0Bo}
\end{align}
and 
\begin{align}
u_{2} & =i\alpha\frac{1}{2\psi}(z-n)\left\{ -2\textrm{Ma}Gn\left[3n(n+1)(z+n)-(m+3n^{2}+4n^{3})\right]\right.\nonumber \\
 & \left.\textrm{+Bo}h\left[-(m+4n+3n^{2})(z+n)+4(n+1)n^{2}\right]\right\} .\label{eq:u2s0Bo}
\end{align}
These velocities can be written in terms of $h$ only by using the
expression for $G$ in terms of $h$ from equation (\ref{eq:G/h for s=00003D00003D0 nonzero Bo}),
in which the two different values of $\gamma$ given by equation (\ref{eq:QuadEqnGamma})
correspond to the two different normal modes for the case $s=0$ and
$\text{Bo}\ne0$.

Using these horizontal velocities, the vertical velocities for the
two branches are obtained by integrating equation (\ref{eq:incompress}):
\begin{equation}
w_{j}(z)=-i\alpha\int_{n_{j}}^{z}u_{j}(z)dz,\label{eq:w in terms of u-1-1}
\end{equation}
(where $n_{1}=-1$ and $n_{2}=n$, defined above for equation (\ref{eq:uj vel})).
The pressures of the two branches can be readily obtained using equation
(\ref{eq:uj momentum}), which yields 
\begin{equation}
i\alpha p_{j}=m_{j}D^{2}u_{j}.\label{eq:p in terms of u-1-1}
\end{equation}

\end{document}